\documentclass{article}
\usepackage{amsthm,amsmath,amssymb,mathrsfs,a4wide,cite,rotating}
\usepackage[all]{xy}
\SelectTips {cm}{}

\DeclareMathOperator{\End}{End}
\DeclareMathOperator{\Spec}{Spec}
\DeclareMathOperator{\ch}{ch}
\DeclareMathOperator{\td}{td}
\DeclareMathOperator{\sign}{sign}
\DeclareMathOperator{\mult}{mult}
\DeclareMathOperator{\Rea}{Re}
\DeclareMathOperator{\Ima}{Im}
\DeclareMathOperator{\Hom}{Hom}
\DeclareMathOperator{\HHom}{\HH om}
\DeclareMathOperator{\Aut}{Aut}
\DeclareMathOperator{\Ext}{Ext}
\DeclareMathOperator{\EExt}{{\EE xt}}
\DeclareMathOperator{\Li}{Li}

\DeclareMathOperator{\Lic}{{\mathcal{L}}i}
\DeclareMathOperator{\Pic}{Pic}
\DeclareMathOperator{\Gr}{Gr}
\DeclareMathOperator{\rk}{rk}

\DeclareMathOperator{\Supp}{Supp}
\DeclareMathOperator{\Tr}{Tr}
\DeclareMathOperator{\Hilb}{Hilb}

\DeclareMathOperator{\Syst}{Syst}
\DeclareMathOperator{\Quot}{Quot}
\DeclareMathOperator{\coker}{coker}
\DeclareMathOperator{\im}{im}
\DeclareMathOperator{\Div}{Div}
\DeclareMathOperator{\Sing}{Sing}

\numberwithin{equation}{section} 

\begin{document}


\def\vev#1{{\langle #1 \rangle}}
\newcommand{\CC}{{\mathbb{C}}}
\newcommand{\DD}{\mathbb{D}}
\newcommand{\EE}{\mathbb{E}}
\newcommand{\FF}{\mathbb{F}}
\newcommand{\LL}{\mathbb{L}}
\newcommand{\RR}{\mathbb{R}}
\newcommand{\QQ}{\mathbb{Q}}
\newcommand{\ZZ}{\mathbb{Z}}
\newcommand{\HH}{\mathbb{H}}
\newcommand{\NN}{\mathbb{N}}
\newcommand{\PP}{\mathbb{P}}

\newcommand{\abs}[1]{\lvert#1\rvert}
\newcommand{\Abs}[1]{\lVert#1\rVert}
\newcommand{\inv}[1]{\frac{1}{#1}}
\newcommand{\vir}{{\chi_{t,\tilde{t}}}}

\theoremstyle{plain}
\newtheorem{definition}[equation]{Definition}
\newtheorem{lem}[equation]{Lemma}
\newtheorem{prop}[equation]{Proposition}
\newtheorem{thm}[equation]{Theorem}
\newtheorem{conj}[equation]{Conjecture}
\newtheorem{cor}[equation]{Corollary}

\theoremstyle{remark}
\newtheorem{rem}[equation]{Remark}
\newtheorem{ex}[equation]{Example}


\title{\bf String Partition Functions and Infinite Products}

\author{
     Toshiya Kawai\\
  {\it Research Institute for Mathematical Sciences,}\\ 
  {\it Kyoto University, Kyoto 606--8502, Japan}\\ 
  {\it toshiya@kurims.kyoto-u.ac.jp} 
\and  
     K{\= o}ta Yoshioka\\
  {\it  Department of Mathematics, Faculty of Science,}\\
  {\it  Kobe University, Kobe 657-8501, Japan}\\
  {\it yoshioka@math.kobe-u.ac.jp}}
  
\date{}
\maketitle

\begin{abstract}
  We continue to explore the conjectural expressions of the
  Gromov-Witten potentials for a class of elliptically and $K3$
  fibered Calabi-Yau 3-folds in the limit where the base $\PP^1$ of
  the $K3$ fibration becomes infinitely large.  At least in this limit
  we argue that the string partition function (= the exponential
  generating function of the Gromov-Witten potentials) can be
  expressed as an infinite product in which the K{\" a}hler moduli and
  the string coupling are treated somewhat on an equal
  footing. Technically speaking, we use the exponential lifting of a
  weight zero Jacobi form to reach the infinite product as in the
  celebrated work of Borcherds. However, the relevant Jacobi form is
  associated with a lattice of Lorentzian signature. A major part of
  this work is devoted to an attempt to interpret the infinite
  product or more precisely the Jacobi form in terms of the bound
  states of $D2$- and $D0$-branes using  a vortex description
  and its suitable generalization.  
\end{abstract}
\bigskip

\section{Introduction}

The Gromov-Witten invariants and their potentials have been vigorously
investigated in recent years mainly due to their mathematical
soundness. See \cite{KM,BM,BF,Beh,LT} for their fundamental
properties.  However the Gromov-Witten potentials emerge somewhat
indirectly in the conventional physical approaches. Indeed, for
Calabi-Yau 3-folds, it is believed \cite{BCOV} that they should appear
in the ``topological limits'' of the naturally defined closed
topological A string amplitudes the explicit evaluations of which are
prohibitively difficult in general.

In the tests of heterotic/type IIA string duality conjectures, it was
desirable to develop the one-loop calculation scheme on the heterotic
string side to extract the objects which might correspond to the
Gromov-Witten potentials on the type IIA string side.  In the
pioneering work of Harvey and Moore \cite{HM1} this task was taken up
and certain integrals involving indefinite theta functions were
explicitly evaluated on the heterotic string side extending the
calculation in \cite{DKL}. In the course of the calculations they
curiously pointed out the relevance of Borcherds' work \cite{Bo1} on
holomorphic infinite products.  The Harvey-Moore method has revealed
the presence of a new interesting subject on the theta correspondence
and has an advantage when discussing automorphic properties.  However
several steps were necessary \cite{HM1} in order to extract the
candidate of the genus zero Gromov-Witten potential from the evaluated
integral.  Recently, the method was extended \cite{MM} to cover the
Gromov-Witten potentials in higher genera for a particular model using
the result of \cite{Bo2} which was itself the extension of the
calculations in \cite{HM1}.  In this case also it was necessary to
take the limit of a relatively  complicated expression to obtain the
candidates of the Gromov-Witten potentials.

Another approach to investigate some features of the Gromov-Witten
potentials of Calabi-Yau 3-folds has been advocated by Gopakumar and
Vafa \cite{GV2} using an $M$-theory interpretation and there have been
some related works \cite{HST,KlZa}.

If the Gromov-Witten potentials are of our sole concern, are there any
possibilities in which we might directly reach their expressions in
all genera?  The previous work \cite{Kaw2} as well as the present one
attempt, albeit in a conjectural and limited sense, to answer this
question in the affirmative for a class of elliptically and $K3$
fibered Calabi-Yau 3-folds in the limit where the base $\PP^1$ of the
$K3$ fibration becomes infinitely large.  In \cite{Kaw2} we tried to
interpret the genus $g$ Gromov-Witten potential in terms of the
lifting of a Jacobi form of weight $2g-2$ so that it can be expressed
in terms of the ``polylogarithm'' $\Li_{3-2g}(\xi)$. There the cases
of genus zero and one were discussed in detail while the higher genus
cases were briefly speculated upon in the concluding section.  The
present work further pursues this line of interpretation.  Our basic
strategy is simple: rather than dealing with the Gromov-Witten
potentials individually we consider the {\it string partition
  function}
\begin{equation}\label{stpf}
  \mathcal{Z}=\exp\left(\sum_{g=0}^\infty x^{2g-2}F_g\right)\,,
\end{equation}
where $F_g$ is the genus $g$ Gromov-Witten potential of the fibered
Calabi-Yau 3-fold $Y$ and $x$ is the string coupling parameter.  We
argue that, in the pertinent limit, $\mathcal{Z}$ can be constructed
by the exponential lifting \cite{Bo1} of a weight zero Jacobi form
associated with a lattice of {\it Lorentzian\/} signature.  Indeed
this construction solves the problem at one blow : $F_g$ can be
expressed as the lifting of a weight $2g-2$ ({\it quasi}) Jacobi form,
thus making the statement in \cite{Kaw2} precise.

More intriguingly and perhaps more significantly, the construction
indicates that $\mathcal{Z}$ can be put (at least in the limit we
consider) into an infinite product which resembles the
Weyl-Kac-Borcherds denominator. As in \cite{Bo1} the most subtle point
in this story is to determine the ``Weyl vector'' which, we find,
should be interpreted as the constant map contributions of the genus
zero and one Gromov-Witten potentials. However we have already
discussed this technically involved problem in \cite{Kaw2} via a
felicitous use of elliptic polylogarithms \cite{BL}.
In fact, one of the motivations for \cite{Kaw2} and the present work
was a desire to better understand the relation between the
Gromov-Witten potentials of the fibered Calabi-Yau 3-folds and the
{\it original} lifting approach of Borcherds \cite{Bo1}.

In our construction and the resulting infinite product representation,
it turns out to be natural to view $\mathcal{Z}$ as a function (or
possibly a section of the appropriate vacuum line bundle) on the
``extended moduli space'' whose tangent space is some domain of
$H^2(Y,\CC)\oplus H^0(Y,\CC)$.  The extended moduli space unifies the
complexified K{\" a}hler moduli and the string coupling constant and
it is natural from the philosophy of ``brane democracy''.  It is also
an appropriate setting for the homological mirror conjecture
\cite{Kon1}.  Thus we should like to have an interpretation of our
proposal in terms of the bound states of $D2$-branes and
$D0$-branes. (In type IIA string theory on Calabi-Yau 3-folds, $D6$-,
$D4$-branes are electro-magnetic duals of $D0$-, $D2$-branes.)  In
this paper we will make some preliminary (and admittedly modest)
efforts toward justification of such an interpretation.  In
particular, we argue that the bound states of a single $D2$-brane and
$D0$-branes are described by abelian vortices and their suitable
generalizations.  We use this interpretation to understand some of the
key expressions. In fact, we are able to give a relatively detailed
and precise description when the $D2$-$D0$ bound system is in a fixed
$K3$ surface.  In such a case, we also point out that the bound state
problem of a $D2$ brane and $D0$-branes is closely related to vertex
operators and their two-point correlation functions.

Presumably the benefit of the lifting procedure employed in this work
resides in the very possibility that we may link together, in a rather
explicit way, the string perturbative theory of the Gromov-Witten
invariants (which is certainly not brane-democratic but relatively
well-understood) and the inherently non-perturbative viewpoint of
$D2$-$D0$-branes about which we have yet to learn more.

The organization of this paper is as follows.  In \S 2, we review the
fundamental properties of the Gromov-Witten potentials for Calabi-Yau
3-folds. In \S 3 we first recall the general definitions and
properties of Jacobi forms as well as those of the Hecke operators.
Then we consider the lifting procedure for a class of weight zero
Jacobi forms associated with certain Lorentzian lattices and discuss
its relation to infinite products.  In \S 4 we give the main
conjecture about the string partition functions of the fibered
Calabi-Yau 3-folds. In \S 5 we attempt to interpret the proposed
expression of the string partition function in terms of the bound
states of $D0$- and $D2$-branes. As mentioned above, we devote most of
this section to the case where the bound system of a single $D2$-brane
and collections of $D0$-branes is in a $K3$ surface.  Technically the
results in \cite{Yos1} turn out to be useful.  In \S 6 we discuss the
relevance of vertex operators and their two-point functions to the
$D2$-$D0$ bound state problem.  As a simple application of our
proposal, we study in \S 7 the behavior of the string partition
function near the conifold point and relate it to the $SU(\infty)$
Chern-Simons theory on $S^3$ thus reproducing the earlier obtained
results \cite{Wit4,P,JP,GV1,GV2,GV3}.  In \S 8 we raise some
directions for further investigations.  Several definitions of the
functions used in this work and their necessary properties are
summarized in Appendix A while Appendix B discusses a conjectural formula
of the elliptic genera of the higher order Kummer varieties introduced
in \cite{Bea0}.

\smallskip

While pursuing the subject of this paper, a paper \cite{KKV} appeared
in which  the authors discuss some relevance of the relative Hilbert
schemes in conjunction with the proposal of \cite{GV2}. In our
approach the relative Hilbert schemes appear naturally in the
$D2$-$D0$-brane bound state interpretation.

\medskip

Part of this work was presented at the 1998 Kinosaki Symposium on
Algebraic Geometry and thenceforth repeated on several occasions. We
are grateful to Max-Planck-Institut f{\" u}r Mathematik in Bonn for
hospitality.  T.K thanks the organizers of the workshop of Activity
``Automorphic Products'' during which he benefited from
conversations with R. Borcherds, R.  Dijkgraaf, V.A. Gritsenko,
L. G{\" o}ttsche, S. Kondo, V.V. Nikulin, K. Saito and K. Yoshikawa.
We also thank M.-H. Saito for discussions.

\bigskip 

\vbox{{\noindent \bf Notation.}

${\bf e}[x]=\exp(2\pi \sqrt{-1} x)$.

$\ZZ_+$: the set of positive  integers.

$\ZZ_-$: the set of negative  integers.

$\NN$: the set of non-negative integers. 
 
$\HH_g$: the Siegel upper space of degree $g$. }

\section{The Gromov-Witten potentials of Calabi-Yau 3-folds}

The Gromov-Witten invariants have been extensively studied in recent
years. For the fundamental properties established so far we refer to
\cite{KM,BM,BF,Beh,LT}.  In this section we review the relevant
materials in the cases of Calabi-Yau 3-folds for later convenience.

\subsection{The Gromov-Witten invariants}

Let $Y$ be a smooth Calabi-Yau 3-fold, {\it i.e.\/} a smooth
3-dimensional projective variety over $\CC$ with $c_1(Y)=0$ and
$h^{1,0}(Y)=h^{2,0}(Y)=0$ where $h^{p,q}(Y)=\dim H^q(Y,\Omega_Y^p)$.
Hence $\Pic(Y)\cong H^2(Y,\ZZ)$ and
$\chi(Y)=2(h^{1,1}(Y)-h^{1,2}(Y))$.  We assume that $H^2(Y,\ZZ)$ is
torsion-free.  Suppose that $\omega_1,\ldots,\omega_l$ generate
$H^2(Y,\ZZ)$ where $l=h^{1,1}(Y)$. Let $D_1,\ldots, D_l$ be divisors
such that $\omega_i=c_1(\mathcal{O}_Y(D_i))$ for $i=1,\dots,l$.  Let
$\iota_i:D_i\hookrightarrow Y$ be the inclusions. Then
$\omega_i\cap[Y]=(\iota_i)_*[D_i]$ where $[\#]$ stands for the
fundamental homology class of $\#$.  We assume that $D_1,\ldots, D_l$
are nef so that $\omega_i \cap\iota_*[C]\ge 0$, $(i=1,\dots,l)$ for
any algebraic curve $C\subset Y$ with the inclusion
$\iota:C\hookrightarrow Y$.

Let $\overline{\mathcal{M}}_{g,n}(Y,\beta)$ be the moduli stack of
stable maps where $g$, $n\ge 0$ and $\beta \in H_2(Y,\ZZ)$.  An
element of $\overline{\mathcal{M}}_{g,n}(Y,\beta)$ is represented by
$(\Sigma_g,p_1,\ldots,p_n,\varphi)$. Here $\Sigma_g$ is a connected
curve of arithmetic genus $g=\dim
H^1(\Sigma_g,\mathcal{O}_{\Sigma_g})$ whose only possible
singularities are ordinary double points while $p_1,\ldots,p_n$ are
distinct nonsingular points on $\Sigma_g$.  The last entry is a
morphism $\varphi:\Sigma_g\rightarrow Y$ such that $\{ \mu\in \Aut
\Sigma_g \mid \varphi \circ \mu=\varphi,\ \mu(p_i)=p_i\}$ is finite
and $\varphi_*[\Sigma_g]=\beta$.

Let
\begin{equation}
  \pi: \overline{\mathcal{C}}_{g,n}(Y,\beta)\rightarrow
  \overline{\mathcal{M}}_{g,n}(Y,\beta)\,,
\end{equation}
be the universal curve over
$\overline{\mathcal{M}}_{g,n}(Y,\beta)$. We have
$\overline{\mathcal{C}}_{g,n}(Y,\beta)=
\overline{\mathcal{M}}_{g,n+1}(Y,\beta)$.  Set
\begin{equation}
  \begin{matrix}
   f:& \overline{\mathcal{C}}_{g,n}(Y,\beta) &\longrightarrow &Y\\
  &(\Sigma_g,p_1,\ldots,p_{n+1},\varphi)&\longmapsto &\varphi(p_{n+1})\,.
\end{matrix}
\end{equation}
The virtual dimension of
$\overline{\mathcal{M}}_{g,n}(Y,\beta)$ is  often smaller than the actual
dimension of $\overline{\mathcal{M}}_{g,n}(Y,\beta)$.  The virtual
fundamental class
$[\overline{\mathcal{M}}_{g,n}(Y,\beta)]^{\text{vir}}$ can be
constructed so that its dimension coincides with the virtual dimension
of $\overline{\mathcal{M}}_{g,n}(Y,\beta)$ \cite{BM,BF,Beh,LT}. This
construction uses the obstruction sheaf $R^1\pi_*f^* {T}_Y$, where $
{T}_Y$ is the tangent sheaf of $Y$, and is given by
\begin{equation}
  [\overline{\mathcal{M}}_{g,n}(Y,\beta)]^{\text{vir}} = e(R^1\pi_*f^*
  {T}_Y) \cap [\overline{\mathcal{M}}_{g,n}(Y,\beta)]\,,
\end{equation}
if $R^1\pi_*f^* {T}_Y$ is locally-free. Here $e(\ )$
represents the Euler class. Intuitively, $e(R^1\pi_*f^*
{T}_Y)$ represents the contribution from the anti-ghost zero
modes.

For a Calabi-Yau 3-fold $Y$, the virtual
dimension of $\overline{\mathcal{M}}_{g,n}(Y,\beta)$ is equal to
$n$. Using the evaluation maps
\begin{equation}
  \begin{matrix}
 \text{ev}_i:&\overline{\mathcal{M}}_{g,n}(Y,\beta)&\longrightarrow &Y\\
&(\Sigma_g,p_1,\ldots,p_n,\varphi)&\longmapsto &\varphi(p_i)\,,
\end{matrix}
\end{equation}
the Gromov-Witten invariants are introduced by
\begin{equation}
  \langle \omega_{i_1}\cdots \omega_{i_n}\rangle_{g,\beta} = (
  \text{ev}^*_1(\omega_{i_1})\cup \cdots \cup
  \text{ev}^*_n(\omega_{i_n})) \cap
  [\overline{\mathcal{M}}_{g,n}(Y,\beta)]^{\text{vir}}\,.
\end{equation}
We extend the Gromov-Witten invariants by $\CC$-linearity:
\begin{equation}
\langle t_{i_1}\omega_{i_1}\cdots t_{i_n}\omega_{i_n}\rangle_{g,\beta} =
t_{i_1}\cdots t_{i_n}\langle \omega_{i_1}\cdots
 \omega_{i_n}\rangle_{g,\beta}\,,
\end{equation}
for $t_{i_1},\ldots, t_{i_n}\in \CC$.

\subsection{The Gromov-Witten potentials and their known general 
properties}

If we write $\omega=\sum_i{t_i\omega_i}\in \mathcal{K}_\CC \subset
H^2(Y,\CC)$ where $\mathcal{K}_\CC$ is the complexified K{\" a}hler
cone, the Gromov-Witten invariants can be compactly organized into the
Gromov-Witten potentials:
\begin{equation}
    F_g=\sum_{\beta\in H_2(Y,\ZZ)}\langle e^\omega\rangle_{g,\beta}\,,
\end{equation}
since
\begin{equation}
  F_g=\sum_{\beta}\sum_{n\geq 0}\frac{1}{n!}\langle \omega^n\rangle_{g,\beta}=
  \sum_{\beta} \sum_{n\geq 0}\sum_{i_1,\ldots,i_n} \frac{t_{i_1}\cdots
    t_{i_n}}{n!}\langle \omega_{i_1}\cdots \omega_{i_n}\rangle_{g,\beta}\,.
\end{equation}

By the fundamental property of topological sigma models or the Divisor
Axiom \cite{KM}, it follows that
\begin{equation}
  \langle \omega^n\rangle_{g,\beta}=(\omega \cap \beta)^n
\langle 1\rangle_{g,\beta}\,,
\end{equation}
for $\beta \not = 0$.
Hence we have
\begin{equation}\label{Fg}
  F_g = \langle e^\omega\rangle_{g,0}+
\sum_{\beta \not =0}\langle 1\rangle_{g,\beta}e^{\omega
    \cap \beta}\,.
\end{equation}  

Let $C_0\subset Y$ be a rigid smooth rational curve $C_0\subset Y$
with normal bundle ${N}=\mathcal{O}_{C_0}(-1)\oplus
\mathcal{O}_{C_0}(-1)$.  Fix a positive integer $h$. Let
$p:\overline{\mathcal{C}}_{g,0}(C_0,h[C_0]) \to
\overline{\mathcal{M}}_{g,0}(C_0,h[C_0])$ be the universal curve and
$\mu:\overline{\mathcal{C}}_{g,0}(C_0,h[C_0]) \to C_0$ the universal
evaluation map.
  It was conjectured in \cite{GP2} and
proved\footnote{See also \cite{AM,M,BCOV}.} in \cite{FP} that the
multiple covering effect of $C_0$ 
can be summarized by
\begin{equation}
  e(R^1p_*\mu^*{N})\cap
  [\overline{\mathcal{M}}_{g,0}(C_0,h[C_0])]^{\text{vir}}= m_g\,
  h^{2g-3}\,,
\end{equation}
where $m_g$ are the rational numbers defined through
\begin{equation}\label{g-to-zero}
  (y^{-1/2}-y^{1/2})^{-2}= -\sum_{g=0}^\infty x^{2g-2}\,m_g\,,\qquad
y=\exp(\sqrt{-1} x)\,.
\end{equation}
Explicitly we have $m_0=1$, $m_1=\frac{1}{12}$ and in general
\begin{equation}
  m_g=\frac{(-1)^{g-1}(2g-1) B_{2g}}{(2g)!}\,.
\end{equation}
For $g>1$, it follows that 
\begin{equation}
  m_g=\frac{(-1)^{g-1}\chi_{g,0}}{(2g-3)!}\,,
\end{equation}
where  we use the formula of the orbifold Euler characteristic of the
moduli space of genus $g(>1)$ curves with $n$ punctures \cite{HZ}:
\begin{equation}
  \chi_{g,n}=(-1)^n\binom{2g-3+n}{n}\frac{B_{2g}}{2g(2g-2)}\,.
\end{equation}

Therefore the multiple coverings of rational curves should contribute
to the second term on the right hand side of \eqref{Fg} in the form
\begin{equation}
  \sum_{\beta' \not = 0} \sum_{h>0}\langle 1\rangle_{0,\beta'} m_g h^{2g-3}
  e^{\omega\cap(h\beta')} = 
\sum_{\beta' \not =  0} \langle 1\rangle_{0,\beta'}m_g
  \Li_{3-2g}( e^{\omega\cap\beta'})\,,
\end{equation}
where the ``polylogarithm'' function $\Li_{3-2g}(\xi)$ is defined in
Appendix A.

The evaluation of the constant map contribution $\langle
e^\omega\rangle_{g,0}$ has been explicitly performed in the literature
\cite{BCOV,GP,KM2}. We briefly recall this.  By the isomorphism
\begin{equation}
  \overline{\mathcal{M}}_{g,n}(Y,0)\cong
  \overline{\mathcal{M}}_{g,n}\times Y\,,
\end{equation}
we have $\pi=\tilde \pi \times \text{id}$ with the universal curve
$\tilde \pi: \overline{\mathcal{C}}_{g,n}\rightarrow
\overline{\mathcal{M}}_{g,n}$. Set $\EE=\tilde
\pi_*\omega_{\overline{\mathcal{C}}_{g,n}/\overline{\mathcal{M}}_{g,n}}$
where
$\omega_{\overline{\mathcal{C}}_{g,n}/\overline{\mathcal{M}}_{g,n}} $
is the relative dualizing sheaf \cite{Mu}.  Thus $\EE^*= R^1 \tilde
\pi_*\mathcal{O}_{\overline{\mathcal{C}}_{g,n}}$ by duality and it
follows that $R^1\pi_*f^* {T}_Y\cong \EE^*\boxtimes
{T}_Y$.  Consequently, we have
\begin{equation}
  [\overline{\mathcal{M}}_{g,n}(Y,0)]^{\text{vir}}=c_{g\dim(Y)}(\EE^*
  \boxtimes {T}_Y)\cap [\overline{\mathcal{M}}_{g,n}(Y,0)]\,,
\end{equation}
where we used $\rk(\EE)=g$.  Then the evaluation of $\langle
e^\omega\rangle_{g,0}$ reduces to the Hodge integrals, {\it i.e.} the
integrals over $\overline{\mathcal{M}}_{g,n}$ of cup products of the
Chern classes $\lambda_i:=c_i(\EE)$.

Set $\mathbb{S}:={\NN}^{l}\smallsetminus \{0\}$. We regard
$\mathbb{S}$ as a poset by the partial ordering: $d'\le d$ $ (d,d' \in
\mathbb{S})$ iff $d_i' \mid d_i$ $ ({}^\forall i)$.  Let us introduce
new variables $q_1=e^{t_1},\ldots,q_l=e^{t_l}$.  If
$d=(d_1,\ldots,d_l)\in \mathbb{S}$ we write $q^d$ for $q_1^{d_1}\cdots
q_l^{d_l}$.  We also introduce
\begin{equation}
  \begin{split}
    \kappa_{ijk}&=D_i\cdot D_j \cdot D_k = (\omega_i \cup \omega_j
    \cup \omega_k)\cap [Y]\,,\\
    \rho_i&=c_2(Y)\cdot D_i=c_2(Y)\cap
    (\iota_i)_*[D_i]=(c_2(Y)\cup \omega_i)\cap [Y]\,.
  \end{split}
\end{equation}
Then combining the above results the Gromov-Witten potentials were
found to have the following expressions:
\begin{align}
  F_0 &= \frac{1}{3!}\sum_{i,j,k} \kappa_{ijk} t_i t_j t_k -
  \frac{\chi(Y)}{2} \zeta(3) + \sum_{d\in \mathbb{S}}  N_0(d)\, m_0 
  \Li_3(q^d)\,,\\
  F_1 &=-\lambda_1 \cap[\overline{\mathcal{M}}_{1,1}]\cdot\sum_i \rho_i
  t_i+ \sum_{d\in \mathbb{S}}\Bigl[ N_0(d)\, m_1+ \sum_{\substack{d'\in
      {\mathbb{S}}\\ d'\le d}} N_1(d')\Bigr] \Li_1(q^d)\,,
\end{align}
and
\begin{equation}
  F_g=(-1)^g \lambda_{g-1}^3 \cap [\overline{\mathcal{M}}_{g,0}] \cdot
  \frac{\chi(Y)}{2}+\sum_{d\in \mathbb{S}}
  N_0(d)\, m_g\Li_{3-2g}(q^d)+\cdots\,,
\end{equation}
for $g>1$. The coefficients $N_0(d)$ and $N_1(d)$ count the primitive
numbers of rational and elliptic curves.

\smallskip 
\begin{rem} 
  In $F_0$ we have inserted the term $-\frac{\chi(Y)}{2} \zeta(3)$ by
  hand. This term seems to lack a satisfactory explanation in the pure
  context of the Gromov-Witten theory but, as well-known, its
  existence has been supported from other approaches. Since
  $\Li_3(\xi)$ and $\Li_1(\xi)$ are multi-valued functions with
  non-trivial monodromy groups (see Appendix A and \cite{Kaw2} for a
  summary) we neglected the terms that can be cancelled by monodromy
  transformations in the expressions of $F_0$ and $F_1$. Recall that
  $\zeta(3)$ is irrational so that $-\frac{\chi(Y)}{2} \zeta(3)$
  cannot be cancelled by a monodromy transformation.
\end{rem}

A basic result due to Mumford \cite{Mu} is:
\begin{equation}
   \lambda_1 \cap[\overline{\mathcal{M}}_{1,1}] =\frac{1}{24}\,.
\end{equation}
Another important result is:
\begin{equation}
   \lambda_{g-1}^3 \cap [\overline{\mathcal{M}}_{g,0}]
=(-1)^{g-1}m_g\, \zeta(3-2g)\,, \qquad (g>1)\,.
\end{equation}
This equation (rewritten in an equivalent form) was conjectured in
\cite{F} and recently proved in \cite{FP}.  See also \cite{MM}
\cite{GV2} for physical justification.

Thus we have seen that $F_g$ contains the constant term proportional to
$\zeta(3-2g)$ and is related to the function $\Li_{3-2g}(\xi)$. (For
$F_1$ we have not considered the term proportional to $\zeta(1)$ since
$\zeta(1)$ is divergent. However, as we will see later, its formal
presence may be preferred from some aesthetic viewpoint.)  In the
following we will see that these features of $F_g$ are indeed realized
in our conjectural  expressions.

\section{Jacobi forms and their liftings}

The purpose of this section is to collect together some fundamental
materials of Jacobi forms whose properties are indispensable for our
construction. In the simplest case a systematic study of Jacobi forms
was initiated in \cite{EZ}. A straightforward extension of \cite{EZ}
leads to the idea of Jacobi forms associated with positive definite
lattices. However, for our present purpose, it is necessary to
consider Jacobi forms associated with lattices of Lorentzian
signature. We note that such possibilities have already been
considered in \cite{GZ} in the context of the Donaldson invariants for
4-manifolds with $b_2^+=1$.

\subsection{Jacobi forms}

Let $(\Pi,\langle\ ,\ \rangle)$ be an even integral lattice, {\it
  i.e.\/} a free $\ZZ$-module $\Pi$ of finite rank endowed with a
symmetric non-degenerate bilinear form $\langle\ ,\ \rangle:\Pi\times
\Pi \rightarrow \ZZ$ satisfying $\langle
\boldsymbol{\lambda},\boldsymbol{\lambda}\rangle\in 2\ZZ$ for all
$\boldsymbol{\lambda}\in \Pi$. Note that we allow $\Pi$ to be
indefinite. As is customary, we write $\Pi$ instead of $(\Pi,\langle\ ,
\ \rangle)$ when the bilinear form is known from the context. We also
write $\Pi(r)$ for $(\Pi,r\langle\ ,\ \rangle)$ where $r \in
\QQ$.  The bilinear form $\langle\ ,\ \rangle$ determines the
canonical embedding $\Pi \subset \Pi^*=\Hom_\ZZ(\Pi,\ZZ)$. By
extending $\langle\ ,\ \rangle$ via $\QQ$-linearity we can regard
$\Pi^*$ as a rational lattice.  We also identify $\Pi_{\CC}$ with $
\Pi^*_{\CC}$ by extending $\langle\ ,\ \rangle$ via $\CC$-linearity.
Given a nonzero rational number $r$, let $\langle r\rangle$ denote the
rank 1 lattice $(\ZZ e,\langle\ ,\ \rangle)$ with the generator $e$
satisfying $\langle e,e\rangle=r$.

We assume that $\Pi^*$ is such that any element of it is either
positive, zero or negative.

\begin{definition}
  A triplet $(\ell,n,\boldsymbol{\gamma})\in \ZZ\times \ZZ\times
  \Pi^*$ is said to be {\em positive} if either of the following three
  cases holds:
\begin{center}
  $ \mathrm{(i)}$ $\ell>0$, $\ 
\mathrm{(ii)}$ $\ell=0\,,\ n>0$,  $\ \mathrm{(iii)}$ $\ell=n=0\,,\ 
\boldsymbol{\gamma}>0$.
\end{center}
We write $(\ell,n,\boldsymbol{\gamma})>0$ if
$(\ell,n,\boldsymbol{\gamma})$ is positive.
\end{definition}

\begin{definition}\label{def:Jacobi}
  A {\em Jacobi form} of weight $k \in \ZZ$ associated with
  $\Pi=(\Pi,\langle\ ,\ \rangle)$ is a meromorphic function
  $\Phi_k:\HH_1\times \Pi_\CC\rightarrow \CC$ satisfying
\begin{enumerate}
\item  For any  $\bigl(\begin{smallmatrix}
    a&b\\c&d
  \end{smallmatrix}\bigr)\in SL_2(\ZZ)$,
  \begin{equation}\label{modular}
    \Phi_k\left(\frac{a\tau+b}{c\tau+d},
      \frac{\boldsymbol{z}}{c\tau+d}\right) = (c\tau+d)^k {\bf
      e}\left[{\frac{c\langle  \boldsymbol{z},
          \boldsymbol{z}\rangle}{2(c\tau+d)}}\right] \Phi_k(\tau,
    \boldsymbol{z})\,.
\end{equation}

\item  For any $ \boldsymbol{\lambda},  \boldsymbol{\mu}\in  \Pi$,
  \begin{equation}
    \Phi_k(\tau, \boldsymbol{z}+ \boldsymbol{\lambda} \tau+
    \boldsymbol{\mu})={\bf e}\left[- \left( \frac{ \langle
          \boldsymbol{\lambda}, \boldsymbol{\lambda}\rangle}{2}\tau +
        \langle \boldsymbol{\lambda}, \boldsymbol{z} \rangle \right)
    \right]\Phi_k(\tau, \boldsymbol{z})\,.
\end{equation}

\item  $\Phi_k$ can be Fourier-expanded in some appropriate
region of $\HH_1\times \Pi_\CC$ as
\begin{equation}\label{Jacobiexp}
  \Phi_k(\tau, \boldsymbol{z}) = \sum_{\substack{n \ge-n_0\\ 
      \boldsymbol{\gamma}\in \Pi^*}}D(n,\boldsymbol{\gamma})\, q^n
  \boldsymbol{\zeta}^{ \boldsymbol{\gamma}}\,,
\end{equation}
where $n_0$ is some non-negative integer and we have introduced the
notation $q= {\bf e}[\tau]$ and
$\boldsymbol{\zeta}^{\boldsymbol{\gamma}} = {\bf
  e}[\langle \boldsymbol{\gamma},\boldsymbol{z}\rangle]$. 
\end{enumerate}
\end{definition}

\begin{rem}
Since $\Phi_k(\tau, -\boldsymbol{z})=(-1)^k \Phi_k(\tau, \boldsymbol{z})$,
we have $D(n,-\boldsymbol{\gamma})=(-1)^k D(n,\boldsymbol{\gamma})$. 
\end{rem}

\begin{definition}
  Suppose that $(\Pi,\langle\ ,\ \rangle)$ is positive definite. Then
  $\Phi_k$ in {\em Definition \ref{def:Jacobi}} is said to be {\em nearly
    holomorphic} if $n_0>0$ while it is said to be {\em weak} if
  $n_0=0$.
\end{definition}

Let $\mathfrak{g}$ be a simple Lie algebra of rank $s$ with a fixed
Cartan subalgebra $\mathfrak{h}$ and $W(\mathfrak{g})$ the Weyl group
of $\mathfrak{g}$. We identify $\mathfrak{h}$ with $\mathfrak{h}^*$
using the Killing form $(\ ,\ ) $. We extend $(\ ,\ ) $ by
${\CC}$-linearity.  We normalize the highest root $\theta$ as
$(\theta,\theta)=2$. Let $Q^\vee=(Q^\vee, (\ ,\ ))$ be the coroot
lattice of $\mathfrak{g}$. Then $Q^\vee$ is a positive definite even
integral lattice of rank $s$ and $P=(Q^\vee)^*$ is the weight lattice
of $\mathfrak{g}$.  With this data we used in \cite{Kaw2} the notion
of Weyl-invariant Jacobi forms following \cite{Wir}:
\begin{definition}
  A {\em Weyl-invariant\/} Jacobi form $\phi_{k,m}$ of weight $k$ and
  index $m$ is a Jacobi form of weight $k$ associated with the lattice
  $Q^\vee(m)$ in the sense of {\em Definition \ref{def:Jacobi}} such
  that it is invariant under the action of $W(\mathfrak{g})$ on
  $Q^\vee(m)_\CC$.
\end{definition}

We note that a weak Jacobi form of even weight in the sense of
\cite{EZ} is a weak Weyl-invariant Jacobi form of $\mathsf{A}_1$.

Let
\begin{equation}
 E_{2k}(\tau)=1-\frac{4k}{B_{2k}}\sum_{n=1}^\infty
  \sigma_{2k-1}(n)q^n,\quad (k\ge 1)\,,
\end{equation}
denote the normalized Eisenstein series of weight $2k$ where
$\sigma_k(n)=\sum\limits_{d \mid n}d^k$.

\begin{definition}
  A meromorphic function on $\HH_1\times \Pi_\CC$ is called a {\em
    quasi} Jacobi form of weight $k$ associated with $\Pi$ if it is
  expressed for some integer $k_0$ as
  $\sum_{k'=k_0}^kp_{k-k'}(E_2,E_4,E_6)\Phi_{k'}$ where $\Phi_{k'}$ is
  a Jacobi form of weight $k'$ associated with $\Pi$ and
  $p_{k-k'}(E_2,E_4,E_6)\in \CC[E_2,E_4,E_6]$ is a quasi modular form
 {\em \cite{KZ}} of weight $k-k'$.
\end{definition}

\subsection{Hecke operators and liftings}
In this section we assume that $\Phi_k$ is a quasi Jacobi form of
weight $k$ associated with an even integral lattice $\Pi$ having
Fourier expansion \eqref{Jacobiexp}.

\begin{definition}
  For $\ell=1,2,\dots$ the action of the  Hecke operator $V_\ell$
  on  $\Phi_k$ is defined, as in {\em
    \cite{EZ}}, by
\begin{equation}\label{Heckedef}
  \Phi_k\vert_{ V_\ell} (\tau,\boldsymbol{z}) :=
  \ell^{k-1}\sum_{\substack{ad=\ell\\ a>0}} \sum_{b=0}^{d-1} d^{-k}
  \Phi_k\left(\frac{a\tau+b}{d},a \boldsymbol{z}\right)\,.
  \end{equation}
\end{definition}

The following relation has already been used in \cite{Kaw2}:
  \begin{lem}
\begin{equation}\label{Heckesum}
  \sum_{\ell=1}^\infty p^\ell  \Phi_k\vert_{V_\ell}
  (\tau,\boldsymbol{z}) = 
\sum_{\substack{\ell,n,\boldsymbol{\gamma}\\ \ell>0}} 
D(\ell n,\boldsymbol{\gamma})
  \Li_{1-k}(p^\ell q^n\boldsymbol{\zeta}^{\boldsymbol{\gamma}})\,.
\end{equation}
\end{lem}
\begin{proof}
From the definition \eqref{Heckedef} the left hand side is equal to
  \begin{equation}
    \sum_{\ell=1}^\infty p^\ell \ell^{k-1}
\sum_{\substack{ad=\ell\\ a>0}} \sum_{b=0}^{d-1} d^{-k} 
\sum_{n,\boldsymbol{\gamma}} D(n,\boldsymbol{\gamma}) 
\mathbf{e}\left[bn/d\right] q^{na/d} 
(\boldsymbol{\zeta}^{\boldsymbol{\gamma}})^a\,.
  \end{equation}
By performing the sum over $b$ we obtain
\begin{equation}
  \begin{split}
  \sum_{\ell=1}^\infty p^\ell \ell^{k-1}
\sum_{\substack{ad=\ell\\ a>0}} d^{-k+1} \sum_{n,\boldsymbol{\gamma}} 
D(nd,\boldsymbol{\gamma}) 
q^{na} (\boldsymbol{\zeta}^{\boldsymbol{\gamma}})^a\\
=\sum_{d=1}^\infty \sum_{n,\boldsymbol{\gamma}} D(nd,\boldsymbol{\gamma})
\sum_{a=1}^\infty 
a^{k-1} ( p^d q^n \boldsymbol{\zeta}^{\boldsymbol{\gamma}})^a\,.
\end{split}
\end{equation}
However, the last expression is equal to the right hand side of
\eqref{Heckesum}.
\end{proof}

This lemma urges us to introduce: 
\begin{definition}
  The action of the Hecke operator $V_0$ on $\Phi_k$ is defined by
\begin{equation}\label{Hecke0}
   \Phi_k\vert_{ V_0} (\tau,\boldsymbol{z}) :=
  \frac{D(0,\boldsymbol{0})}{2}\zeta(1-k) +
  \sum_{(0,n,\boldsymbol{\gamma})>0}D(0,\boldsymbol{\gamma})
  \Li_{1-k}(q^n\boldsymbol{\zeta}^{\boldsymbol{\gamma}})\,.
\end{equation} 
\end{definition}
Combining \eqref{Heckesum} and \eqref{Hecke0} we find that
\begin{lem}\label{liftformula}
\begin{equation}
  \sum_{\ell=0}^\infty p^\ell \Phi_k\vert_{V_\ell}
  (\tau,\boldsymbol{z}) = \frac{D(0,\boldsymbol{0})}{2}\zeta(1-k)
  +\sum_{(\ell,n,\boldsymbol{\gamma})>0} D(\ell n,\boldsymbol{\gamma})
  \Li_{1-k}(p^\ell q^n\boldsymbol{\zeta}^{\boldsymbol{\gamma}})\,.
\end{equation}
\end{lem}

\begin{rem}\label{rem-divergence}
  Since $\zeta(1)$ diverges, the definition \eqref{Hecke0} and hence
  \eqref{liftformula} are meaningless for $k=0$ as they stand.
  Nevertheless the case $k=0$ is the most important one in the next
  subsection. To treat this case adequately one would have to make an
  analytical continuation in $k$ and regularize the divergence
  properly.  However, in the following we will adopt a simple-minded
  approach keeping $\zeta(1)$ as the divergent sum
  $\sum_{h>0}\frac{1}{h}$ in the intermediate process of calculations
  and discard the diverging $\zeta(1)$ in the end.  Hopefully this will
  make the manipulations below transparent although they are
  admittedly formal. 
\end{rem}

\subsection{Lorentzian lattices and Jacobi forms of weight zero}

So far we have been quite general.  In the following we will choose a
specific Lorentzian lattice $\Pi$ and an associated Jacobi form
$\Phi_0$ of weight zero.

Fix a simple Lie algebra $\mathfrak{g}$ of rank $s$ (with the
convention mentioned before) and an associated nearly holomorphic
Weyl-invariant Jacobi form of weight $-2$ and index $m$ denoted
henceforth as $\phi_{-2,m}$. We will focus on the even Lorentzian
lattice of signature $(s,1)$:
\begin{equation}\label{Llattice}
  \Pi=Q^\vee(m)\oplus \langle -2\rangle\,.
\end{equation}
The reason why we select this lattice will become clear in the next
section.  We parametrize the elements of $\Pi_\CC$ as
\begin{equation}
  \Pi_\CC \ni \boldsymbol{z}=z\oplus \nu\, e\,,
\end{equation}
where $z\in Q^\vee(m)_\CC$ and $\nu\in \CC$ with $e$ being the generator
of $\langle -2\rangle$.

Since we have $\Pi^*=P(\inv{m}) \oplus \langle -\inv{2}\rangle$, we write
\begin{equation}
 \Pi^* \ni\boldsymbol{\gamma}= \gamma
\oplus j\,e^*\,,
\end{equation}
where $\gamma\in P({\textstyle\inv{m}})$ and $j\in \ZZ$ with $e^*$
being the generator of $\langle -\inv{2}\rangle$.
  Then we say
$\boldsymbol{\gamma}>0$ if either of the following possibilities holds
\begin{center}
  $ \mathrm{(i)}\ $ $\gamma
>0$,  $\ \mathrm{(ii)}\ $ $\gamma=0$ and $j>0$.
\end{center}
We write
$(\ell,n,\gamma,j)>0$ when $(\ell,n,\boldsymbol{\gamma})>0$.  We also
write $(\ell,n,\gamma)>0$ when $(\ell,n,\gamma,j)>0$ but the
restriction on $j$ is removed.

Consider
\begin{equation}\label{Kdef}
  E(\tau,\nu):=-\sqrt{-1}\,\frac{\vartheta_1(\tau,\nu)}{\eta(\tau)^3}\,,\quad
  (\tau,\nu)\in \HH_1\times \CC\,,
\end{equation}
where 
\begin{equation}
  \vartheta_1(\tau,\nu)= 
\sqrt{-1}(y^{-1/2}-y^{1/2})q^{1/8}
\prod_{n=1}^\infty (1-q^n)(1-q^n y)(1-q^ny^{-1})\,, 
\end{equation}
is the odd Jacobi theta function and
\begin{equation}
 \eta(\tau)=q^{\inv{24}}\prod_{n=1}^\infty (1-q^n)\,,
\end{equation}
is the Dedekind $\eta$ function.
Obviously,
\begin{equation}
  E(\tau,\nu)= (y^{-1/2}-y^{1/2})\prod_{n=1}^\infty
\frac{(1-q^n y)(1-q^n y^{-1})}{(1-q^n)^2}\,.
\end{equation}
Moreover, this can be expressed in terms of the Eisenstein series:
\begin{equation}
  E(\tau,\nu)=-\sqrt{-1}x \exp\left(\sum_{k=1}^\infty
    \frac{(-1)^{k}B_{2k}}{2k(2k)!}x^{2k}E_{2k}(\tau)\right)\,,
\end{equation}
where 
\begin{equation}
  y={\bf e}[\nu]\quad \text{and} \quad x=2\pi \nu\,.
\end{equation}
The function $E(\tau,\nu)$ is essentially the prime form on the
elliptic curve with modulus $\tau$.  It is easy to see that
$E(\tau,\nu)^2$ is a weak Jacobi form of weight $-2$ and index $1$ in
the sense of \cite{EZ} and it actually coincides with
$\tilde\phi_{-2,1}(\tau,\nu)$ in \cite{EZ}, which is one of the two
generators of the ring of weak Jacobi forms with even weights.

We then define 
\begin{equation}\label{Phi0}
  \Phi_0(\tau,\boldsymbol{z})=
  \Phi_0(\tau,z,\nu):=\frac{\phi_{-2,m}(\tau,z)}{E(\tau,\nu)^2}\,,
\end{equation}
which is apparently a Jacobi form of weight zero associated with $\Pi$.
Since we have
\begin{equation}\label{expK}
  \begin{split}
   \frac{1}{E(\tau,\nu)^2}&=\frac{1}{(y^{-1/2}-y^{1/2})^2}\prod_{n=1}^\infty
\frac{(1-q^n)^4}{(1-q^n y)^2(1-q^ny^{-1})^2}\\
&=-\frac{1}{x^2}\exp\left(\sum_{k=1}^\infty
    \frac{(-1)^{k-1}B_{2k}}{k(2k)!}x^{2k}E_{2k}(\tau)\right)\,,
\end{split}
\end{equation}
we can asymptotically expand $\Phi_0(\tau,z,\nu)$ as
\begin{equation}\label{asymp}
  \Phi_0(\tau,z,\nu)=-\sum_{g=0}^\infty x^{2g-2}\varphi_{2g-2,m}(\tau,z)\,,
\end{equation}
where $\varphi_{2g-2,m}$ is a quasi Jacobi form obtained from
$\phi_{-2,m}$ by multiplying a weight $2g$ quasi modular form, {\it
  i.e.\/} an element  of weight $2g$ in $\QQ[E_2,E_4,E_6]$.
Apparently we have
\begin{equation}
  \varphi_{-2,m}(\tau,z)=\phi_{-2,m}(\tau,z)\,.
\end{equation}
We  expand $\varphi_{2g-2,m}$ as
\begin{equation}\label{phiexp}
  \varphi_{2g-2,m}(\tau,z)=\sum_{n,\gamma}c_{g}(n,\gamma)q^n\zeta^\gamma\,,
\end{equation}
where $\zeta^\gamma={\bf e}[(\gamma,z)]$. Then we have
the symmetry property $c_g(n,\gamma)=c_g(n,-\gamma)$.

The expression $(y^{-1/2}-y^{1/2})^{-2}$ appearing in \eqref{expK} has
subtle features and will play an important role later in this
paper. It has   expansions
\begin{equation}\label{wall}
    (y^{-1/2}-y^{1/2})^{-2}=
\sum_{j=1}^\infty jy^{\pm j}\,, \qquad (\abs{y}\lessgtr 1)\,,
\end{equation}
exhibiting a wall-crossing behavior.  On the other hand, precisely on
the wall,  we have
\begin{equation}\label{Fy}
   (y^{-1/2}-y^{1/2})^{-2}=
\frac{1}{2}\sum_{j\in
    \ZZ}\abs{j}\,y^j\,, \qquad (\abs{y}=1,\ y \ne 1)\,.
\end{equation}

In the rest of this section and the next section we will tacitly
assume that we are precisely on the wall, hence the expansion
\eqref{Fy}.  We may thus regard the expression
$(y^{-1/2}-y^{1/2})^{-2}$ as an element of
$\frac{1}{2}\,\ZZ[[y,y^{-1}]]$ by interpreting it as a {\it formal
  distribution\/} \cite{Kac2}.  The reason for assuming \eqref{Fy} is
that the Fourier expansion
\begin{equation}
  \Phi_0(\tau,z,\nu)=\sum_{n,\gamma,j} D(n,\gamma,j)q^n\zeta^\gamma y^j\,,
\end{equation}
has the manifest symmetry properties
\begin{equation}
D(n,\gamma,j)=D(n,-\gamma,j)=D(n,\gamma,-j)\,.  
\end{equation}
Note that we must have $c_0(n,\gamma)\in 2\ZZ$ if we demand
$D(n,\gamma,j)\in \ZZ$. 

However, when we attempt  an interpretation in terms of $D2$-$D0$
bound states in \S \ref{sec:D2D0} we shall be mostly off the wall
using the expansion \eqref{wall}.

\begin{lem}\label{Dc}
\begin{equation}\label{Dcrel}
  \sum_j D(n,\gamma,j) y^j=-\sum_{g=0}^\infty x^{2g-2}c_g(n,\gamma)\,. 
\end{equation}
\end{lem}
\begin{proof}
  This is a direct consequence of \eqref{asymp}.
\end{proof}

Now we would like to consider the actions of the Hecke operators on
$\Phi_0$ and $\varphi_{2g-2,m}$ and compare the results.  For
simplicity we will use the same notation $V_\ell$
$(\ell=0,1,2,\ldots)$ for both $\Phi_0$ and $\varphi_{2g-2,m}$.  For
$\varphi_{2g-2,m}$, the Hecke operator $V_0$ is defined by using the
expansion \eqref{phiexp}. Since we are dealing with (quasi) Jacobi
forms of weight zero we should emphasize again what we have cautioned
in {\it Remark \ref{rem-divergence}}.

The following identity is crucial for our purpose:
\begin{lem}\label{Phivarphi}
  \begin{equation}
\label{relation}
  \Phi_0\vert_{ V_\ell}(\tau,z,\nu) = - \sum_{g=0}^\infty
  x^{2g-2} \varphi_{2g-2,m} \vert_{ V_\ell}(\tau,z)\,,\quad
  (\ell=0,1,2,\ldots)\,.
\end{equation}\end{lem}
\begin{proof}
If $\ell >0$, we find that
\begin{equation}
  \begin{split}
    \Phi_0\vert_{ V_\ell}(\tau,z,\nu)& =
    \ell^{-1}\sum_{\substack{ad=\ell\\ a>0}} \sum_{b=0}^{d-1}
    \Phi_0\left(\frac{a\tau+b}{d},az,a\nu \right)\\ 
    &=\ell^{-1}\sum_{\substack{ad=\ell\\ a>0}} \sum_{b=0}^{d-1}\left( -
      \sum_{g=0}^\infty (ax)^{2g-2}
      \varphi_{2g-2,m}\left(\frac{a\tau+b}{d},az\right)\right)\\ 
    &=-\sum_{g=0}^\infty x^{2g-2} \ell^{2g-3} \sum_{\substack{ad=\ell
        \\ a>0}} \sum_{b=0}^{d-1} d^{-(2g-2)}
    \varphi_{2g-2,m}\left(\frac{a\tau+b}{d},az\right)\\
    &= - \sum_{g=0}^\infty x^{2g-2} \varphi_{2g-2,m}
    \vert_{V_\ell}(\tau,z)\,.
  \end{split}
\end{equation}

As for the case $\ell=0$, we have
\begin{equation}
  \begin{split}
    \Phi_0\vert_{ V_0}(\tau,z,\nu)&= \frac{1}{2}D(0,0,0)\zeta(1)+
    \sum_{(0,n,\gamma,j)>0}D(0,\gamma,j)\Li_1(q^n\zeta^\gamma y^j)\\ 
    &=\frac{1}{2}D(0,0,0)\zeta(1) + \sum_{(0,n,\gamma,j)>0}
    D(0,\gamma,j) \sum_{h>0} \frac{(q^n\zeta^\gamma y^j)^h}{h}\\ 
    &=\frac{1}{2}\sum_{h>0}\frac{1}{h}\sum_j
    D(0,0,j) y^{jh}\\
&\qquad \qquad \qquad +\sum_{h>0} \sum_{(0,n,\gamma)>0}
    \frac{(q^n\zeta^\gamma)^h}{h} \sum_j D(0,\gamma,j) y^{jh}
  \end{split}
\end{equation}
where we used $\zeta(1)=\sum_{h>0}\frac{1}{h}$.  
Thus Lemma \ref{Dc}  shows that
\begin{equation}
  \begin{split}
    \Phi_0\vert_{ V_0}(\tau,z,\nu)&= -
    \frac{1}{2}\sum_{h>0}\frac{1}{h} \sum_{g=0}^\infty
    (hx)^{2g-2}c_g(0,0)\\&\qquad \qquad \qquad 
- \sum_{h>0} \sum_{(0,n,\gamma)>0}
    \frac{(q^n\zeta^\gamma)^h}{h} \sum_{g=0}^\infty
    (hx)^{2g-2}c_g(0,\gamma)\\
   &=-\sum_{g=0}^\infty x^{2g-2}\left( \frac{c_g(0,0)}{2}\zeta(3-2g)
+\sum_{(0,n,\gamma)>0}c_g(0,\gamma)\Li_{3-2g}(q^n\zeta^\gamma)\right)\\
   &= - \sum_{g=0}^\infty x^{2g-2} \varphi_{2g-2,m}
    \vert_{V_0}(\tau,z)\,.
  \end{split}
\end{equation}
This completes the proof of \eqref{relation}.
\end{proof}

Now we set 
\begin{equation}
  \mathcal{F}_g:=\sum_{\ell=0}^\infty p^\ell
  \varphi_{2g-2,m}\vert_{V_\ell}(\tau,z)\,,
\end{equation}
as in \cite{Kaw2}.  We will see in the next section that
$\mathcal{F}_g$ is an important piece of the Gromov-Witten potential
$F_g$ for certain elliptically and $K3$ fibered Calabi-Yau 3-folds.

\begin{rem}
Note however that even for $g\ge 2$, $\mathcal{F}_g$ is not exactly an
automorphic form on the type IV domain but what might be called a
quasi automorphic form since we are using a quasi Jacobi form
$\varphi_{2g-2,m}$ for the lifting. The situation is reminiscent of
that in \cite{Dijk} where the enumerative problem associated with
the Riemann-Hurwitz theory for elliptic curves was discussed and the
connection to quasi modular forms \cite{KZ} were explained.  At
hindsight the encounter with quasi automorphic forms is inevitable and
should be interpreted as the remnant of the holomorphic anomaly
studied in \cite{BCOV}.  It also partially explains why some
extra work is needed  when one uses the Harvey-Moore method \cite{HM1}
to extract the Gromov-Witten potentials: in the Harvey-Moore method
the automorphic properties are always preserved while what we are
after are not precisely automorphic forms.  Although not simply
related to the Gromov-Witten potentials, still it might be possible to
preserve the automorphic property by replacing $\varphi_{2g-2,m}$ by a
genuine Jacobi form $\phi_{2g-2,m}$ as expected in \cite{Kaw2}.  At
least this was already done in the genus one case. 
\end{rem}

Lemma \ref{liftformula}  then tells us that
\begin{prop}
\begin{equation}\label{GWpotential}
 \mathcal{F}_g= \frac{c_g(0,0)}{2}\zeta(3-2g) +\sum_{(\ell,n,\gamma)>0}
  c_g(\ell n,\gamma) \Li_{3-2g}(p^\ell q^n\zeta^\gamma)\,.
\end{equation}
\end{prop}

The following  infinite product is an essential ingredient when we
discuss the string partition function in the next section:
\begin{prop}\label{ingredient}
\begin{equation}\label{tmpZ}
   \exp\left(\sum_{g=0}^\infty
    x^{2g-2}\mathcal{F}_g\right)=e^{-\frac{D(0,0,0)}{2}\zeta(1)}
  \prod_{(\ell,n,\gamma,j)>0}(1-p^\ell q^n\zeta^\gamma y^j)^{D(\ell
    n,\gamma,j)}\,.
\end{equation}
\end{prop}
\begin{proof}
We see that 
  \begin{equation}
    \exp\left(\sum_{g=0}^\infty
    x^{2g-2}\mathcal{F}_g\right)= \exp\left(-\sum_{\ell=0}^\infty
    p^\ell \Phi_0\vert_{V_\ell}(\tau,z,\nu)\right)\,,
  \end{equation}
  by Lemma \ref{Phivarphi}. Then \eqref{tmpZ} follows from Lemma
  \ref{liftformula}.
\end{proof}

\section{String partition function} \label{sec:STP}

By utilizing the results obtained in the above we formulate in this
section the conjectures on the Gromov-Witten potentials and the
string partition function for Calabi-Yau 3-folds endowed with specific
fibration structures. 

\subsection{Fibered Calabi-Yau 3-folds}

Let $Y$ denote a Calabi-Yau 3-fold as in \S 2. In this section we use
the notations introduced there.  We now list up what we will assume on
$Y$.  First, we assume that there exist a $K3$ fibration
$\pi_1:Y\rightarrow W_1$ as well as an elliptic fibration
$\pi_2:Y\rightarrow W_2$. The two fibrations are assumed to be
compatible. This implies that a generic fiber of $\pi_1$ is an elliptic
$K3$ surface.  We mostly assume that $\pi_2:Y\rightarrow W_2$ has a
section.  Next we assume that all the singular fibers of
$\pi_1:Y\rightarrow W_1$ are irreducible.  Then $W_1\cong \PP^1$ and
$W_2\cong \FF_a$ $(a=0,1,\ldots,12)$ where
$\FF_a=\PP_{\PP^1}(\mathcal{O}_{\PP^1}\oplus \mathcal{O}_{\PP^1}(a))$
is a Hirzebruch surface.  See for instance, \cite{HS}. (In general,
the allowed possibilities of the base of an elliptic Calabi-Yau 3-fold
with a section are del Pezzo, Enriques, Hirzebruch or blown-up
Hirzebruch surfaces \cite{MV}.)

Furthermore the Picard lattice of a generic fiber of
$\pi_1:Y\rightarrow W_1$, which is necessarily an elliptic $K3$, is
assumed to coincide with $H\oplus Q^\vee(-m)$ where $H$ is the
hyperbolic plane, {\it i.e.} the even unimodular indefinite lattice of
rank $2$ with intersection matrix $\left(
\begin{smallmatrix}
  0&1\\ 1&0
\end{smallmatrix}\right)$,  $m$ is some positive 
integer and $Q^\vee$ is the coroot lattice of some simple Lie algebra
$\mathfrak{g}$ of rank $s=l-3$.

With these assumptions we express the complexified K{\" a}hler
parameters as in \cite{Kaw2}:
\begin{equation}\label{chamber}
  \begin{split}
    t_1&=\log u-\log q-\frac{a}{2}(\log p-\log q)\,,\\
    t_2&=\log p-\log q, \\
    t_3&=\log q-(\gamma_0,\log\zeta)\,,\\
    t_{i+3}&=(\Lambda_i,\log\zeta)\,,\quad (i=1,\ldots,s)\,,
  \end{split}
\end{equation}
where $\Lambda_i$ $(i=1,\ldots,s)$ are the fundamental weights of
$\mathfrak{g}$ and $\gamma_0$ is some positive weight. This
parametrization is such that $\omega_1$ is the pullback via $\pi_1$ of
the fundamental cohomology class of $W_1$.
We  have
\begin{equation}
  \rho_1=24\,,\quad \rho_2=24+12a\,,\quad \rho_3=92\,.
\end{equation}

The parametrization \eqref{chamber} should allow us to
 fix a particular ``fundamental chamber'' in
which we work in the following.

\subsection{The main conjectures}

As in \cite{Kaw2} we assume that there exists a nearly holomorphic
Jacobi form of weight $-2$ and
index $m$ associated with $Y$ in the form
\begin{equation}\label{genus0Jacobi}
  \phi_{-2,m}(\tau,z)=\frac{\Psi_{10,m}(\tau,z)}{\eta(\tau)^{24}}\,,
\end{equation}
where $\Psi_{10,m}(\tau,z)$ is a weak Weyl-invariant (with respect to
$\mathfrak{g}$) Jacobi form of weight $10$ and index $m$ satisfying
$\Psi_{10,m}(\tau,0)=-2E_4(\tau)E_6(\tau)$.  Sadly, we are aware of
neither a general algorithm to determine the precise form of
$\Psi_{10,m}(\tau,z)$ from the geometric information of $Y$ nor
whether there are additional conditions on $Y$ for $\phi_{-2,m}$ to
exist. However at least we must have
\begin{equation}
    c_{0}(-1,\gamma)=0\,,\quad \text{for $\gamma \ne 0$}\,, \quad
    c_{0}(-1,0)=-2\,,\quad 
    c_{0}(0,0)=-\chi(Y)\,,
\end{equation}
for the following conjectures to make sense.  We substitute
\eqref{genus0Jacobi} in the definition \eqref{Phi0} of $\Phi_0$.  We
assume that the coefficients $D(n,\gamma,j)$ are integers so that
$c_0(n,\gamma)$ are even integers.

Now we can state our conjectures on the Gromov-Witten potentials:
\begin{conj}\label{F_gconj}
  The Gromov-Witten potentials behave as
\begin{equation}
  \begin{split}
&F_0=F_0^{(0)} +\mathcal{F}_0 +O(q_1)\,,\\
&F_1=F_1^{(0)}
  +\mathcal{F}_1 +O(q_1)\,,\\
&F_g=\mathcal{F}_g+O(q_1)\,,\quad \text{for $g>1$}\,, 
\end{split}
\end{equation}
where $\mathcal{F}_g$ are given by \eqref{GWpotential} and
\begin{equation}
  \begin{split}
 F_0^{(0)}&= \frac{1}{3!}\sum_{i,j,k} \kappa_{ijk} t_i t_j t_k\,, \\
F_1^{(0)}&= -\lambda_1
\cap[\overline{\mathcal{M}}_{1,1}]\cdot\sum_i \rho_i
t_i=-\frac{1}{24}\sum_i \rho_i t_i\,.
\end{split}
\end{equation}
\end{conj}

In the fundamental chamber, we gave in \cite{Kaw2} conjectural
formulas of $F_0^{(0)}$ and $F_1^{(0)}$ expressed in terms of the data
of $\phi_{-2,m}$.  This was achieved by employing the elliptic
polylogarithm of Beilinson and Levin \cite{BL} which is the
holomorphic version of that of Zagier \cite{Z}. The result reads as
follows:
\begin{conj}\label{yukawaconj} In the fundamental chamber we have
\begin{equation}
  \begin{split}
    F_0^{(0)}&=\log u\left\{\log p\log q-\frac{m}{2}(\log
      \zeta,\log\zeta)\right\}\\
    &\quad +\left(\frac{m}{2} - \frac{\mathcal{I}_2}{24s}\right) \log
    p\cdot(\log
    \zeta,\log\zeta)\\
    &\quad + \frac{1}{3}(\log q)^3-\frac{\mathcal{I}_2}{24s}\log
    q\cdot(\log\zeta,\log\zeta)
    +\frac{1}{12}\sum_{\gamma>0}c_0(0,\gamma)(\gamma,\log \zeta)^3\,,
  \end{split}
\end{equation}
and\footnote{The normalization of $F_1$ differs from that in
  \cite{Kaw2} by $m_1=\frac{1}{12}$.}
\begin{equation}
 F_1^{(0)}= -\frac{1}{24}(24\log u+ 24\log p + 44\log q)
 + \frac{1}{24}\sum_{\gamma>0}c_0(0,\gamma)(\gamma,\log\zeta)\,,
\end{equation}
where $\log \zeta= 2\pi \sqrt{-1} z$ and
$\mathcal{I}_2=\sum\limits_{\gamma>0} c_0(0,\gamma)(\gamma,\gamma)$.
\end{conj}
{\noindent As} shown in \cite{Kaw2} these formulas are such that if we
replace $\Li_r$ $(r=1,3)$ by $\Lic_r$ $(r=1,3)$, we have (at least to
the first order in $q_1$)
\begin{equation}
  F_g(u,p,q,\zeta)=F_g(u',q,p,\zeta)\,,
\end{equation}
for all $g \ge 0$, where
\begin{equation}
  \log u'=\log u -(\log p-\log q)\,.
\end{equation}
For the definition of $\Lic_r$ see Appendix A.

Since we have 
\begin{equation}\label{mone}
  c_{g}(-1,0)=m_g\, c_{0}(-1,0)\,,
\end{equation}
and 
\begin{equation}
  c_{g}(0,0)=m_g\, c_{0}(0,0)\,,\quad(g\neq 1)\,, \qquad
  c_{1}(0,0)=m_1\, c_{0}(0,0)-2 c_{0}(-1,0)\,,
\end{equation}
it is easy to check that the conjectured expressions of $F_g$ are
consistent with the general results reviewed in \S 2.

For concrete examples of $Y$ and some corroboration of Conjectures
\ref{F_gconj} and \ref{yukawaconj} for $g=0,1$, see \cite{Kaw2} and
references therein.

If we translate these conjectures on the Gromov-Witten potentials into
the language of the string partition function using Proposition
\ref{ingredient}, we reach the main conjecture of this paper:
\begin{conj}
  The string partition function behaves as
\begin{equation}\label{mainconj}
  \mathcal{Z} = \exp\left(x^{-2}F_0^{(0)} + F_1^{(0)}\right)\left[
    \prod_{\left(\ell,n,\gamma,j\right)>0} (1-p^\ell q^n\zeta^\gamma
    y^j)^{D(\ell n,\gamma,j)} +O(q_1)\right]\,,
\end{equation}
where we neglected $\zeta(1)$ appearing in {\em \eqref{tmpZ}}.
\end{conj}

Eq.\eqref{mainconj} bears strong resemblance to Borcherds' infinite
product formulas \cite{Bo1}\footnote{ When making this analogy, it
  should be born in mind that there is a conventional ambiguity: we
  could replace $\mathcal{Z}$ by $\mathcal{Z}^{-1}$ in the definition of the
  string partition function.}. This should be so since we have
employed more or less the same kind of lifting procedure.  However,
the important difference lies in that we used the lifting of a weight
zero Jacobi form associated with a {\it Lorentzian\/} lattice.  This
has entailed a more complicated expression of the ``Weyl vector''
$x^{-2}F_0^{(0)} + F_1^{(0)}$ which exhibits chamber dependence as in
the case of the ordinary Weyl vector. It should be noted that since
$F_0^{(0)}$ and $F_1^{(0)}$ are respectively homogeneously cubic and
linear in $t_i$, the homogeneous degree of $x^{-2}F_0^{(0)} +
F_1^{(0)}$ as a function of $x$ and $t_i$ is one, lending further
support to the interpretation of $x^{-2}F_0^{(0)} + F_1^{(0)}$ as the
``Weyl vector''.

\section{An interpretation in terms of $D2$--$D0$ bound states}
\label{sec:D2D0}

In eq.\eqref{mainconj} we observe that the complexified
K{\" a}hler moduli and the string coupling $x$ are unified in a rather
nice way.  In fact, the geometrical origin of the Lorentzian lattice
$\Pi$ we have used may be attributed to the following relation:
\begin{equation}
  H^2(Y,\ZZ)\oplus H^0(Y,\ZZ)\supset H\oplus  Q^\vee(-m)\oplus
  \langle 2\rangle = H \oplus \Pi(-1)\,,
\end{equation}
where we identified $H^0(Y,\ZZ)$ with the lattice $\langle 2\rangle$.  The
(analytically continued) string coupling $x$ parametrizes $H^0(Y,\CC)$
just as the complexified K{\" a}hler moduli parametrize (a cone of)
$H^2(Y,\CC)$.  Thus it seems natural to view the string partition
function as a function (or a section of the appropriate vacuum line
bundle) over the extended moduli space whose tangent space is some
domain of $H^2(Y,\CC)\oplus H^0(Y,\CC)$.  This fact immediately
suggests that there should be an interpretation of the infinite
product \eqref{mainconj} in terms of the bound states of $D2$- and
$D0$-branes.  In the following we wish to develop some arguments
supporting this picture.

\smallskip
\begin{rem}
  As mentioned in Introduction, $D6$- and $D4$-branes are duals of
  $D0$- and $D2$-branes.  Thus the above extended moduli space is a
  {\it half\/} of the usual extended moduli space \cite{Witmirror}
  whose tangent space is contained in $\oplus_{i=0}^3
  H^{2i}(Y,\CC)$. 
\end{rem}

In the conjectured expression \eqref{mainconj} all the information is
encoded in the Jacobi form $\Phi_0$ or its Fourier coefficients
$D(n,\gamma,j)$. Our basic expectation in the following is that
$\Phi_0$ should be interpreted as the function counting the bound
states of a single $D2$-brane and $D0$-branes moving inside the fibers
of the $K3$ fibration.

\subsection{Preliminaries}

\subsubsection{Notations}

For a smooth complex projective variety $V$, we define the Hodge
polynomial by
\begin{equation}
  \chi_{t,\tilde t}(V):=\sum_{p,q=0}^{\dim(V)}(-1)^{p+q}h^{p,q}(V)t^p
  {\tilde t}^q\,,
\end{equation}
where $h^{p,q}(V)=\dim H^q(V,\Omega_V^p)$. We also introduce
\begin{equation}
    \chi_{t}(V):=\chi_{t,1}(V)\,,
\end{equation}
which is essentially the Hirzebruch $\chi_y$ genus of $V$. Note that
the Euler characteristic of $V$ is given by $\chi(V)=\chi_1(V)$.

The $r^\mathrm{th}$ symmetric group $\mathfrak{S}_r$ naturally acts on
$V^r=V\times \dots \times V$ ($r$ times) as permutations on $r$
letters. The quotient is the $r^\mathrm{th}$ symmetric product
$V^{(r)}:=V^r/\mathfrak{S}_r$. In general $V^{(r)}$ has orbifold
singularities when $\dim(V)>1$ while it is smooth when $\dim(V)=1$. We
set $V^{(0)}=\{\mathrm{pt}\}$.

Let $V$ be an even dimensional Calabi-Yau manifold. Then the elliptic
genus  $\mathcal{E}_{V}(\tau,\nu)$ is a weak
Jacobi form of weight $0$ and index $\dim(V)/2$ \cite{KYY}
with the expansion
\begin{equation}
  \mathcal{E}_{V}(\tau,\nu)=y^{-\dim(V)/2}\chi_y(V)+O(q)\,.
\end{equation}
We have the duality relation $\chi_{y^{-1}}(V)=y^{-\dim(V)}\chi_y(V)$.

Let $V$ be a complex algebraic variety of dimension $n$ and $E$ a
coherent sheaf on $V$. We denote by $\Supp(E)$ the support of $E$. The
dimension of $E$, denoted as $\dim(E)$, is defined to be that of
$\Supp(E)$ and $E$ is called of pure dimension $m$ if $\dim(F)=m$ for
all nontrivial coherent subsheaves $F \subset E$.  

In the following, by $H_k(V,\mathbb{A})$ we always mean the $k^{\rm
  th}$ Borel-Moore homology group \cite{Bre} with coefficients in a
commutative ring $\mathbb{A}$.  The fundamental homology class of $V$,
which is an element of $H_{2n}(V,\ZZ)$, is denoted by $[V]$. If $V$ is
smooth, the operation $\cap[V]$ gives the Poincar{\' e} duality
isomorphism: $H^k(V,\ZZ)\cong H_{2n-k}(V,\ZZ)$. If $V$ is smooth and
compact, the Borel-Moore homology coincides with the ordinary one.

If $W\subset V$ is a closed subvariety with the inclusion $\iota:
W\hookrightarrow V$, we frequently write $[W]$ instead of
$\iota_*[W]$.
 
In this section we usually denote by $X$ a projective $K3$ surface.

\subsubsection{$D$-brane charges}

We recall some generalities on $D$-brane charges. It has been argued
\cite{HM2} that $D$-brane charges in $X$ are associated with Mukai
vectors.  The Mukai lattice of $X$ is the total integer cohomology
group
\begin{equation}
  H^{2*}(X,\ZZ)=H^0(X,\ZZ)\oplus H^2(X,\ZZ)\oplus
  H^4(X,\ZZ)\,,
\end{equation}
endowed with the symmetric bilinear form
\begin{equation}
  \langle {v},{v}'\rangle_X=c_1\cdot c_1'-ab'-a'b\,,
\end{equation}
for any ${v}=(a,c_1,b)\in H^{2*}(X,\ZZ)$ and ${v}'=(a',c_1',b')\in
H^{2*}(X,\ZZ)$. Here the notation $v=(a,c_1,b)$ means $v=a\oplus c_1
\oplus b$ with $a\in H^0(X,\ZZ)$, $c_1\in H^2(X,\ZZ)$ and $b\in
H^4(X,\ZZ)$.  We have $H^{2*}(X,\ZZ)\cong\mathsf{E}_8(-1)^{\oplus
  2}\oplus H^{\oplus 4}$ where $\mathsf{E}_8$ is the positive definite
even unimodular lattice of rank $8$.

The Grothendieck group $K_0(X)$ is defined to be the quotient of the
free abelian group generated by all the coherent sheaves (up to
isomorphisms) on $X$ by the subgroup generated by the elements
${F}-{E}-{G}$ for each short exact sequence
\begin{equation}\label{eseq}
  0\to {E}\to {F} \to {G} \to 0\,
\end{equation}
of coherent sheaves on $X$. In what follows, we shall use the same
notation $E$ for both a coherent sheaf on $X$ and its image in
$K_0(X)$.

Let ${v}:K_0(X)\to \oplus_i H^{2i}(X,\QQ)$ be the module
homomorphism defined by Mukai vectors \cite{Muk1,Muk2,Muk3}, namely
$E \mapsto {v}({E}):=\ch({E}) \sqrt{\td(X)}$.  Explicitly
we have
\begin{equation}
  {v}({E})=\left(\rk({E}),\,c_1({E}),\,
    \rk({E})\,
    \varrho+\inv{2}c_1({E})^2-c_2({E}) \right)\,,
\end{equation}
where $\varrho\in H^4(X,\ZZ)$ is the fundamental cohomology class of
$X$ so that $\varrho\cap[X]=1$.  Thus actually we have
${v}(K_0(X))\subset H^{2*}(X,\ZZ)$ since $H^2(X,\ZZ)$ is even.
This definition is such that
\begin{equation}
  \chi({E},{F}):=\sum_{i=0}^2 (-1)^i \dim
  \Ext^i({E},{F})=-
  \langle {v}({E}),{v}({F})\rangle_X\,,
\end{equation}
by the Hirzebruch-Riemann-Roch theorem.  

The Mukai lattice has several distinguished isometries. For instance,
for an invertible sheaf $L$ on $X$, the map
\begin{equation}\label{isometry}
  \begin{split}
  {v}({E})\mapsto& \ch(L) {v}({E})\\
&\quad =  {v}({E})+ \left(0, \rk({E})\, c_1(L)\,,
    c_1({E})\cup c_1(L)+\frac{\rk({E})}{2}c_1(L)^2\right)\,,
\end{split}
\end{equation}
gives an isometry of $H^{2*}(X,\ZZ)$.

Let ${Q}:K_0(X)\to
\oplus_iH_{2i}(X,\QQ)$ be defined by
\begin{equation}
\begin{split}
 E \mapsto {Q}({E}) 
:=&{v}({E})\cap[X]\\ 
=&
  \left(\rk({E})\,[X],\,c_1({E})\cap[X],\,
    \chi(X,{E})- \rk({E})\, \right)\,.
\end{split}
\end{equation}
  We call
${Q}({E})$ the $D$-brane charge of ${E}$ with
its component in $H_{2i}$ representing the $D2i$-brane charge.
Since ${Q}$ is a module homomorphism, it follows that
\begin{equation}
  {Q}({F})={Q}({E})+{Q}({G})\,,
\end{equation}
for each exact sequence \eqref{eseq} of coherent sheaves on $X$.  This
may be interpreted as the charge conservation law when
making the D-brane state associated with ${F}$ out of those
associated with ${E}$ and ${G}$.

Let $C$ be a curve on $X$ and let $\iota:C\hookrightarrow X$ be the
inclusion. If ${E}$ is a coherent sheaf on $C$, the direct
image $\iota_*{E}$ is a torsion sheaf on $X$ obtained by
``extending by zero''.  Suppose that we have the exact sequence
\eqref{eseq} now for coherent sheaves on $C$. Since $\iota_*$ is an
exact functor, we have
\begin{equation}
  {Q}(\iota_*{F})=
{Q}(\iota_*{E})+{Q}(\iota_*{G})\,.
\end{equation}
This may also be regarded as the charge conservation law for D-brane
states without $D4$-branes. Similar formula holds for a 0-dimensional
subscheme   instead of $C$.

The above consideration can be extended almost verbatim to a smooth
Calabi-Yau manifold $Y$ of any dimension. We define ${Q}:K_0(Y)\to
\oplus_i H_{2i}(Y,\QQ)$ by $E\mapsto {Q}({E}):= {v}({E})\cap [Y]$ where
${v}({E})=\ch({E})\sqrt{\td(Y)}$.

\begin{rem} In the above, we have defined ${Q}$ on
  the Grothendieck group $K_0(Y)$.  However, in more general contexts
  like homological mirror conjecture \cite{Kon1} or Fourier-Mukai
  transforms, the domain of ${Q}$ must be (naturally) extended
  from $K_0(Y)$ to the bounded derived category $\mathbf{D}^b(Y)$ of
  coherent sheaves on $Y$. 
\end{rem}

Let $V$ be a smooth variety of dimension $n$ and $W$ an
$m$-dimensional irreducible and reduced subvariety of $V$ with the
inclusion map $\iota:W\hookrightarrow V$. Then by using a resolution
of singularities $\pi:\tilde W\to W$ and the Grothendieck-Riemann-Roch
theorem for singular varieties \cite{Ful}, one can show [\citen{CG},
\S 5.8--5.9] that
\begin{lem}
  \begin{equation}
    \begin{split}
      \ch_k(\iota_*\mathcal{O}_W)\cap[V]&=0\,,\quad \text{for
        $k<n-m$},\\ \ch_{n-m}(\iota_*\mathcal{O}_W)\cap[V]&=[W]\,.
   \end{split}
 \end{equation}
\end{lem}
Suppose that a coherent sheaf $E$ on $V$ is of pure dimension $m$. Let
$\Supp(E)=\cup_i S_i$ be the support of $E$, where $S_i$ are
irreducible and reduced.  We shall define the multiplicity of $E$
along $S_i$.  Let $\mathcal{O}_{V,S_i}$ be the stalk of
$\mathcal{O}_V$ at $S_i$ and $E_{V,S_i}$ the stalk of $E$ at $S_i$.
Let $I_{S_i,S_i}$ be the stalk of the ideal sheaf of $S_i$.  Then
$(\mathcal{O}_{V,S_i},I_{S_i,S_i})$ is a local ring with the residue
field $\mathcal{O}_{V,S_i}/I_{S_i,S_i} \cong K(S_i)$, where $K(S_i)$
is the function field of $S_i$.  Since $(I_{S_i,S_i})^k E_{V,S_i}=0$
for some $k$, there is a filtration
\begin{equation}\label{eq:length}
 0 \subset F_i^1 \subset F_i^2 \subset \dots \subset F_i^{s_i}=E_{V,S_i}
\end{equation}
such that
$F_i^{j}/F_i^{j-1} \cong K(S_i)$.
We define the multiplicity of $E$ along $S_i$ by
$\mult_{S_i}(E):=s_i$.
Namely, $\mult_{S_i}(E)$ is the length of $E_{V,S_i}$
as an $\mathcal{O}_{V,S_i}$-module.

\begin{lem}\label{lem:fund}
\begin{equation}
  \begin{split}
\ch_k(E)\cap[V]&=0\,,\quad \text{for $k<n-m$},\\
     \ch_{n-m}(E)\cap [V]&=\sum_i \mult_{S_i}(E) [S_i]\,.
  \end{split}
\end{equation}
\end{lem}
\begin{proof}
  See [\citen{CG}, \S 5.8-- 5.9].
\end{proof}

\begin{prop} Suppose that $Y$ is  a (smooth) Calabi-Yau manifold.
  \begin{enumerate}
  \item[(i)] Let $\iota: Z \hookrightarrow Y$ be a 0-dimensional
    subscheme of length $d$. Then,
    \begin{equation}
      Q(\iota_*\mathcal{O}_Z)=(0,\ldots,0,d)\,.
    \end{equation}
  \item[(ii)] Let $E$ be a coherent sheaf of pure dimension $1$ on $Y$. If
    $\Supp(E)=\cup_i C_i$ with $C_i$ being irreducible and reduced,
    then
  \begin{equation}
    Q(E)=(0,\ldots,0,\sum_i \mult_{C_i}(E)[C_i],\chi(Y,E))\,.
  \end{equation}
\end{enumerate}
\end{prop}
\begin{proof}
 {\em (i)} is an easy consequence of Lemma \ref{lem:fund} while 
{\em (ii)} follows from Lemma \ref{lem:fund}, $c_1(Y)=0$ and Riemann-Roch.
\end{proof}

\begin{cor}\label{D2D0charge}
  Let $\iota:C\hookrightarrow Y$ be an irreducible and reduced curve
  on $Y$ and $F$ a torsion-free sheaf on $C$. Then,
\begin{equation}
 Q(\iota_*F)=(0,\dots,0,\rk(F)[C],\chi(C,F))\,.
\end{equation}
\end{cor}
\begin{proof}
  Recall  that
\begin{equation}\label{eulerCX}
  \chi(C,{F})=\chi(Y,\iota_*{F})\,,
\end{equation}
since $H^i(C,{F}) \cong H^i(Y,\iota_*{F})$ for the
inclusion $\iota$.
\end{proof}

\begin{rem}
  If $C$ is smooth, Corollary \ref{D2D0charge} can also be seen (as
  done in \cite{HM2} for the case $n=2$) by directly using the
  Grothendieck-Riemann-Roch theorem for nonsingular varieties:
\begin{equation}
    \ch(\iota_!{F})\td(Y)=\iota_*(\ch({F})\td(C))\,,
\end{equation}
where $\iota_!{F}:=\sum_i
(-1)^iR^i\iota_*{F}=\iota_*{F}$ since
$R^i\iota_*{F}$ vanishes for $i>0$. 
\end{rem}
\begin{rem}\label{rem:deg}
  We define the degree of $F$ by
  \begin{equation}
 \deg(F):=\chi(C,F)-\rk(F)\chi(C,\mathcal{O}_C)\,.   
\end{equation}
Then, we have
\begin{equation}\label{RRC}
  \chi(C,F)=\deg(F)+\rk(F)(1-p_a(C))\,,
\end{equation}
where $p_a(C)$ is the arithmetic genus of $C$. If $F$ is locally-free,
$\deg(F)$ reduces to the ordinary degree of $F$  with
\eqref{RRC} being the Riemann-Roch theorem for a singular curve.
\end{rem}

\begin{ex}
Let $Z\subset X$ be a 0-dimensional subscheme of length $d$ and let
$\iota:Z\hookrightarrow X$ be the inclusion.  If we denote the ideal
sheaf of $Z$ by ${I}_Z$ we have an exact sequence,
\begin{equation}\label{Hilbeseq}
  0\to {I}_Z\to \mathcal{O}_X \to
  \mathcal{O}_X/{I}_Z=\iota_*\mathcal{O}_Z
  \to 0\,.
\end{equation}
Since ${Q}(\iota_*\mathcal{O}_Z)=(0,0,d)$ and
${Q}(\mathcal{O}_X)=([X],0,1)$ we obtain
\begin{equation}\label{DchargeofI_Y}
  {Q}({I}_Z)=([X],0,1-d)\,.
\end{equation}
\end{ex}

\subsubsection{Hilbert polynomials}

Let ${H}_X$ be an ample invertible sheaf on $X$. The Hilbert
polynomial $P_{{E}/X}(n)\in \QQ[n]$ of a coherent sheaf
${E}$ on $X$ is defined by
\begin{equation}\label{HPM}
  P_{{E}/X}(n):=\chi(X,{E}\otimes
  {H}_X^{\otimes n}) =\frac{\rk({E})\,\deg(X)}{2}\,
  n^2+\deg({E})\, n+ \chi(X,{E})\,,
\end{equation}
where $\deg({E}):=(c_1({E})\cup c_1({H}_X) )
\cap[X]$ and $\deg(X):=\deg({H}_X)$.  Since $[{E}]
\mapsto P_{{E}/X}(n)$ is a module homomorphism, we obtain
\begin{equation}
P_{{F}/X}(n)=P_{{E}/X}(n)+P_{{G}/X}(n)\,,
\end{equation}
for each exact sequence \eqref{eseq} of coherent sheaves on $X$.

Let $C$ be a projective irreducible curve polarized by an ample
invertible sheaf ${H}_C$ on $C$. Let ${F}$ be a coherent sheaf on $C$.
The Hilbert polynomial of ${F}$ is similarly given by
\begin{equation}\label{HPC}
  P_{{F}/C}(n):=\chi(C,{F}\otimes
  {H}_C^{\otimes n})=\rk({F})\, \deg(C)\,
  n+\chi(C,{F})\,,
\end{equation}
where $\deg(C):=\deg({H}_C)$. Suppose that we have an inclusion
$\iota:C\hookrightarrow X$.  Since $\iota^*{H}_X$ is also
ample, one may choose ${H}_C=\iota^*{H}_X$. Then it follows that
\begin{equation}
  P_{{F}/C}(n)=P_{\iota_*{F}/X}(n)\,.
\end{equation}
This can also be directly checked by using  \eqref{eulerCX} and 
\begin{equation}
  \begin{split}
\deg(\iota_*{F})&=\rk({F})\,(c_1(\mathcal{O}_X(C))\cup
c_1({H}_X))\cap [X]\\
&=\rk({F})\,c_1({H}_X)\cap
\iota_*[C]=\rk({F}) \deg({H}_C)\,.
\end{split}
\end{equation}

Comparing \eqref{HPM} and \eqref{HPC} with the expressions of
$D$-brane charges one finds that the coefficients of Hilbert
polynomials are, in a sense, scalar projections of $D$-brane
charges. In particular, if $D4$-brane charges vanish, $D0$-brane
charges coincide with the constant terms of Hilbert polynomials. This
fact may be a useful observation later in this section.

\subsubsection{Some moduli spaces}

Let ${E}$ be a coherent sheaf on $X$. Fix an ample invertible sheaf
${H}_X$ on $X$ and expand $P_{{E}/X}(n)$ in the form $P_{{E}/X}(n)=
\sum_{i=0}^{\dim({E})}\alpha_i({E})n^i/i!$.  A coherent sheaf ${E}$ on
$X$ is called semi-stable (stable) if it is pure and satisfies
\begin{equation}
  \frac{P_{{E}'/X}(n)}{\alpha_{\dim({E}')}({E}')}
  \le
  \frac{P_{{E}/X}(n)}{\alpha_{\dim({E})}({E})}\quad
 \quad \left(
\frac{P_{{E}'/X}(n)}{\alpha_{\dim({E}')}({E}')}
  <
  \frac{P_{{E}/X}(n)}{\alpha_{\dim({E})}({E})}\right)\,,
  \qquad(n\gg 0)\,,
\end{equation}
for any proper subsheaf ${E}'\subset {E}$.
There is another notion of stability due to Mumford:
\begin{enumerate}
\item
The case where $\rk(E)>0$: A coherent sheaf $E$ is slope semi-stable
(stable) if it is torsion-free and
satisfies
\begin{equation}
  \frac{\deg({E}')}{\rk({E}')}
  \le
  \frac{\deg(E)}{\rk (E)}\quad
 \quad \left(
\frac{\deg({E}')}{\rk({E}')}
  <
\frac{\deg(E)}{\rk (E)} \right)\,,
\end{equation}
for any subsheaf $E'$ of $0<\rk ({E}')<\rk (E)$. 
\item
The case where $\rk(E)=0$: A coherent sheaf $E$ is slope semi-stable
(stable) if it is of pure dimension 1 and
\begin{equation}
  \frac{\chi(X,{E}')}{\deg({E}')}
  \le
  \frac{\chi(X,E)}{\deg (E)}\quad
 \quad \left(
\frac{\chi(X,{E}')}{\deg({E}')}
  <
\frac{\chi(X,E)}{\deg (E)} \right)\,,
\end{equation}
for any subsheaf $E'$ of $0<\deg ({E}')<\deg (E)$. 
\end{enumerate}
By \eqref{HPM} and \eqref{HPC}, we have the following relations:
\begin{equation}
\text{slope stable} \Rightarrow  
\text{stable} \Rightarrow  
\text{semi-stable}\Rightarrow  
\text{slope semi-stable}.
\end{equation}

Let $\mathcal{M}_{H_X}(v)$ be the moduli space of semi-stable (with
respect to ${H}_X$) sheaves on $X$ with $D$-brane charge $v\cap[X]$.
Let $\mathcal{M}^s_{H_X}(v)\subset \mathcal{M}_{H_X}(v)$ be the subset
parametrizing stable sheaves. If $\mathcal{M}^s_{H_X}(v)$ is not
empty, it is smooth of dimension $\langle {v},{v}\rangle_X+2$.  If $v$
is primitive and ${H}_X$ is a general point of the ample cone of $X$,
$\mathcal{M}_{H_X}(v)=\mathcal{M}^s_{H_X}(v)$ and
$\mathcal{M}_{H_X}(v)$ is irreducible symplectic (hence hyperk{\"
  a}hler). Since the choice of $H_X$ is not so important, we usually
denote $\mathcal{M}_{H_X}(v)$ by $\mathcal{M}(v)$.  In the following,
we deal with the cases where $\mathcal{M}(v)=\mathcal{M}^s(v)$.  If
$v\cap[X]$ is expressed as $(r[X],[C],a)$, we frequently use the
notation $\mathcal{M}(r,C,a)$ for $\mathcal{M}(v)$. When the
isomorphism class $[E]$ of a coherent sheaf $E$ belongs to
$\mathcal{M}(v)$, we simply write $E\in\mathcal{M}(v)$ instead of
$[E]\in\mathcal{M}(v)$.  For more details on $\mathcal{M}(v)$ see the
original works \cite{Muk1,Muk2,Muk3} or an exposition \cite{HL}, and
for recent developments on $\mathcal{M}(v)$ see
\cite{O'Grady1,O'Grady2, Yos1,Yos2}.

Let $V$ be a projective scheme polarized by an ample invertible sheaf
${H}_V$. Fix a coherent sheaf ${F}$ on $V$.
Informally speaking, the Grothendieck Quot-scheme
$\Quot^{P(n)}_{{F}/V}$ parametrizes quotient sheaves of
${F}$ having a common Hilbert polynomial $P(n)$ or
equivalently exact sequences $0\to {E}\to {F} \to
{G} \to 0$ such that $P_{{G}/V}(n)=P(n)$. If $V$ is an
$S$-scheme we can similarly consider relative Quot-schemes
$\Quot^{P(n)}_{{F}/V/S}$. See \cite{Grot, HL} for more
details.

\begin{ex}
  A fundamental case is the Hilbert scheme
  $X^{[d]}:=\Hilb^d_X=\Quot^d_{\mathcal{O}_X/X}$ of 0-dimensional
  subschemes of length $d$ in $X$. In this case relevant short exact
  sequences are in the form \eqref{Hilbeseq} and $X^{[d]}$ is an
  irreducible symplectic manifold of dimension $2d$. If $C \subset X$
  is an irreducible curve we have
\begin{equation}
  X^{[d]}\cong\mathcal{M}(1,0,1-d)\cong 
\mathcal{M}(1,C,p_a(C)-d)\,,
\end{equation}
where the first isomorphism is obtained by sending a 0-dimensional
subscheme $Z\subset X$ to its ideal sheaf
${I}_Z$ ({\it cf.\/} \eqref{DchargeofI_Y}) while the
second one reflects the isometry \eqref{isometry} and is obtained by
sending a 0-dimensional subscheme $Z$ to
${I}_Z(C):={I}_Z\otimes
\mathcal{O}_X(C)$.
\end{ex}

\subsection{$D0$-branes bound to a rigid smooth $D2$-brane in
  a Calabi-Yau manifold}
\label{subsec:rigidD@D0}

We begin by considering a single $D2$-brane wrapping around a fixed
closed (nonsingular) Riemann surface $C_h$ of genus $h$ so that the
world-volume of the $D2$-brane is $C_h\times \RR$ with the time
running in the direction of $\RR$. Let us imagine that this $D2$-brane
is bound to collections of $D0$-branes. Taking into account the fact
that $D0$-branes are the pure magnetic sources as seen from the $D2$
brane, we may regard $D0$-branes as vortices. To concretely realize
vortices one may consider, as the effective world-volume theory of the
combined system,  $N=2$ abelian Higgs model \cite{ENF} or more
generally  $N=2$ abelian Born-Infeld type theory \cite{CNF} on
$C_h\times \RR$ . The precise form of the effective theory does not
matter since the BPS conditions are universal \cite{CNF} and are given
by the so-called (abelian) {\it vortex equations\/} on $C_h$. Thus the
moduli space of the relative configuration of $D0$-branes with respect
to the fixed $D2$ brane should coincide with the moduli space of
vortices.

The mathematics of the vortex equations on closed Riemann surfaces has
been much investigated in the literature \cite{Brad,BDGW}\footnote{The
  vortex equations have also appeared as the BRST fixed configurations
  in Witten's analysis of {\it two} dimensional linear sigma models
  \cite{Witphase}.}.  We now review this subject rather in detail
since it is conceptually important in what follows.

Suppose that a hermitian $C^\infty$ line bundle ({\it i.e.\/}
$U(1)$-bundle) $L \to C_h$ is given.  Let $\mathcal{A}$ be the space
of unitary connections on $L$ and $\Omega$ the space of $C^\infty$
sections of $L$.  Our convention is such that $\sqrt{-1}A$ is a {\it
  real}-valued 1-form on $C_h$ if $A\in \mathcal{A}$. The curvature
two-form is given by $F_A=dA$ and the covariant derivative $D_A=d+A$
can be decomposed as $D_A=\partial_A+\bar\partial_A$ where
$\partial_A$ and $\bar\partial_A$ are respectively the $(1,0)$ and
$(0,1)$ part of $D_A$. Since $\bar\partial_A$ determines a holomorphic
structure on $L$, we can view $\mathcal{A}$ as the space of
holomorphic structures on $L$.

Let $\omega$ denote the K{\" a}hler form on $C_h$.  Then the vortex
equations are the equations for $(A,\phi)\in \mathcal{A}\times \Omega$
given by:
\begin{equation}\label{vortex}
  \begin{split}
    &\bar \partial_A\phi=0\,,\\ 
&\Lambda_\omega F_A-\sqrt{-1}(\abs{\phi}^2-c^2)=0\,,
  \end{split}
\end{equation}
where $\Lambda_\omega$ is the adjoint of $\wedge\, \omega$ and $c$ is
a real constant.  The first equation of \eqref{vortex} means that the
section $\phi$ is holomorphic with respect to the holomorphic
structure determined by $\bar \partial_A$. Thus, in order to have a
solution for $\phi$ we must have $d:=\deg(L)\ge 0$.  The integration
of the second equation of \eqref{vortex} gives the {\it stability}
condition
\begin{equation}
  d < \frac{c^2}{2\pi}\mathrm{Area}(C_h)\,,
\end{equation}
which is necessary for the existence of solutions.
The sufficiency was also shown in \cite{Brad}.

The space $\mathcal{A}\times \Omega$ is equipped with a natural K{\"
  a}hler, hence symplectic structure. The action of the $U(1)$ gauge
group $\mathcal{G}$ on $\mathcal{A}\times \Omega$ is symplectic and
has a moment map given by $\mu(A,\phi)=\Lambda_\omega F_A -
\sqrt{-1}\abs{\phi}^2$.  Let $\mathcal{S}=\{(A,\phi)\in
\mathcal{A}\times \Omega \mid \phi \not \equiv 0\ \text{and}\ 
\bar\partial_A \phi=0\}$ be the set of solutions to the first equation
of the vortex equations. Then the moduli space of vortices is given by
the symplectic quotient
\begin{equation}\label{squot}
  \left\{\mu^{-1}\left(-\sqrt{-1}c^2\right)\cap
    \mathcal{S}\right\}/\mathcal{G}\,.
\end{equation}

The complex gauge group $\mathcal{G}^\CC$ acts on $\mathcal{A}\times
\Omega$ leaving $\mathcal{S}$ invariant. We can identify the complex
quotient
\begin{equation}\label{cquot}
  \mathcal{S}/\mathcal{G}^\CC\,,
\end{equation}
with the set of effective divisors of degree $d$ on $C_h$, hence with
the $d^{\text{th}}$ symmetric product $C_h^{(d)}$ which is a
smooth $d$ dimensional K{\" a}hler manifold. This is so since
every nonzero holomorphic section of an invertible sheaf determines an
effective divisor and vice versa up to scalars.  Indeed, there is a
natural morphism (the Abel-Jacobi map) 
\begin{equation}\label{AJmap}
\mathscr{A}^d:C_h^{(d)}\to
\Pic^d_{C_h}\,,  
\end{equation}
taking an effective divisor $D$ of degree $d$ to the invertible sheaf
$\mathcal{O}_{C_h}(D)$ such that every fiber
$(\mathscr{A}^d)^{-1}(\mathcal{O}_{C_h}(D))$ is a projective space
$\PP H^0(C_h,\mathcal{O}_{C_h}(D))\cong\abs{D}$. In other words,
\begin{equation}
  C_h^{(d)}\cong \{(L,U)\mid L\in \Pic^d_{C_h},\ U\subset
  H^0(C_h,L),\ \dim U=1\}\,.
\end{equation}
Let $K$ be a
canonical divisor of $C_h$.  If $d>2h-2$, the morphism $\mathscr{A}^d$
makes $C_h^{(d)}$ a projective bundle over $\Pic^d_{C_h}$ since
\begin{equation}
\Ext^1(\mathcal{O}_{C_h}(-D),\mathcal{O}_{C_h})\cong
H^1(C_h,\mathcal{O}_{C_h}(D))\cong
H^0(C_h,\mathcal{O}_{C_h}(K-D))^*=0\,,
\end{equation}
so that we have $\dim H^0(C_h,\mathcal{O}_{C_h}(D))=d+1-h$ by the
Riemann-Roch theorem.  This can be rephrased in the following way.
Let $\mathcal{P}$ be the Poincar{\' e} line bundle over
$\Pic^d_{C_h}\times C_h$ with $\mathcal{P}\vert_{\{{L}\}\times
  C_h}\cong {L}$ for every ${L}\in \Pic^d_{C_h}$ and let $\nu:
\Pic^d_{C_h}\times C_h\to \Pic^d_{C_h}$ be the projection. Then, if
$d>2h-2$, $\nu_*\mathcal{P}$ is a vector bundle of rank $d+1-h$ and we
have $C_h^{(d)}\cong \PP(\nu_*\mathcal{P})$.

If $d\ge 0$ and the stability condition is satisfied, the two
quotients \eqref{squot} and \eqref{cquot} are isomorphic. This is a
story familiar in the context of the Kobayashi-Hitchin correspondence
\cite{LuTe}.  Therefore, the moduli space of vortices can be
identified with the symmetric product $C_h^{(d)}$ with $d$ being the
number of vortices.

The cohomology of $C_h^{(d)}$ was studied by Macdonald \cite{Mac}. In
particular we have
\begin{equation}
  \sum_{d=0}^\infty \chi_{t,\tilde t}(C_h^{(d)})y^d=
  \frac{(1-ty)^h(1-\tilde ty )^h}{(1-y)(1- t {\tilde
      ty})}\,,\qquad(\abs{y}<1\,,\ \abs{t\tilde t y}<1)\,.
\end{equation}
This immediately leads to
\begin{equation}
  \sum_{d=0}^\infty \chi_{t}(C_h^{(d)})y^d=(1-ty)^{h-1}(1-y)^{h-1}\,,
  \qquad(\abs{y}<1\,,\ \abs{ty}<1)\,,
\end{equation}
and
\begin{equation}\label{eulergen}
  \sum_{d=0}^\infty \chi(C_h^{(d)})y^d=(1-y)^{2h-2}\,,\qquad (\abs{y}<1)\,.
\end{equation}
Thus we find that 
$\chi(C_0^{(d)})=d+1$, which is consistent with the isomorphism
$C_0^{(d)}\cong \PP^d$.
For $h\ge 1$ it follows that
\begin{equation}
  \chi(C_h^{(d)})=
\begin{cases}(-1)^d\binom{2h-2}{d} &\text{if $d \le 2h-2$},\\
  0 &\text{if $d>2h-2$}.
\end{cases}
\end{equation}
The vanishing of $\chi(C_h^{(d)})$ for $d>2h-2$ can also be seen as
follows.  As mentioned $C_h^{(d)}$ is a projective bundle over
$\Pic^d_{C_h}$. Since $\Pic^d_{C_h}$ is homeomorphic to $T^{2h}$, we see
that $\chi(C_h^{(d)})=\chi(\PP^{d-h})\chi(T^{2h})=0$.

Now going back to our problem, we suppose that the smooth Riemann
surface $C_h$ can be embedded in $X$ (or more generally a smooth
Calabi-Yau manifold $Y$).  In view of Corollary \ref{D2D0charge},
Remark \ref{rem:deg} and \eqref{eulergen}, the appropriate state
counting function of the bound system of a $D2$-brane wrapping once
around $C_h$ and $D0$-branes sticked to $C_h$ may be given by
\begin{equation}\label{twsteulergen}
  \sum_{d=0}^\infty
  \chi(C_h^{(d)})y^{d+1-h}=(y^{-1/2}-y^{1/2})^{2h-2}\,, \qquad
  (\abs{y}<1)\,.
\end{equation}
This expression obviously enjoys the symmetry property under the
exchange $y \leftrightarrow y^{-1}$. This is gratifying since the
variable $y$ will be identified with the one in the previous sections
and in that case the symmetry is required from the fact we are
considering a closed string theory.

\begin{rem}
  The shift of $D0$-brane charge can formally be incorporated by
  considering the line bundle $\tilde L=L\otimes
  \mathcal{O}_{C_h}(K)^{-1/2}$ instead of $L$ since $\deg(\tilde
  L)=\chi(C_h,L)$. This twisting of the line bundle is very much
  reminiscent of that in the theory of the Seiberg-Witten monopole
  equations for 4-manifolds \cite{Witmon}.  This should not be too
  much surprising since it is known that the vortex equations and the
  monopole equations are closely related \cite{Gar}. The vortex
  equations can be considered as the dimensional reduction of the
  monopole equations.  Indeed, the expression
  $(y^{-1/2}-y^{1/2})^{2h-2}$ is also equal to the Donaldson or
  Seiberg-Witten series of $C_{h}\times T^2$ \cite{MS} with the
  symmetry under the exchange $y\leftrightarrow y^{-1}$ being the
  charge conjugation symmetry of the monopole equations. For $h \ge
  1$, this was shown also in \cite{BJSV}.  There is subtlety when
  $h=0$ since $b_2^+(C_0\times T^2)=1$ and there is a wall-crossing
  phenomena ({\it cf.} \eqref{wall}). In this case a path integral
  justification requires the evaluation of the $u$-plane integral
  \cite{MW} which has been done in \cite{MM2}. 
\end{rem}

\begin{rem}
  Given a real 3-dimensional manifold $M$ we associate the variables
  $y_i$ to the generators of the free part of $H_1(M,\ZZ)$.  Then the
  Reidemeister torsion\footnote{The Reidemeister torsion is
    essentially equal to the Ray-Singer torsion \cite{RS, Chee,
      Mull}.} $\tau(M;y_i)$ of $M$ is closely related to the Alexander
  polynomial \cite{Tur}: If $b_1(M)>1$, $\tau(M;y_i)$ coincides with
  the Alexander polynomial $\Delta_M(y_i)\in \ZZ[y_i,y_i^{-1}]$ which
  can be made symmetric under the exchange $y_i \leftrightarrow
  y_i^{-1}$. If, on the other hand, $b_1(M)=1$ and $\partial
  M=\emptyset$, we have
\begin{equation}\label{Reidemeister}
  \tau(M;y)=\frac{\Delta_M(y)}{(y^{-1/2}-y^{1/2})^2}\,,\qquad
  \Delta_M(y)\in \ZZ[y,y^{-1}]\,,
\end{equation}
where $\Delta_M(y)$ is the Alexander polynomial symmetric under
the exchange $y \leftrightarrow y^{-1}$.
In particular, we have
\begin{equation}
 \tau(C_{h}\times S^1;y)=(y^{-1/2}-y^{1/2})^{2h-2}\,, 
\end{equation}
where $y$ is associated with $[S^1]$. See for instance,
\cite{Chee,Mull}.  According to Meng and Taubes \cite{MT},
$\tau(M;y_i)$ coincides with the Seiberg-Witten series of $M$ defined
through the 3-dimensional version of the Seiberg-Witten monopole
equations.  See also a recent work \cite{MM2} for the connection
between the Donaldson-Witten partition function and the Reidemeister
torsion. See also \cite{FS} for a relation between the Seiberg-Witten
series of 4-manifolds and knot theory. It is rather curious to note
that, in the following, we will encounter expressions quite similar to
\eqref{Reidemeister}.
\end{rem}

\begin{rem}
  Another reason for the significance of $(y^{-1/2}-y^{1/2})^{2h-2}$
  is the following.  Let $C_{h}$ be a rigid \cite{Pan} smooth curve of
  genus $h$ in a Calabi-Yau 3-fold $Y$ where ``rigid'' means that the
  normal bundle ${N}={N}_{C_h/Y}$ satisfies $H^0(C_{h},{N})=0$.  
Let
$p:\overline{\mathcal{C}}_{g,0}(C_h,[C_h]) \to
\overline{\mathcal{M}}_{g,0}(C_h,[C_h])$ be the universal curve and
$\mu:\overline{\mathcal{C}}_{g,0}(C_h,[C_h]) \to C_h$ the universal
evaluation map.
Then
  it was proved in \cite{Pan} that
\begin{equation}\label{Pid}
  (y^{-1/2}-y^{1/2})^{2h-2}= (-1)^{h-1}\sum_{g=h}^\infty
  x^{2g-2}\,m_{g \to h}\,\,,
\end{equation}
where $y=\exp(\sqrt{-1} x)$ and
\begin{equation}
  m_{g \to h}:= e(R^1p_*\mu^*{N})\cap [
  \overline{\mathcal{M}}_{g,0}(C_{h},[C_{h}]) ]^{\text{vir}}\,.
\end{equation}
Note that $m_g$ in \S 2 is equal to $m_{g\to 0}$ and $m_{h\to h}=1$.
Eq.\eqref{Pid} is important in the sense that it plays a key role in
relating the $D2$-$D0$ state counting and the Gromov-Witten
invariants.
\end{rem}

\subsection{A $D2$-brane moving in $K3$} \label{subsec:D2inK3}

As a warm-up for the next subsection we briefly recall the situation
where a single $D2$-brane (not bound to any $D0$-branes) moves in $X$.
This case was first studied in \cite{YZ}.

Let $C$ be an  irreducible and reduced curve
(which is not necessarily smooth) in $X$.  One can consider the
(component of) generalized Picard scheme $\Pic^d_C$ parametrizing
invertible sheaves of degree $d$ on $C$ up to isomorphisms. Although
$\Pic^d_C$ is not complete in general, one can consider its
compactification $\overline{\Pic}{}^d_C$ as the set of isomorphism
classes of rank-1 torsion-free sheaves of degree $d$ on $C$ where the
degree of a rank-1 torsion-free sheaf ${L}$ is defined by
$\chi(C,{L})-\chi(C,\mathcal{O}_C)$. Tensoring with an invertible
sheaf of degree $k$ gives an isomorphism :
$\overline{\Pic}{}^d_C\xrightarrow{\sim} \overline{\Pic}{}^{d+k}_C$.

Let $C_h\subset X$ be a connected nonsingular curve of genus $h$. Then
the complete linear system $\abs{C_h}$ is the set of all effective
divisors linearly equivalent to $C_h$ and $\abs{C_h}\cong \PP^h$. The
latter statement can be seen as follows. First we have $H^2(X,
\mathcal{O}_X(C_h))\cong H^0(X, \mathcal{O}_X(-C_h))^*=0$ by vanishing
theorem.  The exact sequence $0\to \mathcal{O}_X \to
\mathcal{O}_X(C_h)\to \omega_{C_h} \to 0$, where
$\omega_{C_h}:=\mathcal{O}_{C_h}(C_h)$ is a canonical sheaf on $C_h$,
leads, by using $H^1(X,\mathcal{O}_X)=0$, to an exact sequence $0\to
H^1(X,\mathcal{O}_X(C_h))\to H^1(C_h,\omega_{C_h}) \to
H^2(X,\mathcal{O}_X)\to 0$. Since the map $H^1(C_h,\omega_{C_h}) \to
H^2(X,\mathcal{O}_X)$ is surjective and both spaces are 1-dimensional,
the kernel $H^1(X,\mathcal{O}_X(C_h))$ must vanishes. Then, applying
the Riemann-Roch theorem, we obtain the desired result.

Setting $\mathcal{S}_h:=\abs{C_h}$, let $\mathcal{C}_h\subset 
\mathcal{S}_h\times X$ be the universal curve. For the flat family
$\mathcal{C}_h/\mathcal{S}_h$ we assume that all the fibers of the
structure morphism $p:\mathcal{C}_h\to \mathcal{S}_h$ are 
irreducible and reduced curves (of arithmetic genus $h$).

A $\mathcal{S}_h$-flat $\mathcal{O}_{\mathcal{C}_h}$ module
$\mathcal{F}$ is called a relative rank-1 torsion-free (resp.\ 
invertible) sheaf of degree $d$ on $\mathcal{C}_h/\mathcal{S}_h$ if at
each point $s\in \mathcal{S}_h$ the fiber $\mathcal{F}_s$ is a rank-1
torsion-free (resp.\ invertible) sheaf of degree $d$ on the fiber
$(\mathcal{C}_h)_s$.  

Denote by $j:\bar{\mathcal{J}}^d_h \to \mathcal{S}_h$ 
the relative compactified Picard
scheme $\overline{\Pic}{}^d_{\mathcal{C}_h/\mathcal{S}_h} \to \mathcal{S}_h$ 
of degree $d$ which is the set of isomorphism classes of relative rank-1
torsion-free sheaves of degree $d$ on $\mathcal{C}_h/\mathcal{S}_h$.

As before, tensoring with a relative invertible sheaf of degree $k$
provides an isomorphism
\begin{equation}
  \sigma_k:\bar\mathcal{J}^d_h \overset{\sim}{\longrightarrow}
  \bar\mathcal{J}^{d+k}_h\,.
\end{equation}
Since the fibers of $p$ are Gorenstein, the relative dualizing sheaf
$\omega_{\mathcal{C}_h/\mathcal{S}_h}$ is a relative invertible sheaf
of degree $2h-2$ on $\mathcal{C}_h/\mathcal{S}_h$. Thus we can use
this for the construction of $\sigma_{2h-2}$.

Also the map $F \mapsto F^*=
\mathcal{H}\!om_{\mathcal{O}_{(\mathcal{C}_h)_s}}
(F,\mathcal{O}_{(\mathcal{C}_h)_s})$, $s \in \mathcal{S}_h$,
$F \in (\bar\mathcal{J}^d_h)_s$ determines an isomorphism
\begin{equation}
  \epsilon:\bar\mathcal{J}^d_h \overset{\sim}{\longrightarrow}
  \bar\mathcal{J}^{-d}_h\,,
\end{equation}
Especially, if we set $\epsilon_\omega:= \sigma_{2h-2}\circ \epsilon$,
we obtain
\begin{equation}\label{epsilon_omega}
  \epsilon_\omega:\bar\mathcal{J}^d_h \overset{\sim}{\longrightarrow}
  \bar\mathcal{J}^{2h-2-d}_h\,.
\end{equation}
This map is obtained by $F\mapsto F^\star :=
\mathcal{H}\!om_{\mathcal{O}_{(\mathcal{C}_h)_s}}(F,
(\omega_{\mathcal{C}_h/\mathcal{S}_h})_s)=F^*\otimes
(\omega_{\mathcal{C}_h/\mathcal{S}_h})_s$. We note that
$\deg(F^*)=-\deg(F)$ and
$\chi((\mathcal{C}_h)_s, F^\star)=-
\chi((\mathcal{C}_h)_s, F)$.

It is known \cite{Muk1,Muk2,Muk3} that $\bar\mathcal{J}^d_h$ is an
irreducible symplectic manifold of dimension $2h$ and
\begin{equation}\label{Jacobianmoduli}
  \bar\mathcal{J}^d_h \cong \mathcal{M}(0,C_h,d+1-h)\,.
\end{equation}

Yau and Zaslow \cite{YZ} proposed that the state counting function of
a single $D2$-brane moving in $X$ is given by
\begin{equation}\label{YZformula}
  \sum_{h=0}^\infty \chi\left(\bar\mathcal{J}^0_h\right)
q^{h-1}=\inv{\eta(\tau)^{24}}\,.
\end{equation}
This proposal and its implication for the enumeration of nodal
rational curves in $X$ were further studied in \cite{Bea,FGS}.

\subsection{$D0$-branes bound to a $D2$-brane moving in $K3$} 
\label{subsec:D0D2inK3}

In order to extend the results in \S \ref{subsec:rigidD@D0} and
describe the bound states of $D0$-branes and a $D2$-brane moving in
the $K3$ surface $X$, there are two basically different but equivalent
points of view. As we saw in \S \ref{subsec:rigidD@D0}, the moduli
spaces of vortices are isomorphic to the symmetric products of
(smooth) curves.  Going to the relative situation, we are led to
consider relative Hilbert schemes of points on curves. This gives the
first approach. On the other hand, we also observed that the moduli
spaces of vortices are those of pairs consisting of line bundles on
curves and their sections. This latter viewpoint can be generalized
and we are led to consider the so-called coherent systems
\cite{LeP}.

\subsubsection{Relative Hilbert schemes}\label{subsubsec:relHilb}

We start with the first viewpoint.  We assume the same setting as in
\S \ref{subsec:D2inK3}. In particular all the fibers of
$p:\mathcal{C}_h\to \mathcal{S}_h$ are irreducible and reduced.

Let $X$ be polarized by an ample invertible sheaf $H_X$.
Each fiber $(\mathcal{C}_h)_s$ is polarized by
$(\iota_s)^*H_X$ where
$\iota_s:(\mathcal{C}_h)_s\hookrightarrow X$ is the inclusion.

Now fix a relative rank-1 torsion-free sheaf $\mathcal{F}$ of degree
$k$ on $\mathcal{C}_h/\mathcal{S}_h$.  Since $\mathcal{F}$ is
$\mathcal{S}_h$-flat and $\mathcal{S}_h$ is connected, the 
Hilbert polynomial of a fiber,
\begin{equation}
  P_{\mathcal{F}_s/(\mathcal{C}_h)_s}(n)= \deg((\mathcal{C}_h)_s)\, n+
\chi((\mathcal{C}_h)_s, \mathcal{F}_s)=\deg((\mathcal{C}_h)_s)\, n+k+1-h\,,
\end{equation}
is constant as a function of $s\in \mathcal{S}_h$.  Fix a positive
integer $d$. Then  the relative Quot-scheme
$q:\Quot^d_{\mathcal{F}/\mathcal{C}_h/\mathcal{S}_h} \to
\mathcal{S}_h$ parametrizes
\begin{equation}\label{quot:seq}
  0\longrightarrow E \longrightarrow \mathcal{F}_s\longrightarrow
  G\longrightarrow 0\,, \qquad (s\in \mathcal{S}_h)\,,
\end{equation}
where $E$ and $G$ are coherent sheaves on $(\mathcal{C}_h)_s$
satisfying $P_{G/(\mathcal{C}_h)_s}(n)=d$. Let $\mathcal{E}$ be the
universal subsheaf and $\mathcal{G}$ the universal quotient sheaf
corresponding respectively to $E$ and $G$ in \eqref{quot:seq}:
\begin{equation}
  0 \longrightarrow \mathcal{E} \longrightarrow (q \times
  id_X)^*\mathcal{F} \longrightarrow \mathcal{G} \longrightarrow 0.
\end{equation}
For simplicity, we set $q^{\#}\mathcal{F}:=(q \times id_X)^*\mathcal{F}$.  
Notice
that
\begin{equation}
  P_{\mathcal{E}_u/(\mathcal{C}_h)_{q(u)}}(n)=
 \deg((\mathcal{C}_h)_{q(u)})\,
  n+k-d+1-h\,,\quad u \in \Quot^d_{\mathcal{F}/\mathcal{C}_h/\mathcal{S}_h}.
\end{equation}

As for the $D$-brane charges, we see that
\begin{equation}\label{Dchargeoffibers}
  \begin{split}
    {Q}((\iota_{q(u)})_*\mathcal{E}_u)&=(0,[C_h],k-d+1-h)\,,\\ 
    {Q}((\iota_{q(u)})_*q^{\#}\mathcal{F}_u)&=(0,[C_h],k+1-h) \,,\\ 
    {Q}((\iota_{q(u)})_*\mathcal{G}_u)&=(0,0,d)\,,
  \end{split}
\end{equation}
where $u \in \Quot^d_{\mathcal{F}/\mathcal{C}_h/\mathcal{S}_h}$.
 Note that the $D$-brane charges
\eqref{Dchargeoffibers} are constant as functions of
$u \in \Quot^d_{\mathcal{F}/\mathcal{C}_h/\mathcal{S}_h}$.
This is intuitively plausible since
``charges'' must be conserved for a continuous family of curves.

An important case is $\mathcal{F}=\mathcal{O}_{\mathcal{C}_h}$.  By a
slight abuse of notation, we denote by $\mathcal{C}_h^{[d]}$ the
relative Hilbert scheme $\Hilb^d_{\mathcal{C}_h/\mathcal{S}_h}=
\Quot^d_{\mathcal{O}_{\mathcal{C}_h}/\mathcal{C}_h/\mathcal{S}_h}$
parametrizing $\mathcal{S}_h$-flat subschemes of $\mathcal{C}_h$
relatively of dimension $0$ and length $d$. Obviously in this case we
have ${Q}((\iota_{q(u)})_*\mathcal{E}_u)=(0,[C_h],-d+1-h)$. As we will
see later, $\mathcal{C}_h^{[d]}$ is projective and smooth of dimension
$d+h$.

One can construct the (degree $d$ component of) Abel-Jacobi map
\cite{AK} which is the forgetful morphism
\begin{equation}\label{AJrel}
  \mathscr{A}^d_\mathcal{F}:
\Quot^d_{\mathcal{F}/\mathcal{C}_h/\mathcal{S}_h}\longrightarrow
   \bar\mathcal{J}^{k-d}_h\,,
\end{equation}
obtained by sending $u\in
\Quot^d_{\mathcal{F}/\mathcal{C}_h/\mathcal{S}_h}$ to the isomorphism
class of $\mathcal{E}_u$.  ({\it cf.\/} \eqref{Jacobianmoduli} and
\eqref{Dchargeoffibers}.)  The fiber of $\mathscr{A}^d_\mathcal{F}$ at
$t\in \bar\mathcal{J}^{k-d}_h$ is isomorphic to
$\PP\Hom_{(\mathcal{C}_h)_{j(t)}}(I,\mathcal{F}_{j(t)})$ where $I$ is
a rank-1 torsion-free $\mathcal{O}_{(\mathcal{C}_h)_{j(t)}}$-module
representing $t$. The map $\mathscr{A}^d_\mathcal{F}$ is smooth over
$t$ if $\Ext^1_{(\mathcal{C}_h)_{j(t)}}(I,\mathcal{F}_{j(t)})=0$.  See
\cite{AK} for more details.

By tensoring with a relative invertible sheaf $\mathcal{L}$ of degree
$\ell$ we obtain a commutative diagram \cite{AK}:
\begin{equation}\label{CD1}
\xymatrix@R+3mm@C+8mm{
\Quot^d_{ {\mathcal{F}}/{\mathcal{C}}_h/{\mathcal{S}}_h }
\ar@*{[thinner]}[r]^-{\sim}_-{\otimes \mathcal{L}}
\ar[d]_-{{\mathscr{A}}^d_{\mathcal{F}}} &
\Quot^d_{ {\mathcal{F}}\otimes{\mathcal{L}}/{\mathcal{C}}_h/{\mathcal{S}}_h }
\ar[d]^-{{\mathscr{A}}^d_{ {\mathcal{F}}\otimes{\mathcal{L}} }}\\
{\bar{\mathcal{J}}}^{k-d}_h\ar[r]^-{\sim}_-{\sigma_\ell}&
{\bar{\mathcal{J}}}^{\ell+k-d}_h}
\end{equation}

Since $\omega_{\mathcal{C}_h/\mathcal{S}_h}=:\omega$ is a relative
invertible sheaf of degree $2h-2$ 
(indeed it is isomorphic to
$\mathcal{O}_{\mathcal{C}_h}(\mathcal{C}_h)$), we may take
$\mathcal{L}=\omega$ in \eqref{CD1}. Thus we obtain a commutative
diagram

\begin{equation}\label{CD:Hilb}
 \xymatrix@R-2mm@C+5mm{ 
{\mathcal{C}}_h^{[d]}\ar[rr]^-{\sim}_-{\otimes\omega}
   \ar[dd]_-{\mathscr{A}^d_\mathcal{O}} \ar[dr]& &
   \Quot^d_{ \omega/\mathcal{C}_h/\mathcal{S}_h }
   \ar[dd]^-{\mathscr{A}^d_{ \omega}}\ar[dl]\\
  &{\bar\mathcal{J}^{d}_h}&\\
 {\bar\mathcal{J}^{-d}_h} \ar[ur]^>>>>>>>>>>{\text{
   \begin{rotate}{45}
     $\sim$
   \end{rotate}}
}_-{\epsilon}
 \ar[rr]^-{\sim}_-{\sigma_{2h-2}} & &
{\bar\mathcal{J}^{2h-2-d}_h}
\ar[ul]_>>>>>>>>>{\text{
   \begin{rotate}{-40}
     $\sim$
   \end{rotate}}}^-{\epsilon_\omega}}
\end{equation}
where $\mathcal{O}:=\mathcal{O}_{\mathcal{C}_h}$.  The down diagonal
arrows may be viewed as extensions of \eqref{AJmap}.  In particular
the south-east arrow $\mathcal{C}_h^{[d]}\to \bar\mathcal{J}^{d}_h$ is
obtained by sending $u\in \mathcal{C}_h^{[d]}$ to (the isomorphism
class of) $\mathcal{E}^*_u$ where $\mathcal{E}$ is the universal
subsheaf of $\mathcal{O}$.  We note that
${Q}((\iota_{q(u)})_*\mathcal{E}^*_u)=(0,[C_h],d+1-h)$.

When $\mathcal{F}=\omega$, the smoothness condition of
$\mathscr{A}^d_\omega$ over  $t\in \bar
\mathcal{J}_h^{2h-2-d}$ becomes 
\begin{equation}
\Ext^1_{(\mathcal{C}_h)_{j(t)}}(I,\omega_{j(t)})\allowbreak\cong \allowbreak
H^0((\mathcal{C}_h)_{j(t)},I)^*\allowbreak=\allowbreak 0\,.  
\end{equation}
Since $\deg(I)=2h-2-d$, we see that if $d>2h-2$,
$\mathscr{A}^d_\omega$ is smooth over every point of $\bar
\mathcal{J}_h^{2h-2-d}$.  Since
$\Hom_{(\mathcal{C}_h)_{j(t)}}(I,\omega_{j(t)})\cong
H^1((\mathcal{C}_h)_{j(t)},I)^* $ and $\chi(
(\mathcal{C}_h)_{j(t)},I)=h-1-d$, the fibers of $\mathscr{A}^d_\omega$
for $d>2h-2$ are isomorphic to
$\PP\Hom_{(\mathcal{C}_h)_{j(t)}}(I,\omega_{j(t)})\cong \PP^{d-h}$.
Precisely the same result holds for $\mathscr{A}^d_\mathcal{O}$ since
we have the commutative diagram \eqref{CD:Hilb}.  We refer again to
\cite{AK} for more details.

It is natural to set $\mathcal{C}_h^{[0]}:=\mathcal{S}_h\cong
\PP^h$. We also have an isomorphism $\mathcal{C}_h^{[1]}\cong
\mathcal{C}_h$.

With these preliminaries, we may regard
$\mathcal{C}_h^{[d]}$ as the  moduli space of the $D2$-$D0$
bound states in $X$.
In order to count the $D2$-$D0$ bound states, we are naturally led to
consider a combination
\begin{equation}
  \sum_{h=0}^\infty\sum_{d=0}^\infty \chi(\mathcal{C}_h^{[d]})\,
  q^{h-1}y^{d+1-h}\,,
\end{equation}
where we stress that the exponent of $y$ measures the $D0$-brane
charge.  In the rest of this sub-subsection, we assume for every $h\ge
0$ that $C_h$ satisfies the condition ($\star$1) to be explained in \S
\ref{subsubsec:coh.sys}.  Then we have
\begin{thm}\label{relHilb}
  For $0<\abs{q}<\abs{y}<1$,
\begin{equation}\label{relHilbeq}
  \sum_{h=0}^\infty\sum_{d=0}^\infty \chi(\mathcal{C}_h^{[d]})\,
  q^{h-1}y^{d+1-h}=\inv{\chi_{10,1}(\tau,\nu)}\,,
  \end{equation}
where
\begin{equation}\label{FJcoeff} 
  \begin{split}
   \chi_{10,1}(\tau,\nu)&=\eta(\tau)^{24} E(\tau,\nu)^2\\[1mm]
&=(y^{-1/2}-y^{1/2})^2\, q\prod_{n=1}^\infty(1-q^n)^{20}(1-q^n y)^2
(1-q^n y^{-1})^2\,.
  \end{split}
\end{equation}
\end{thm}
This result may be viewed as an amalgamation of \eqref{twsteulergen}
and \eqref{YZformula}.  The proof of this theorem is given later when
we reformulate the problem in terms of coherent systems.

By putting $w=q/y$ we can
cast Theorem \ref{relHilb} in a more symmetric form:
\begin{cor}
For $0<\abs{w}<1$, $0<\abs{y}<1$, 
\begin{equation}
  \begin{split}
&\sum_{h=0}^\infty\sum_{d=0}^\infty \chi(\mathcal{C}_h^{[d]})\, w^h
  y^d  \\
&\qquad = \prod_{n=1}^\infty
  \inv{(1-(wy)^n)^{20}(1-(wy)^{n-1} w)^2(1-(wy)^{n-1} y)^2}\,.
\end{split}
\end{equation}
\end{cor}
Since the right hand side is symmetric under the exchange of $w$ and
$y$ we readily obtain
\begin{cor}
  (degree-genus duality)
  \begin{equation}\label{dg-duality}
    \chi(\mathcal{C}_h^{[d]})=\chi(\mathcal{C}_d^{[h]})\,.
  \end{equation}
\end{cor}

We note that $\chi_{10,1}(\tau,\nu)$ is the (unique up to a
multiplicative constant) cusp Jacobi form of weight $10$ and index $1$
and can alternatively be expressed in terms of the Eisenstein(-Jacobi)
series \cite{EZ}:
\begin{equation}
 \chi_{10,1}(\tau,\nu)= \frac{E_6(\tau)\,E_{4,1}(\tau,\nu)
-E_4(\tau)\,E_{6,1}(\tau,\nu)}{144}\,.
\end{equation}
In fact $\chi_{10,1}(\tau,\nu)$ is the first Fourier-Jacobi coefficient
of the Igusa cusp form of weight 10:
\begin{equation}
 \chi_{10}(\Omega)=\sum_{n=1}^\infty q^n \chi_{10,n}(\tau',\nu)\,,
\end{equation}
where $\Omega=\left(
\begin{smallmatrix}
  \tau & \nu \\ \nu& \tau'
\end{smallmatrix}\right) \in \HH_2$. The infinite product representation
\eqref{FJcoeff} has a beautiful extension to $\chi_{10}(\Omega)$ as
found by Gritsenko and Nikulin \cite{GN}. They applied the exponential
lifting procedure by Borcherds to a particular weak Jacobi form of
weight $0$ and index $1$ which, as we have observed in \cite{Kaw0},
happens to be the elliptic genus of $K3$ surfaces. For a partial
review on the relations between $\chi_{10}(\Omega)$ and $K3$ surfaces,
see \cite{Kaw1}.

Now we quote the following result from \cite{GS,Che,EGL},
\begin{lem}
For $\abs{q}<\min(1,\abs{y},\abs{y}^{-1})$,
\begin{equation}
  \frac{1}{\chi_{10,1}(\tau,\nu)} = \sum_{h=0}^\infty q^{h-1}
  \frac{y^{-h} \chi_y(X^{[h]})}
  {(y^{-1/2}-y^{1/2})^2}\,.
\end{equation}  
\end{lem}
\begin{rem}
  This is not precisely in the form presented in \cite{GS,Che,EGL} but
  is trivially related to it. 
\end{rem}

One may interpret this result in an alternative way since the inverse
of $\chi_{10}(\Omega)$ has also a very nice expansion.
Indeed, the result of \cite{DMVV} can be rephrased as
\begin{equation}\label{invIgusa}
  \frac{1}{\chi_{10}(\Omega)} =\sum_{h=0}^\infty q^{h-1}\,
  \frac{\mathcal{E}^{\text{orb}}_{X^{(h)}}(\tau',\nu)}{
\chi_{10,1}(\tau',\nu)}\,,
\end{equation}
where $\mathcal{E}^{\text{orb}}_{X^{(h)}}(\tau,\nu)$ is the orbifold
elliptic genus of $X^{(h)}$. 
Comparing the limits $\Ima \tau' \rightarrow \infty$ on both sides of
\eqref{invIgusa} one finds that
\begin{equation}\label{chiy}
  \frac{1}{\chi_{10,1}(\tau,\nu)} = \sum_{h=0}^\infty q^{h-1}
  \frac{y^{-h} \chi^{\text{orb}}_y ( X^{(h)} )}
  {(y^{-1/2}-y^{1/2})^2}\,,
\end{equation}
where $\mathcal{E}^{\text{orb}}_{X^{(h)}}(\tau ,\nu)=y^{-h}
  \chi^{\text{orb}}_y (X^{(h)})+O(q)$. It thus follows that
\begin{lem}\label{equality-of-chiy}
  \begin{equation}
   \chi^{\text{\em orb}}_y ( X^{(h)} )=\chi_y ( X^{[h]} )=
\chi_y(\bar\mathcal{J}^d_h)\,.
  \end{equation}
\end{lem}
The second equality will be proved in Theorem \ref{thm:deform}.

\begin{rem}
Naturally we are led to the conjecture:
\begin{equation}
  \mathcal{E}^{\text{orb}}_{X^{(h)}}(\tau,\nu) =
  \mathcal{E}_{X^{[h]}}(\tau,\nu) =
  \mathcal{E}_{\bar\mathcal{J}^d_h}(\tau,\nu)\,.
\end{equation}
Unfortunately this 
does not follow simply from Lemma \ref{equality-of-chiy}. 
\end{rem}

Hence, as a corollary to Theorem \ref{relHilb},  we find that
\begin{cor} \label{relHilbeulergen}
For any nonnegative integer $h$ and $\abs{y}<1$,
\begin{equation}\label{eq:relHilbeulergen}
  \begin{split}
    \sum_{d=0}^\infty \chi(\mathcal{C}_h^{[d]})\, y^{d+1-h}
    &=\frac{y^{-h} \chi_y(\bar\mathcal{J}^d_h)}
    {(y^{-1/2}-y^{1/2})^2}\\ 
&= (h+1)(y^{-1/2}-y^{1/2})^{2h-2}
    +\cdots\\
&\qquad\qquad \quad+ \chi(\bar\mathcal{J}^d_h)
(y^{-1/2}-y^{1/2})^{-2}\,,
\end{split}
\end{equation}
where the last expression  represents  the expansion in    
$(y^{-1/2}-y^{1/2})^{2k-2}$ for $k=h,h-1,\dots,1,0$.
\end{cor}
This should be considered as a generalization of
\eqref{twsteulergen} and it immediately implies
\begin{cor}
For $d>2h-2$, 
  \begin{equation}
    \chi(\mathcal{C}_h^{[d]})=(d+1-h) 
\chi(\bar\mathcal{J}^d_h)\,.
  \end{equation} 
\begin{proof} 
  If $d>2h-2$ the only relevant part for the calculation of
  $\chi(\mathcal{C}_h^{[d]})$ is the term
  $$\chi(\bar\mathcal{J}^d_h) (y^{-1/2}-y^{1/2})^{-2}$$ in
  \eqref{eq:relHilbeulergen}. Then use the series expansion
  \eqref{wall} for $\abs{y}<1$.
\end{proof}
\end{cor}
This result is consistent with the earlier mentioned condition for the
smoothness of the Abel-Jacobi map $\mathscr{A}^d_\mathcal{O}$ since
$\chi(\PP^{d-h})=d+1-h$.  Also comparing the coefficients of $y^{1-h}$
in \eqref{eq:relHilbeulergen} one finds that
$\chi(\mathcal{C}_h^{[0]})=h+1$, which is consistent with
$\mathcal{C}_h^{[0]}\cong \PP^h$.

\begin{rem}
As remarked before, 
the expression
\begin{equation}
  \frac{y^{-h} \chi_y(\bar\mathcal{J}^d_h)}
  {(y^{-1/2}-y^{1/2})^2}=\frac{y^{-h} \chi_y(X^{[h]})}
  {(y^{-1/2}-y^{1/2})^2}\,,
\end{equation}
is reminiscent of \eqref{Reidemeister} with the numerators playing the
role of the symmetrized Alexander polynomial. We are not sure whether
this is merely a coincidence or suggests the existence of certain
theories on $X\times T^2$ or $X\times S^1$ which give rise to the {\it
  relative\/} versions of the Seiberg-Witten invariants.
\end{rem}

\subsubsection{Coherent systems} \label{subsubsec:coh.sys}

Now we turn to the second viewpoint of $D2$-$D0$ bound system on
$X$. In the following we simplify notations by using $\langle\ ,\ 
\rangle$ for $\langle\ ,\ \rangle_X$.

Suppose that we are given a coherent sheaf $E$ on $X$ and a vector
subspace $U$ of $ H^0(X,E)\cong \Hom(\mathcal{O}_X,E)$.  The pair
$(E,U)$ is called a coherent system \cite{LeP}. A coherent system
$(E,U)$ is called of dimension $m$ if $\dim(E)=m$.
One may equivalently define a
coherent system as a sheaf homomorphism $f:U\otimes \mathcal{O}_X \to
E$, where $U$ is a finite dimensional vector space and $E$ is a
coherent sheaf, with the property that $H^0(f):U\hookrightarrow
H^0(X,E)$ is injective.
Throughout this sub-subsection, 
we assume that $\mathcal{M}(v)$ consists of slope stable sheaves.

\begin{rem}
Assume that $v \cap [X]=(r[X],[C],a)$ with a primitive $[C]$.
Then $\mathcal{M}(v)$ consists of slope stable sheaves $E$ for a general
$H_X$ if (i) $r>0$, (ii) $r=0$ and $a \ne 0$, or
(iii) $r=0$ and $|C|$ consists of irreducible and reduced members.
\end{rem}

\vspace{1pc}
   
Let\footnote{As remarked before, we use the same notations for
  isomorphism classes and their representatives.}
\begin{equation}
  \Syst^n(v):=\{(E,U) \mid E \in \mathcal{M}(v), U \subset H^0(X,E),
  \dim U=n \}
\end{equation}
denote the coarse moduli space of coherent systems constructed by Le
Potier \cite{LeP}.  Thus $\Syst^n(v)$ is a projective scheme.

\begin{definition}
We set
\begin{equation}
\begin{split}
 \mathcal{M}(v)_i &:=\{E \in \mathcal{M}(v)\mid \dim H^0(X,E)=i \},\\
 \Syst^n(v)_i &:=p^{-1}_v(\mathcal{M}(v)_i),
\end{split}
\end{equation}
where $p_v:\Syst^n(v) \to \mathcal{M}(v)$ is the natural
projection.
\end{definition}

Let $C_h$ be an effective divisor on $X$ satisfying $C_h^2=2h-2$.  We
consider the following two conditions on $C_h$:
\smallskip
\begin{enumerate}
\item[($\star$1)]
{\em There is an ample line bundle $H$ such that}
\begin{equation}\label{eq:condition1}
 C_h\cdot H=\min \{L\cdot H \mid L \in \Pic(X),\ L\cdot H>0 \}.
\end{equation}
\item[($\star$2)]
{\em Every member of $|C_h|$ is irreducible and reduced.}
\end{enumerate}
\smallskip
\begin{rem}
 Obviously, the condition ($\star$1) implies the condition ($\star$2). 
\end{rem}

\begin{rem}\label{rem:C_h-cond}
  If $\Pic(X)={\ZZ}\, C_h$ with $h>1$, then $C_h$ satisfies
  ($\star$1).  In the moduli space of polarized $K3$ surfaces of degree
  $2h-2$, the locus of $(X,C_h)$ with $\rk(\Pic(X))>1$ is countable
  union of hypersurfaces. Hence for a general point $(X,C_h)$,
  $\Pic(X)={\ZZ}\, C_h$.  If $\pi:X \to {\PP}^1$ is an elliptic $K3$
  surface with a section such that $\Pic(X)={\ZZ}\,\sigma \oplus
  {\ZZ}\,f$, where $\sigma$ is a section of $\pi$ and $f$ a fiber of
  $\pi$, then $C_h=f$ satisfies ($\star$1) with $h=1$ and $C_h=\sigma$
  satisfies ($\star$1) with $h=0$.  Indeed, $\sigma+3f$ is ample and
  $f\cdot (\sigma+3f)=\sigma\cdot(\sigma+3f)=1$.
\end{rem}

\begin{rem}
Under ($\star$2), $|C_h|$ always contains a smooth curve 
[\citen{Fri}, p.133--p.135].
\end{rem}

\begin{rem}
Whenever we assume the condition ($\star$1), we use $H$ in 
\eqref{eq:condition1} for the polarization of $X$.
\end{rem}
Let $\Gr(k,l)$ denote the Grassmannian parametrizing $l$-dimensional
vector subspaces of $\CC^k$.  The following is a consequence of
[\citen{Yos1}, Lem.~2.1,\ Lem.~2.4]:
\begin{lem}
  Assume that $C_h$ satisfies {\em ($\star$1)} and $n \leq r$. Define
  $v, w \in H^{2*}(X,{\mathbb Z})$ by $v \cap [X]=(r[X],[C_h],a)$ and
  $w \cap [X]=((r-n)[X],[C_h],a-n)$. Any element $f:U\otimes
  \mathcal{O}_X \to E$ of $\Syst^n(v)$ is an injection and $\coker f$
  is a (slope) stable sheaf.  Hence we have a morphism
\begin{equation}
\begin{matrix}
q_v:\Syst^n(v) & \longrightarrow & \mathcal{M}(w)\\
&\\
(f:U \otimes \mathcal{O}_X \hookrightarrow E) & \longmapsto & \coker f\,.
\end{matrix}
\end{equation}
Moreover, by setting $m=n-(r+a)$,  we obtain the following diagram:
\begin{equation}\label{eq:corresp-diagram}
\xymatrix@R=5pt{&\Syst^n(v)_i\ar[dl]_-{p_v}\ar[dr]^-{q_v}&\\
      \mathcal{M}(v)_i& &  \mathcal{M}(w)_{i-n} } 
\end{equation}
where $p_v$ is an \'{e}tale locally trivial $\Gr(i,n)$-bundle and
$q_v$ is an \'{e}tale locally trivial $\Gr(m+i,n)$-bundle.
\end{lem}

More precisely, we proved Lemma 2.1 in \cite{Yos1} under the assumption
$\Pic(X)={\ZZ}\,C_h$.  Since the same proof as there works under the
assumption ($\star$1), we obtain the diagram
\eqref{eq:corresp-diagram}.

The following is well-known.
\begin{lem}\label{lem:proj-dim}
  Let $E$ be a torsion-free sheaf or a coherent sheaf of pure
  dimension 1 on $X$.  Let $\phi:V_0 \to E$ be a surjective
  homomorphism from a locally-free sheaf $V_0$.  Then $\ker \phi$
   is a locally-free sheaf or $\ker \phi=0$.
\end{lem}
Indeed for a torsion-free or a pure dimension 1 sheaf $E$,
$\mathrm{depth}_{\mathcal{O}_{X,x}} E_x \geq 1$ for all point $x \in
X$, where $\mathcal{O}_{X,x}$ and $E_x$ are the stalks of
$\mathcal{O}_X$ and $E$ at $x$ respectively ({\it cf.\/} [\citen{HL},
1.1]). Since $X$ is smooth of dimension 2, the homological dimension
$\mathrm{hd}_{\mathcal{O}_{X,x}}(E_x)$ of $E_x$ satisfies an equality
$\mathrm{hd}_{\mathcal{O}_{X,x}}(E_x)+
\mathrm{depth}_{\mathcal{O}_{X,x}} E_x=2$.  Hence
$\mathrm{hd}_{\mathcal{O}_{X,x}}(E_x) \leq 1$, which implies our
claim.

\vspace{1pc}
 
The next Lemma is an extension of [\citen{Yos1}, Lem 5.2].
\begin{lem}\label{lem:smoo}
  Under the condition {\em($\star$1)}, $\Syst^n(v)$ is a smooth scheme
  of dimension $\langle v,v \rangle+2-n(n+\langle v_1,v \rangle)$,
  where $v_1\cap [X]=([X],0,1)$, namely $v_1=v(\mathcal{O}_X)$.
\end{lem}

\begin{proof}
Let $\Lambda=(E,U)$ be a point of $\Syst^n(v)$.
By He \cite{He},
the Zariski tangent space of $\Syst^n(v)$ at $\Lambda$
is given by
$
\EExt^1(\Lambda,\Lambda),
$
the obstruction of infinitesimal liftings 
belong to the kernel of the composition of homomorphisms
\begin{equation}
\tau:\EExt^2(\Lambda,\Lambda)\to
\Ext^2(E,E) \overset{tr}{\to} H^2(X,\mathcal{O}_X)\,,
\end{equation}
and 
\begin{equation}
\EExt^2(\Lambda,\Lambda)
\cong \EExt^2(U \otimes \mathcal{O}_X \to E,E),
\end{equation}
where $\EExt^*(U \otimes \mathcal{O}_X \to E, \ast)$
is the hypercohomology associated to the complex
$U \otimes \mathcal{O}_X \to E$. 
Moreover there is an exact sequence 
\begin{equation}
\xymatrix@R=0pt{ 0\ar[r] & \EExt^0(\Lambda,\Lambda)\ar[r] &  
\Hom(E,E) \ar[r] &\Hom(U \otimes \mathcal{O}_X,E)/V\\
 \ar[r] & \EExt^1(\Lambda,\Lambda)\ar[r] & \Ext^1(E,E)\ar[r] & 
 \Ext^1(U \otimes \mathcal{O}_X,E)\\
\ar[r] & \EExt^2(\Lambda,\Lambda)\ar[r] & \Ext^2(E,E)
\ar[r] & \Ext^2(U \otimes \mathcal{O}_X,E)=0 }
\end{equation}
where $V:=\im(\Hom(U \otimes \mathcal{O}_X,U \otimes \mathcal{O}_X) \to 
\Hom(U \otimes \mathcal{O}_X,E))$. 
Then the Serre dual of $\tau$ is the composition of homomorphisms
\begin{equation}\label{eq:Serre-dual}
H^0(X,\mathcal{O}_X) \to \Hom(E,E) \hookrightarrow
\HHom(E,U \otimes \mathcal{O}_X \to E)\,.
\end{equation}

So we shall prove that
$\HHom(E,U \otimes \mathcal{O}_X \to E) \cong \CC$.
Let 
\begin{equation}
0 \longrightarrow \mathcal{O}_X \otimes \Ext^1(E,\mathcal{O}_X)^*
\longrightarrow G \longrightarrow E \longrightarrow 0
\end{equation}
be the universal extension, {\it i.e.\/}
the extension class corresponds to
 the identity element in
 \begin{equation}
\End(\Ext^1(E,\mathcal{O}_X)) \cong \Ext^1(E,\mathcal{O}_X \otimes
\Ext^1(E,\mathcal{O}_X)^*)\,.
\end{equation}
We set $i:=\dim \Ext^1(E,\mathcal{O}_X)$.  Since $\dim \HHom(E,U \otimes
\mathcal{O}_X \to E)\ge 1$  by
\eqref{eq:Serre-dual}, it is sufficient to prove that
\begin{enumerate}
\item[(1)] $ \HHom(E,U \otimes \mathcal{O}_X \to E ) \longrightarrow
 \HHom(G,U \otimes \mathcal{O}_X \to E)$
is injective,
\item[(2)] $\HHom(G,U \otimes \mathcal{O}_X \to E) \cong \CC$.
\end{enumerate}

{\noindent $\bullet$} Proof of (1):
Since there is an exact sequence
\begin{equation}
 \EExt^{-1}(\mathcal{O}_X^{\oplus i},U \otimes \mathcal{O}_X \to E) 
\longrightarrow
 \HHom(E,U \otimes \mathcal{O}_X \to E) \longrightarrow
 \HHom(G,U \otimes \mathcal{O}_X \to E),
\end{equation}
it is sufficient to prove that
$\EExt^{-1}(\mathcal{O}_X^{\oplus i},U \otimes \mathcal{O}_X \to E)=0$.
We note that
\begin{equation}
  \EExt^{-1}(\mathcal{O}_X^{\oplus i},U \otimes \mathcal{O}_X \to E)=
  \ker (\Hom(\mathcal{O}_X^{\oplus i},U \otimes \mathcal{O}_X ) \to
  \Hom(\mathcal{O}_X^{\oplus i},E)).
\end{equation}
Since $U$ is a subspace of $\Hom(\mathcal{O}_X,E)$,
$\EExt^{-1}(\mathcal{O}_X^{\oplus i},U \otimes \mathcal{O}_X \to E)=0$.
Hence (1) holds.

\smallskip

{\noindent $\bullet$} Proof of (2): It follows from [\citen{Yos1},
Thm. 2.5] that $G \in \mathcal{M}(v+iv_1)_{-\langle
  v_1,v+iv_1\rangle}$, {\it i.e.\/} $H^1(X,G)=0$. Hence
$\Ext^1(G,\mathcal{O}_X)=0$ by Serre duality.  By the stability of
$G$, we also have $\Hom(G,\mathcal{O}_X)=0$.  By the exact sequence
\begin{equation}
\Hom(G,U \otimes \mathcal{O}_X) \longrightarrow \Hom(G,E) \longrightarrow
\HHom(G,U \otimes \mathcal{O}_X \to E)\longrightarrow
\Ext^1(G,U \otimes \mathcal{O}_X),
\end{equation}
$\Hom(G,E) \cong \HHom(G,U \otimes \mathcal{O}_X \to E)$.
Since $\Hom(G,E)$ fits in an exact sequence
\begin{equation}
 \Hom(G,\mathcal{O}_X^{\oplus i}) \longrightarrow \Hom(G,G) 
\longrightarrow \Hom(G,E) \longrightarrow
 \Ext^1(G,\mathcal{O}_X^{\oplus i}),
\end{equation}
and $\Hom(G,G)\cong \CC$,  we have $\Hom(G,E) \cong \CC$.  Thus (2)
holds.
\end{proof}

The proposition below was first shown by Markman [\citen{Mar},
Thm. 39].
\begin{prop}\label{prop:dual}
Assume that $C_h$ satisfies the condition {\em ($\star$1)}.
For $n \geq r$, we have an isomorphism
\begin{equation}
\delta: \Syst^n(r,C_h,a) \overset{\sim}{\longrightarrow} \Syst^n(n-r,C_h,n-a).
\end{equation}
If $n=1$ and $r=0$, then the same assertion holds 
under the condition {\em($\star$2)}.
\end{prop}

\begin{proof}
  For a coherent system $f:U \otimes \mathcal{O}_{X} \to E$ belonging
  to $\Syst^n(r,C_h,a)$, our assumptions and [\citen{Yos1}, Lem. 2.1]
  imply that
\begin{enumerate}
\item[(i)]
 $f$ is surjective 
in codimension 1 (and hence $\dim \coker f=0$)
and $\ker f$ is a (slope) stable sheaf, or 
\item[(ii)]
$f$ is injective and $\coker f$ is a (slope) stable sheaf
\end{enumerate}
according as (i) $n>r$ or (ii) $n=r$.
For the second case, $f$ is also generically surjective.
There is an exact sequence 
\begin{equation*}
\xymatrix@R=0pt{
0\ar[r] &{\mathcal{H}}om_{\mathcal{O}_X}(U \otimes \mathcal{O}_X \to E,
\mathcal{O}_X)\ar[r]
        &{\mathcal{H}}om_{\mathcal{O}_X}(E,\mathcal{O}_X)\ar[r]
        &{\mathcal{H}}om_{\mathcal{O}_X}(U \otimes \mathcal{O}_X,
\mathcal{O}_X)\\
\ar[r]^-{g} 
        &{\mathcal{E}}xt^1_{\mathcal{O}_X}(U \otimes \mathcal{O}_X \to E,
\mathcal{O}_X)\ar[r]
        &{\mathcal{E}}xt^1_{\mathcal{O}_X}(E,\mathcal{O}_X)\ar[r]
        &{\mathcal{E}}xt^1_{\mathcal{O}_X}(U \otimes \mathcal{O}_X,
\mathcal{O}_X) \\
  \ar[r]&{\mathcal{E}}xt^2_{\mathcal{O}_X}(U \otimes \mathcal{O}_X \to E,
\mathcal{O}_X)\ar[r]
        &{\mathcal{E}}xt^2_{\mathcal{O}_X}(E,\mathcal{O}_X)\ar[r]
        &{\mathcal{E}}xt^2_{\mathcal{O}_X}(U \otimes \mathcal{O}_X,
\mathcal{O}_X) }
\end{equation*}
Since $f$ is generically surjective,
${\mathcal{H}}om_{\mathcal{O}_X}(E,\mathcal{O}_X) \to
{\mathcal{H}}om_{\mathcal{O}_X}(U \otimes
\mathcal{O}_X,\mathcal{O}_X)$ is injective.  Hence we obtain
${\mathcal{H}}om_{\mathcal{O}_X}(U \otimes \mathcal{O}_X\to
E,\mathcal{O}_X)=0$.  Since $E$ is torsion-free or of pure dimension
1, Lemma \ref{lem:proj-dim} implies that
$\mathcal{E}xt^2_{\mathcal{O}_X}(E,\mathcal{O}_X)=0$.  Since $U\otimes
\mathcal{O}_X$ is a free module, ${\mathcal{E}}xt^k_{\mathcal{O}_X}(U
\otimes \mathcal{O}_X,\mathcal{O}_X)=0$ for all $k>0$. Thus we obtain
$\mathcal{E}xt^2_{\mathcal{O}_X}(U \otimes \mathcal{O}_X \to
E,\mathcal{O}_X)=0$.  We set $D(E):={\mathcal{E}}xt^1_{\mathcal{O}_X}
(U \otimes \mathcal{O}_X \to E,\mathcal{O}_X)$.  We shall prove that
$D(E)$ is a (slope) stable sheaf of $v(D(E)) \cap [X]=
((n-r)[X],[C_h],n-a)$.  We first compute $v(D(E))$: In the
Grothendieck group $K_0(X)$, we have
\begin{equation}
  \begin{split}
&\sum_i(-1)^i{\mathcal{E}}xt^i_{\mathcal{O}_X}(U \otimes \mathcal{O}_X \to E,
\mathcal{O}_X)
\\&\qquad=\sum_i(-1)^i{\mathcal{E}}xt^i_{\mathcal{O}_X}(E,\mathcal{O}_X)-
\sum_i(-1)^i{\mathcal{E}}xt^i_{\mathcal{O}_X}(U \otimes \mathcal{O}_X,
\mathcal{O}_X).
\end{split}
\end{equation}
For $(a,c_1,b) \in H^{2*}(X,\ZZ)$, we set $(a,c_1,b)^*:=(a,-c_1,b)$.
Then we get 
\begin{equation}
\begin{split}
v(\sum_i(-1)^i{\mathcal{E}}xt^i_{\mathcal{O}_X}(E,\mathcal{O}_X))&=v(E)^*\\
v(\sum_i(-1)^i{\mathcal{E}}xt^i_{\mathcal{O}_X}(U \otimes \mathcal{O}_X,
\mathcal{O}_X))
&=v(U \otimes \mathcal{O}_X)^*.
\end{split}
\end{equation}
Hence we see that $v(D(E)) \cap [X]=
((n-r)[X],[C_h],n-a)$.
We next show that $D(E)$ is (slope) stable:
By using the diagram
\begin{equation*}
  \xymatrix{
   \ker f \ar[r]\ar[d] &  U \otimes \mathcal{O}_X \ar[d]\ar[r] & 
   \im f \ar[d]\\
     0\ar[r]           &   E\ar[r]                          & E   }
\end{equation*}
we have an exact sequence
\begin{equation*}
\xymatrix@R=0pt{
0 \ar[r]&{\mathcal{E}}xt^1_{\mathcal{O}_X}(\coker f,\mathcal{O}_X)\ar[r]
    &D(E)\ar[r]
    &{\mathcal{H}}om_{\mathcal{O}_X}(\ker f,\mathcal{O}_X)\\
  \ar[r]& {\mathcal{E}}xt^2_{\mathcal{O}_X}(\coker f,\mathcal{O}_X)\ar[r]
    &0 & &}
\end{equation*}
Hence $D(E)$ is torsion-free or of pure dimension 1
according as $n>r$ or $n=r$.
If $n>r$, then $\ker f$ is a (slope) stable vector bundle.
Hence $(\ker f)^*$ is also stable,
which implies that $D(E)$ is also (slope) stable.
Thus $g: U^* \otimes \mathcal{O}_X \to D(E)$ is an element of 
$\Syst^n(n-r,C_h,n-a)$.
If $n=r$, then $\ker f=0$, and hence
$D(E) \cong {\mathcal{E}}xt^1_{\mathcal{O}_X}(\coker f,\mathcal{O}_X)$.
Since $\Supp(\coker f)$ is irreducible and reduced,
$D(E)$ is a stable sheaf. 
Therefore $g:U^{*} \otimes \mathcal{O}_X \to D(E)$ also belongs to 
$\Syst^n(n-r,C_h,n-a)$.
Hence we obtain a map 
\begin{equation}
\delta:
\Syst^n(r,C_h,a) \longrightarrow \Syst^n(n-r,C_h,n-a).
\end{equation}

We shall prove that this
map is holomorphic.  For this purpose, we consider a family
$\mathbf{f}:\mathcal{U} \boxtimes \mathcal{O}_X \to \mathcal{E}$ of
coherent systems parametrized by a scheme $S$ such that $\mathcal{E}$
is flat over $S$ and $\mathcal{U}$ is a vector bundle of rank $n$ on
$S$.  Let $\lambda:\mathcal{W}_0 \to \mathcal{E}$ be a surjective
homomorphism from a locally-free sheaf $\mathcal{W}_0$ to $\mathcal{E}$.
We set $\mathcal{W}_1:= \ker(\mathcal{W}_0 \oplus \mathcal{U} \boxtimes
\mathcal{O}_X \to \mathcal{E})$.  
Since ${\mathcal{E}}_s$, $s \in S$ is torsion-free or a coherent sheaf
of pure dimension 1,
Lemma \ref{lem:proj-dim} implies that
$\mathcal{W}_1$ is a locally-free sheaf.
We consider a homomorphism
$\psi:\mathcal{W}_1 \oplus \mathcal{U} \boxtimes \mathcal{O}_X \to
\mathcal{W}_0 \oplus \mathcal{U} \boxtimes \mathcal{O}_X$
sending $(x,y) \in \mathcal{W}_1 \oplus \mathcal{U} \boxtimes \mathcal{O}_X$
to $-x+y \in \mathcal{W}_0 \oplus \mathcal{U} \boxtimes \mathcal{O}_X$,
where we regard $\mathcal{W}_1$ and $\mathcal{U} \boxtimes \mathcal{O}_X$
as subsheaves of $\mathcal{W}_0 \oplus \mathcal{U} \boxtimes \mathcal{O}_X$.
Then we obtain a morphism of complex which is
quasi-isomorphic:
\begin{equation*}
\xymatrix@+2mm{ 
\mathcal{W}_1 \oplus {\mathcal{U}} \boxtimes {\mathcal{O}_X} \ar@{->>}[d]
\ar[r]^-{\psi}
& \mathcal{W}_0 \oplus {\mathcal{U}} \boxtimes {\mathcal{O}_X} 
\ar[d]^-{(\lambda,\mathbf{f})}\\
{\mathcal{U}} \boxtimes {\mathcal{O}_X} \ar[r]^-{\mathbf{f}} &{\mathcal{E}} }
\end{equation*}
Since the construction of $\psi$ is compatible with base change and
$\psi_s, s \in S$ is generically surjective, $\psi^*_s$ is injective,
where $\psi^*:(\mathcal{W}_0 \oplus \mathcal{U} \boxtimes
\mathcal{O}_X)^* \to (\mathcal{W}_1 \oplus \mathcal{U} \boxtimes
\mathcal{O}_X)^*$ is the dual of $\psi$.  Hence $\coker
\psi^*=\mathcal{E}xt^1_{\mathcal{O}_{S \times X}} (\mathcal{U}
\boxtimes \mathcal{O}_X \to \mathcal{E}, \mathcal{O}_{S \times X})$ is
flat over $S$ and $(\coker \psi^*)_s \cong
\mathcal{E}xt^1_{\mathcal{O}_X} (\mathcal{U}_s \otimes \mathcal{O}_X
\to \mathcal{E}_s, \mathcal{O}_X)$.  Let ${\mathbf g}:\mathcal{U}^{*}
\boxtimes \mathcal{O}_X \to \coker \psi^{*}$ be the homomorphism
induced by the natural inclusion $i:\mathcal{U}^{*} \boxtimes
\mathcal{O}_X \hookrightarrow \mathcal{W}_1^* \oplus \mathcal{U}^*
\boxtimes \mathcal{O}_X$.  Then ${\mathbf g}:\mathcal{U}^{*} \boxtimes
\mathcal{O}_X \to \coker \psi^{*}$ is a family of coherent systems.
Therefore $\delta$ is a holomorphic map.  In the same way, we can
construct a holomorphic map $\delta':\Syst^n(n-r,C_h,n-a) \to
\Syst^n(r,C_h,a)$.  Then $\delta'$ is the inverse of $\delta$.
Indeed, by using the diagram
\begin{equation*}
\xymatrix@+2mm{ & & \mathcal{U}^{*} \boxtimes \mathcal{O}_X 
\ar@{^{(}->}[d]_-{i}\ar@{=}[r]& \mathcal{U}^{*} 
\boxtimes \mathcal{O}_X\ar[d]^-{\mathbf{g}} & \\
0\ar[r]&\mathcal{W}_0^{*} \oplus \mathcal{U}^{*}
 \boxtimes \mathcal{O}_X\ar[r]^-{\psi^*}&
\mathcal{W}_1^{*} \oplus \mathcal{U}^{*} 
\boxtimes \mathcal{O}_X\ar[r]&\coker \psi^{*}\ar[r]&0}
\end{equation*}
we obtain the following diagram
\begin{equation*}
\xymatrix@+2mm{ 
& & \mathcal{U} \boxtimes \mathcal{O}_X \ar@{^{(}->}[d]_-{(0,id)}\ar@{=}[r]& 
\mathcal{U}
\boxtimes \mathcal{O}_X \ar[d]^-{ \delta'({\mathbf g}) } & \\
0\ar[r] & \mathcal{W}_1 \oplus \mathcal{U} \boxtimes \mathcal{O}_X
\ar[r]^-{(- \psi,i^{*})}&
 (\mathcal{W}_0 \oplus \mathcal{U} \boxtimes \mathcal{O}_X)
\oplus \mathcal{U} \boxtimes \mathcal{O}_X
\ar[r] &  \coker(- \psi,i^{*})\ar[r] & 0}
\end{equation*}
Then we can easily show that $\delta'({\mathbf g}):\mathcal{U}
\boxtimes \mathcal{O}_X \to \coker(- \psi,i^{*})$ is identified with
${\mathbf f}:\mathcal{U} \boxtimes \mathcal{O}_X \to {\mathcal{E}}$.
Thus $\delta' \circ \delta=id$.  $\delta \circ \delta'=id$ also
follows from the same argument.
\end{proof}

\begin{cor}\label{cor:dual}
By the above isomorphism, we have the following diagram:
\begin{equation}\label{eq:corresp-diagram1}
\xymatrix@R=10pt{&\Syst^n(v)_{r+a+i}  
\cong \Syst^n(w)_{n+i}\ar[dl]_>>>>>>>{p_v}\ar[dr]^>>>>>>{p_w}&\\
\mathcal{M}(v)_{r+a+i}& & \mathcal{M}(w)_{n+i}}
\end{equation}
where $v \cap [X]=(r[X],[C_h],a)$ and $w \cap [X]=((n-r)[X],[C_h],n-a)$.
\end{cor}

\begin{proof}
  Let $U \otimes \mathcal{O}_X \to E$ be an element of
  $\Syst^n(v)_{r+a+i}$.  Since $\mathcal{E}xt^k_{\mathcal{O}_X}( U
  \otimes \mathcal{O}_X \to E,\mathcal{O}_X) =0$ for $k \ne 1$, we
  obtain
\begin{equation}\label{eq:spectral1}
  \EExt^{k+1}(U \otimes \mathcal{O}_X \to E,\mathcal{O}_X) \cong
  H^k(X,\mathcal{E}xt^1_{\mathcal{O}_X}( U \otimes \mathcal{O}_X \to
  E,\mathcal{O}_X)).
\end{equation}
Since $\mathcal{E}xt^1_{\mathcal{O}_X}( U \otimes \mathcal{O}_X \to
E,\mathcal{O}_X)$ is a stable sheaf of positive degree, Serre duality
and \eqref{eq:spectral1} imply that
\begin{equation}
  \EExt^3(U \otimes \mathcal{O}_X \to E,\mathcal{O}_X) = H^2(X,
  \mathcal{E}xt^1_{\mathcal{O}_X}( U \otimes \mathcal{O}_X \to
  E,\mathcal{O}_X))=0.
\end{equation}
By using the canonical exact sequence
\begin{equation}
  \begin{split}
 0=\Ext^1(U \otimes \mathcal{O}_X,\mathcal{O}_X) &\longrightarrow
 \EExt^{2}(U \otimes \mathcal{O}_X \to E,\mathcal{O}_X) \\
&\longrightarrow
 \Ext^2(E,\mathcal{O}_X) 
\longrightarrow \Ext^2(U \otimes \mathcal{O}_X,\mathcal{O}_X)
  \to 0,
\end{split}
\end{equation}
we see that
\begin{equation}
 \begin{split}
   \dim H^1(X,\mathcal{E}xt^1_{\mathcal{O}_X}( U \otimes \mathcal{O}_X
   \to E,\mathcal{O}_X)) &=\dim \EExt^{2}(U \otimes \mathcal{O}_X \to
   E,\mathcal{O}_X)\\ &=\dim \Ext^2(E,\mathcal{O}_X)-n\\ &=\dim
   H^0(X,E)-n=r+a+i-n.
 \end{split}
\end{equation}
 \end{proof}

\begin{rem}
  We can easily generalize Lemma \ref{lem:smoo}, Proposition
  \ref{prop:dual} and Corollary \ref{cor:dual} to $N(mv_1,v)$ in
  \cite{Yos1}.
\end{rem}

We now explain the equivalence between relative Hilbert schemes of points
on curves and coherent systems under the condition $(\star 2)$.  
First we remark that 
\begin{lem} Under the condition {\em ($\star$2)},
\begin{equation}\label{eq:X-C}
\Syst^1(0,C_h,d+1-h) \cong 
\Syst_{\mathcal{C}_h/\mathcal{S}_h}
(1,\bar{\mathcal{J}}^d_h)\,,
\end{equation}
where
\begin{equation}
\Syst_{\mathcal{C}_h/\mathcal{S}_h}
(1,\bar{\mathcal{J}}^d_h):=\{\mathcal{O}_C \to L\mid
C \in \mathcal{S}_h=|C_h|,\, L \in \overline{\Pic}^d_C \}\,,
\end{equation}
is the relative moduli space of coherent systems on 
$p:\mathcal{C}_h \to \mathcal{S}_h$.
\end{lem}
\begin{proof}
  Let $\mathcal{L} \boxtimes \mathcal{O}_X \to {\mathcal{E}}$ be a
  family of coherent systems parametrized by a scheme $S$ such that
  ${\mathcal{E}}_s \in \mathcal{M}(0,C_h,d+1-h)$ for all $s \in S$,
  where $\mathcal{L}$ is a line bundle on $S$.  Replacing
  ${\mathcal{E}}$ by $(\mathcal{L} \boxtimes \mathcal{O}_X)^{*}
  \otimes {\mathcal{E}}$, we may assume that
  $\mathcal{L}=\mathcal{O}_S$.  We consider a locally-free resolution
  (Lemma \ref{lem:proj-dim})
\begin{equation}
  0 \longrightarrow V_1 \overset{\phi}{\longrightarrow} V_0
  \longrightarrow {\mathcal{E}} \longrightarrow 0.
\end{equation}
Then $\det \phi:\det V_1 \to \det V_0$ is injective and it defines an
effective Cartier divisor $\Div({\mathcal{E}})$ on $S \times X$.
$\Div({\mathcal{E}})$ is called the scheme-theoretic support of
${\mathcal{E}}$ and ${\mathcal{E}}$ is an
$\mathcal{O}_{\Div({\mathcal{E}})}$-module.  Thus we can regard
${\mathcal{E}}$ as a sheaf on ${\Div({\mathcal{E}})}$ and we get a
homomorphism $\psi:\mathcal{O}_{\Div({\mathcal{E}})} \to
{\mathcal{E}}$.  Since the construction of ${\Div({\mathcal{E}})}$ is
compatible with the base change, ${\Div({\mathcal{E}})}$ is flat over
$S$ and $\psi_s \ne 0$ for all $s \in S$. Thus we get a morphism
$\alpha:\Syst^1(0,C_h,d+1-h) \to \Syst_{\mathcal{C}_h/\mathcal{S}_h}
(1,\bar\mathcal{J}_h^d)$.  Conversely, for a flat family of Cartier
divisors $\mathcal{D} \subset S \times X$ and a family of coherent
systems $\psi:\mathcal{O}_\mathcal{D} \to {\mathcal{E}}$,
$\mathcal{O}_{S \times X} \to \mathcal{O}_\mathcal{D} \to
{\mathcal{E}}$ gives a family of coherent systems on $S \times X$,
where we regard ${\mathcal{E}}$ as a sheaf on $S \times X$.  Hence we
have a morphism $\beta:\Syst_{\mathcal{C}_h/\mathcal{S}_h}
(1,\bar\mathcal{J}_h^d) \to \Syst^1(0,C_h,d+1-h)$.  Clearly $\beta
\circ \alpha=id$.  Since every member $C \in |C_h|$ is irreducible and
reduced, set-theoretically $\alpha \circ \beta =id$.  In particular,
$\Syst^1(0,C_h,d+1-h)$ is isomorphic to the reduced subscheme
$\Syst_{\mathcal{C}_h/\mathcal{S}_h}
(1,\bar\mathcal{J}_h^d)_{\text{red}}$ of
$\Syst_{\mathcal{C}_h/\mathcal{S}_h} (1,\bar\mathcal{J}_h^d)$.
Therefore it is sufficient to prove that $\beta$ induces an injective
homomorphism
\begin{equation}
\beta_x:T_x(\Syst_{\mathcal{C}_h/\mathcal{S}_h}(1,\bar\mathcal{J}_h^d))
\longrightarrow T_{\beta(x)}(\Syst^1(0,C_h,d+1-h))
\end{equation}
of Zariski tangent spaces for all $x \in
\Syst_{\mathcal{C}_h/\mathcal{S}_h}
(1,\bar\mathcal{J}_h^d)$.
Let $\psi:\mathcal{O}_C \to E$ be a coherent system corresponding 
to a point  $x \in \Syst_{\mathcal{C}_h/\mathcal{S}_h}
(1,\bar\mathcal{J}_h^d)$.
Assume that $\beta_x(\xi)=0$ for a tangent vector $\xi \in T_x(
\Syst_{\mathcal{C}_h/\mathcal{S}_h}(1,\bar\mathcal{J}_h^d))$.
Let $\Psi:\mathcal{O}_\mathcal{D} \to {\mathcal{E}}$
be a family of coherent systems corresponding to
$\xi$, where
$\mathcal{D} \subset S \times X$ is a flat family of Cartier divisors
over $S:=\Spec({\CC}[t]/(t^2))$.
We claim that $\Div({\mathcal{E}})=\mathcal{D}$.
Then $\alpha(\beta(\Psi))=\Psi$,
which implies that $\xi=0$.

{\noindent $\bullet$} Proof of the claim:
Our assumption implies that ${\mathcal{E}} \cong
\mathcal{O}_S \boxtimes E$.
In particular, $\Div({\mathcal{E}})=S \times \Div(E)=S \times C$.
Since ${\mathcal{E}}$ is generated by one element
on $S \times (X \smallsetminus \Sing(C))$,
by the construction of
$\Div({\mathcal{E}})$,
we get 
\begin{equation}
\label{eq:div}
\Div({\mathcal{E}})_{|S \times (X \smallsetminus \Sing(C))}
=\mathcal{D}_{|S \times (X \smallsetminus \Sing(C))}.
\end{equation}
Since first order deformations of $\Div(E)=C$ are classified by
$H^0(C,\mathcal{O}_C(C))$ and the map $H^0(C,\mathcal{O}_C(C)) \to H^0(C
\smallsetminus \Sing(C),\mathcal{O}_C(C))$ is injective, it follows from
\eqref{eq:div} that $\Div({\mathcal{E}})=\mathcal{D}$.

This completes the proof of  \eqref{eq:X-C}.
\end{proof}
\begin{rem}
  See Lemma \ref{lem:smooth2} below.
\end{rem}

Let $\mathcal{O}_X \to L$ be an
element of $\Syst^1(0,C_h,a)$ and set $C:=\Supp(L)$.  
We have an exact sequence
\begin{equation}
  0 \longrightarrow \mathcal{O}_X \longrightarrow
  \mathcal{E}xt^1_{\mathcal{O}_X}(\mathcal{O}_X \to L,\mathcal{O}_X)
  \longrightarrow \mathcal{E}xt^1_{\mathcal{O}_X}(L,\mathcal{O}_X)
  \longrightarrow 0,
\end{equation}
since ${\mathcal{H}}om_{\mathcal{O}_X}(L,\mathcal{O}_X)=0$.  Hence we
obtain the following commutative diagram under the condition
($\star$2):
\begin{equation}\label{CD:Hilb-Syst}
\xymatrix@+2mm{{\mathcal{C}}_h^{[d]}=\Hilb_{\mathcal{C}_h/\mathcal{S}_h}^d 
\ar[d]_-{\mathscr{A}^d_\mathcal{O}}
&\Syst^1(0,C_h,d+1-h)\ar[l]^-{\tilde{\epsilon}}_-{\sim}
\ar[d]^-{p_v}\ar[r]^-{\sim}_-{\delta}&
\Syst^1(1,C_h,-d+h)\ar[d]^-{q_w}\\
{\mathcal{M}}(0,C_h,-d+1-h)&{\mathcal{M}}(0,C_h,d+1-h)
\ar[l]_-{\sim}^-{\epsilon}\ar[r]_-{\zeta}^-{\sim}&{\mathcal{M}}(0,C_h,-d+h-1)
}
\end{equation}
where $v \cap [X]=(0,[C_h],d+1-h)$,
$w \cap [X]=([X],[C_h],-d+h)$, and $\zeta$ and 
$\tilde{\epsilon}$ are isomorphisms defined by
\begin{equation}
\begin{matrix}
 \zeta:&L & \longmapsto &\mathcal{E}xt^1_{\mathcal{O}_X}(L,\mathcal{O}_X),\\
 \tilde{\epsilon}:&(\mathcal{O}_C \to L_{|C}) & \longmapsto& 
 ({\mathcal{H}}om_{\mathcal{O}_C}(L_{|C},\mathcal{O}_C) \subset \mathcal{O}_C).
\end{matrix}
\end{equation}
For $L \in \mathcal{M}(0,C_h,d+1-h)$,
$\mathcal{E}xt^1_{\mathcal{O}_X}(L,\mathcal{O}_X)$ is also supported
on $C$ and we have
\begin{equation}
 \mathcal{E}xt^1_{\mathcal{O}_X}(L,\mathcal{O}_X)_{|C} \cong \mathcal{
  H}om_{\mathcal{O}_C}(L_{|C},\mathcal{O}_C) \otimes \omega_C\,. 
\end{equation}
Hence we may identify $\zeta$ with $\epsilon_\omega$ in
\eqref{epsilon_omega}. The diagram \eqref{CD:Hilb-Syst} then implies
that we can also adopt $\Syst^1(0,C_h,d+1-h)$ as the pertinent moduli
space of $D2$-$D0$ bound states.

The following was first proved by Huybrechts \cite{Huy1}
based on the description of moduli spaces in \cite{O'Grady1}.
We can find a more direct proof in \cite{Yos1}.
\begin{thm}\label{thm:deform}
If $C_h$ is ample or satisfies the condition $(\star1)$, then
$\mathcal{M}(r,C_h,a)$ is deformation equivalent to $X^{[h-ra]}$.
In particular,
$\vir(\mathcal{M}(r,C_h,a))=\vir(X^{[h-ra]})$.
Moreover, if $r>0$ and $\xi \in \Pic(X)$ is primitive, then
the same assertions hold for $\mathcal{M}(r,\xi,a)$.
\end{thm} 

\begin{proof}
  That the assertions hold is
guaranteed by [\citen{Yos1}, Thm.~0.2] unless
$r=0$ and $C_h$ is not ample.  
Hence we may assume that $r=0$ and $C_h$ satisfies $(\star1)$.
The following argument is very similar to the last part 
of the proof of [\citen{Yos1}, Thm. 3.6]. 
Let $H$ be an ample line bundle in \eqref{eq:condition1}.
Replacing $E \in \mathcal{M}(0,C_h,a)$ by
$E \otimes H^{\otimes n} \in \mathcal{M}(0,C_h,a+n \deg(C_h))$, $n \gg 0$,
we may assume that the evaluation map
$\phi:H^0(X,E) \otimes \mathcal{O}_X \to E$ is surjective
for all $E \in \mathcal{M}(0,C_h,a)$.
By [\citen{Yos1}, Lem. 2.1], $\ker \phi$ is a stable sheaf.
Then the correspondence 
\begin{equation}
\begin{matrix}
R:&\mathcal{M}(0,C_h,a) & \longrightarrow & \mathcal{M}(a,-C_h,0)\\
&E & \longmapsto & \ker \phi
\end{matrix}
\end{equation}
gives an immersion.
Since $\mathcal{M}(a,-C_h,0)$ is irreducible (indeed deformation equivalent
to $X^{[h]}$),
$R$ is an isomorphism.
Therefore $\mathcal{M}(0,C_h,a)$ is also deformation equivalent
to $X^{[h]}$.
\end{proof}

\begin{rem}
  The isomorphism $R$ is called the reflection by $v(\mathcal{O}_X)$
  ({\it cf.\/} \cite {Muk2, Yos1}).  Indeed $v(\mathcal{O}_X)$ is a
  $(-2)$-vector and $v(R(E))\allowbreak=\allowbreak\dim
  H^0(X,E)v(\mathcal{O}_X)-v(E)\allowbreak =\allowbreak-(\langle
  v(\mathcal{O}_X),v(E) \rangle v(\mathcal{O}_X)\allowbreak
  +\allowbreak v(E))$.  Hence $-R(E)$ is the reflection of $v(E)$ by
  $v(\mathcal{O}_X)$. Since a $(-2)$ reflection is an important piece
  of the isometry group of the Mukai lattice, it is important to
  analyze its geometric realization.
\end{rem}

\vspace{1pc}

Let us  set
\begin{equation}
  (\xi)_\infty:=\prod_{n=0}^\infty(1-\xi q^n)\,,\qquad \text{and} \qquad
  \Theta(\xi):=(\xi)_\infty(q/\xi)_\infty(q)_\infty\,.
\end{equation}
For each $n\in \ZZ$  we define $\sign(n)$  by
\begin{equation}
  \sign(n)=
  \begin{cases}
    +1&\text{if $n\ge 0$},\\
    -1&\text{if $n <0$}.
  \end{cases}
\end{equation}
Then, the following is well-known:
\begin{lem}\label{lem:kronecker}
 For $0<\abs{q}<\abs{\xi_1}<1$ and $0<\abs{q}<\abs{\xi_2}<1$,
\begin{equation}
  \frac{(q)_\infty^3 \Theta(\xi_1\xi_2)}{\Theta(\xi_1)\Theta(\xi_2)}=
  \sum_{\sign(i)=\sign(j)}\sign(i)q^{ij}\xi_1^i\xi_2^j\,.
\end{equation}
\end{lem}
\begin{proof}
  See \cite{Wei,Hic,Z2}.
\end{proof}

Now we are in a position to  state the main assertion:
\begin{thm}\label{thm:hodge}
Assume that $C_h$ satisfies {\em ($\star$1)} for all $h \geq 0$. 
Then, for
$0<\abs{q}<\abs{y}<1$,
  \begin{equation}\label{eq:mainformula}
\begin{split}
\sum_{h=0}^\infty
&\sum_{d=0}^\infty \chi_{t,\tilde{t}}(\Syst^1(0,C_h,d+1-h))(t\tilde t)^{1-h} 
q^{h-1}y^{d+1-h}\\
&\qquad \qquad=\frac{-1}{q (y)_\infty
(q/y)_\infty
((t\tilde t y)^{-1})_\infty
(t\tilde t y q)_\infty
(t \tilde t^{-1} q)_\infty
(q)_\infty^{18}
(t^{-1}\tilde t q)_\infty}\,. 
\end{split}
\end{equation}  
In particular, by setting $t=\tilde t=1$, we obtain
\begin{equation}
\sum_{h=0}^\infty
\sum_{d=0}^\infty \chi(\Syst^1(0,C_h,d+1-h))
q^{h-1}y^{d+1-h}=\inv{\chi_{10,1}(\tau,\nu)}\,.
\end{equation}
Moreover, if $C_h$ is ample and satisfies {\em ($\star$2)}, then
$\chi_{t,\tilde{t}}(\Syst^1(0,C_h,d+1-h))$ is meaningful and can be
obtained from \eqref{eq:mainformula} as if $C_h$ satisfied {\em
  ($\star$1)}.
\end{thm}

For the proof of this theorem, we need the notion of virtual Hodge
polynomials. For a scheme $V$ over $\CC$, cohomology with compact
support $H^*_c(V,{\QQ})$ has a natural mixed Hodge structure
\cite{Del}.  Let $e^{p,q}(V):=\sum_k(-1)^k h^{p,q}(H_c^k(V))$ be the
virtual Hodge numbers and $e(V):=\sum_{p,q}e^{p,q}(V)t^p
{\tilde{t}}^q$ the virtual Hodge polynomial of $V$.  The following
properties are useful for the computation of $e(V)$. (For more details
on virtual Hodge polynomials, see [\citen{Che}, 0.1].)
\begin{lem}\label{vHodge}
{\noindent }
\begin{enumerate}
\item[(a)] Suppose that $V$ has a decomposition $V=\cup_{i=1}^k V_i$ into
  mutually disjoint locally closed subsets.  Then
$$
 e(V)=\sum_{i=1}^k e(V_i).
$$
\item[(b)]
If $V$ is a smooth projective variety, then
$e(V)=\vir(V)$. 
\end{enumerate}
\end{lem}

For each integer $n$, we set
\begin{equation}
 [n]:=\frac{(t \tilde{t})^n-1}{t\tilde t-1}\,.
\end{equation}
Then,
\begin{lem}\label{lem:etale}
  Let $\pi:V \to W$ be an \'{e}tale locally trivial $\PP^n$-bundle
  over $W$. Assume that $V$ is projective over $W$. Then
\begin{equation}
 e(V)=[n+1]e(W).
\end{equation}
\end{lem}

\begin{proof}
  We may assume that $W$ is smooth by applying Lemma \ref{vHodge}
  ({\em a}) successively.  Since $\pi$ is a projective morphism, the
  Leray spectral sequence for $\pi$ degenerates.  Moreover we obtain
  $R^2 \pi_* \QQ \cong \QQ$, and hence $R^{2i} \pi_* \QQ \cong \QQ$
  for $1 \leq i \leq n$.  Since $H^*_c(V,\QQ)$ is the Poincar\'{e}
  dual of $H^*(V,\QQ)$, we obtain our claim.
\end{proof}
{\it Proof of Theorem \ref{thm:hodge}}:
By Lemma \ref{lem:smoo},
$\Syst^1(0,C_h,d+1-h)$ is smooth.
Hence it is sufficient to compute the virtual Hodge polynomial
$e(\Syst^1(0,C_h,d+1-h))$.

We start with the computation of
$e(\Syst^1(r,C_h,a))$, $r+a \geq 0$. 
Under the condition $r+a\geq 0$,
 \eqref{eq:corresp-diagram} gives the following diagram:

$$
\xymatrix@=15pt{&\Syst^1(r+1,C_h,a+1)_{r+a+1+i}\ar[ld]_>>>>>>>>>{p_1}
 \ar[rd]^>>>>>>>{p_2}&\\
{\mathcal{M}}(r+1,C_h,a+1)_{r+a+1+i}& & {\mathcal{M}}(r,C_h,a)_{r+a+i}}
$$
where $p_1$ is an \'{e}tale locally trivial $\PP^{r+a+i}$-bundle
and $p_2$ is an \'{e}tale locally trivial $\PP^{i-1}$-bundle.
By Lemma \ref{lem:etale}, we have a relation
\begin{equation}\label{eq:1}
 \begin{split} 
  \sum_{i \geq 0}[i]e(\mathcal{M}(r,C_h,a)_{r+a+i})&=
  \sum_{i \geq 0}[r+a+2+i]e(\mathcal{M}(r+1,C_h,a+1)_{r+a+2+i})\\
  &=[r+a+2]e(\mathcal{M}(r+1,C_h,a+1))\\
  & \quad \quad +
  (t \tilde{t})^{r+a+2}
  \sum_{i \geq 0}[i]e(\mathcal{M}(r+1,C_h,a+1)_{r+a+2+i}).
 \end{split}
\end{equation}
Applying this  successively, we see that
\begin{equation}
 \begin{split}
  &\sum_{i \geq 0}[i]e(\mathcal{M}(r,C_h,a)_{r+a+i})\\
  &=[r+a+2]e(\mathcal{M}(r+1,C_h,a+1))+
  (t \tilde{t})^{r+a+2}
\sum_{i \geq 0}[i]e(\mathcal{M}(r+1,C_h,a+1)_{r+a+2+i})\\
  &= \cdots\\
  &= [r+a+2]e(\mathcal{M}(r+1,C_h,a+1))+(t \tilde{t})^{r+a+2}[r+a+4]
  e(\mathcal{M}(r+2,C_h,a+2))\\
  & \quad +\dots+
  (t \tilde{t})^{\sum_{j=1}^{k-1}(r+a+2j)}[r+a+2k]
  e(\mathcal{M}(r+k,C_h,a+k))+\cdots.
 \end{split}
\end{equation}
Since 
\begin{equation}
 e(\Syst^1(r,C_h,a))=\sum_{i \geq 0}[r+a+i]e(\mathcal{M}(r,C_h,a)_{r+a+i})
\end{equation}
and
$\sum_{j=0}^{k-1}(r+a+2j)=(r+a+k-1)k$,
we obtain  that
\begin{equation}\label{eq:h>0}
 \begin{split}
  e(\Syst^1(r,C_h,a))&=
  [r+a]e(\mathcal{M}(r,C_h,a))+(t \tilde{t})^{r+a}
 \sum_{i \geq 0}[i]e(\mathcal{M}(r,C_h,a)_{r+a+i})\\
 &=\sum_{k \geq 0}
  (t \tilde{t})^{(r+a+k-1)k}[r+a+2k]
  e(\mathcal{M}(r+k,C_h,a+k)).
 \end{split}
\end{equation}

Now using \eqref{eq:h>0} with $r=0$, we find that
\begin{equation}
 \begin{split}
  \sum_{h \geq 0}
  \sum_{a \geq 0}&e(\Syst^1(0,C_h,a) )y^a 
  (t\tilde t)^{1-h}q^{h-1} \\
   &=
  \sum_{h \geq 0}\sum_{a \geq 0}\sum_{k \geq 0}
  (t \tilde{t})^{(a+k-1)k}[a+2k]
  e(\mathcal{M}(k,C_h,a+k))y^a (t\tilde t)^{1-h}q^{h-1}\\
  &=\sum_{h \geq 0}
  \sum_{j \geq i }\sum_{i \geq 0}(t \tilde{t})^{(j-1)i}[i+j]
  e(\mathcal{M}(i,C_h,j))y^{j-i} (t\tilde t)^{1-h}q^{h-1}\\
  &=\sum_{h \geq 0} \sum_{j \geq i}\sum_{i \geq 0}(t \tilde{t})^{(j-1)i}[i+j]
  e(X^{[(C_h^2)/2-ij+1]})y^{j-i} (t\tilde t)^{1-h} q^{h-1}\\
  &=\frac{t \tilde t}{q}\left(
  \sum_{j \geq i}\sum_{i \geq 0}(t \tilde{t})^{-i}[i+j]
  y^{j-i}q^{ij}\right) \left(\sum_n e(X^{[n]})(t\tilde t)^{-n}q^{n} \right),
 \end{split}
\end{equation}
where we applied Theorem \ref{thm:deform} to
$e(\mathcal{M}(i,C_h,j))$.

For $\Syst^1(0,C_h,-a)$, $a>0$, we use Corollary \ref{cor:dual} to
find a relation
\begin{equation}
 \sum_{i \geq 1}[i]e(\mathcal{M}(0,C_h,-a)_i)=
 \sum_{i \geq 1}[a+1+i]e(\mathcal{M}(1,C_h,1+a)_{a+i+1}).
\end{equation}
By using \eqref{eq:1} successively and performing a similar
calculation as above, we see that
\begin{equation}
 \begin{split}
  \sum_{h \geq 0}
  \sum_{a > 0}&e(\Syst^1(0,C_h,-a))y^{-a} (t\tilde t)^{1-h}
  q^{h-1}\\
&=\sum_{h \geq 0}
  \sum_{a > 0}\sum_{i \geq 0}[i]e(\mathcal{M}(0,C_h,-a)_{i})y^{-a}
 (t\tilde t)^{1-h}
  q^{h-1}\\
  &=
  \sum_{h \geq 0}\sum_{a > 0}\sum_{k \geq 1}
  (t \tilde{t})^{(a+k-1)k-a}[a+2k]
  e(\mathcal{M}(k,C_h,a+k))y^{-a} (t\tilde t)^{1-h}q^{h-1}\\
&=\sum_{h \geq 0}
\sum_{i > j }\sum_{j \geq 1}(t \tilde{t})^{(i-1)j-(i-j)}[i+j]
e(\mathcal{M}(j,C_h,i))y^{j-i} (t\tilde t)^{1-h}q^{h-1}\\
&=\sum_{h \geq 0} \sum_{i> j}\sum_{j \geq 1}(t \tilde{t})^{(j-1)i}[i+j]
e(X^{[(C_h^2)/2-ij+1]})y^{j-i} (t\tilde t)^{1-h} q^{h-1}\\
&=\frac{t\tilde t}{q}\left(
\sum_{i > j}\sum_{j \geq 1}(t \tilde{t})^{-i}[i+j]
y^{j-i}q^{ij}\right) \left(\sum_n e(X^{[n]})(t\tilde t)^{-n} q^{n} \right).
\end{split}
\end{equation}

Combining the above results we obtain that
\begin{equation}
\begin{split}
\sum_{h \geq 0}
\sum_{a }&e(\Syst^1(0,C_h,a))y^a 
(t\tilde t)^{1-h}q^{h-1}\\ 
&=\frac{t\tilde t}{q}\left(
\sum_{i \geq 0, j>0}(t \tilde{t})^{-i}[i+j]
y^{j-i}q^{ij}\right) \left(\sum_n e(X^{[n]}) (t\tilde t)^{-n}q^{n} \right)\\
&=\frac{t\tilde t}{q(t \tilde{t}-1)}
\left(\sum_{i \geq 0, j>0}
((t \tilde{t}y)^j y^{-i}-(t \tilde{t}y)^{-i} y^{j} )
q^{ij} \right)
\left(\sum_n e(X^{[n]}) (t\tilde t)^{-n} q^{n} \right)\\
&=\frac{-1}{q(1-(t \tilde{t})^{-1})}
   \frac{(q)_\infty^3 \Theta((t\tilde
    t)^{-1})}{ \Theta(y) \Theta((t\tilde
    ty)^{-1})}
\left(\sum_n e(X^{[n]})(t\tilde t)^{-n}q^{n} \right)\,,
\end{split}
\end{equation}
where we used Lemma \ref{lem:kronecker} in the last step.
Since
\begin{equation}
  \begin{split}
   \frac{(q)_\infty^3 \Theta((t\tilde
    t)^{-1})}{ \Theta(y) \Theta((t\tilde
    ty)^{-1})}&=
\frac{(q)_\infty^2
((t\tilde t)^{-1})_\infty
(t\tilde t q)_\infty}{(y)_\infty
(q/y)_\infty
((t\tilde t y)^{-1})_\infty
(t\tilde t y q)_\infty}\\
&=\frac{(1-(t\tilde t)^{-1})(q)_\infty^2
((t\tilde t)^{-1}q)_\infty
(t\tilde tq)_\infty}{(y)_\infty
(q/y)_\infty
((t\tilde t y)^{-1})_\infty
(t\tilde t y q)_\infty}
\end{split}
  \end{equation}
and
\begin{equation}
  \sum_{n=0}^\infty e(X^{[n]})(t \tilde t)^{-n}q^n=
\frac{1}{
((t \tilde t)^{-1}q)_\infty
(t \tilde t^{-1} q)_\infty
(q)_\infty^{20}
(t^{-1}\tilde t q)_\infty
(t\tilde t q)_\infty}     
\end{equation}
by \cite{Che,GS},
we reach the desired result.
The last assertion of the theorem follows from the following two lemmas.
({\it cf.\/} {\it Remark \ref{rem:C_h-cond}}.)
\qed

\begin{lem}\label{lem:smooth2}
Under the condition {\em ($\star$2)},
$\Syst^1(0,C_h,a)$ is smooth of dimension $2h+a-1$.
\end{lem}

\begin{proof}
  By Proposition \ref{prop:dual}, $\Syst^1(0,C_h,a)$ is isomorphic to
  $\Syst^1(1,C_h,1-a)$.  Hence we shall prove that
  $\Syst^1(1,C_h,1-a)$ is smooth.  Let $f:\mathcal{O}_X \to I_Z(C)$ be
  an element of $\Syst^1(1,C_h,1-a)$.  Then condition ($\star$2)
  implies that $f$ is injective and $L:=\coker f$ is a rank-1
  torsion-free sheaf when restricted to its support $C$.  In order to
  prove the smoothness of $\Syst^1(1,C_h,1-a)$ at $f:\mathcal{O}_X \to
  I_Z(C)$, it is sufficient to prove that $\Hom(I_Z(C),L) \cong \CC$.
  Since $I_Z(C)_{|C}/(\text{torsion}) \cong L_{|C}$ and $L$ is simple, we
  obtain our claim.
\end{proof}

\begin{lem}\label{lem:deform2}
Let $(X_i,H_i)$, $i=1,2$ be polarized $K3$ surfaces
such that
\begin{enumerate}
\item[(i)]
$H_1^2=H_2^2$.
\item[(ii)]
Every member of $|H_i|$ is irreducible and reduced.
\end{enumerate}
Then $\Syst^1(0,H_1,a)$ is deformation equivalent to
$\Syst^1(0,H_2,a)$.
\end{lem}

\begin{proof}
  It is sufficient to prove the deformation equivalence of
  $\Syst^1(1,H_i,1-a)$ $(i=1,2)$.  By the connectedness of the moduli
  space of polarized $K3$ surfaces, there is a family of polarized
  $K3$ surfaces $\pi:(\mathcal{X},\mathcal{H}) \to S$ such that $S$ is
  irreducible and there are two points $s_1,s_2 \in S$ which satisfy
  $(\mathcal{X}_{s_i},\mathcal{H}_{s_i})=(X_i,H_i)$.  Then there is a
  family of moduli spaces of coherent systems
  $\phi:\Syst^1(1,\mathcal{H},1-a) \to S$ such that
  $\Syst^1(1,\mathcal{H},1-a)_s=\Syst^1(1,\mathcal{H}_s,1-a)$ and
  $\phi$ is a projective morphism.  Assume that every member of
  $|\mathcal{H}_s|$ is irreducible and reduced for a point $s \in S$.
  Let $\mathcal{O}_{\mathcal{X}_s} \to I_Z(\mathcal{H}_s)$ be a point
  of $\Syst^1(1,\mathcal{H}_s,1-a)$.  By the proof of Lemma
  \ref{lem:smooth2}, $\tau:\EExt^2(\mathcal{O}_{\mathcal{X}_s} \to
  I_Z(\mathcal{H}_s),I_Z(\mathcal{H}_s)) \to
  \Ext^2(I_Z(\mathcal{H}_s),I_Z(\mathcal{H}_s)) \to
  H^2(\mathcal{X}_s,\mathcal{O}_{\mathcal{X}_s})$ is injective.  By a
  standard argument, the obstruction of infinitesimal lifting lives in
  $\EExt^2(\mathcal{O}_{\mathcal{X}_s} \to
  I_Z(\mathcal{H}_s),I_Z(\mathcal{H}_s))$.  Let $c_1(\mathcal{H}) \in
  R\pi^2_* {\ZZ}$ be the relative cohomology class of $\mathcal{H}$.
  Since $\Pic_{\mathcal{X}/S}^{c_1(\mathcal{H})} \to S$ is smooth
  (indeed isomorphic), the injectivity of $\tau$ implies that
  infinitesimal deformations of $\mathcal{O}_{\mathcal{X}_s} \to
  I_Z(\mathcal{H}_s)$ are unobstructed. Hence $\phi$ is a smooth
  morphism at $s$.  In particular, $W:=\{s \in S \mid\text {$\phi$ is
    not smooth at $s$}\}$ is a proper closed subset of $S$.  Since
  $s_1,s_2 \in S \smallsetminus W$ and $\phi_{|\phi^{-1}(S
    \smallsetminus W)}$ is smooth, we obtain our claim.
\end{proof}

\subsection{$D0$-branes bound to a $D2$-brane moving in the 
fibers of  the $K3$ fibration}

In the above we have been studying the case where the $D2$-$D0$ bound
system is moving in a fixed $K3$ surface $X$. Similarly, we should
like to investigate the case where the bound system of a single
$D2$-brane and collections of $D0$-branes is moving in the fibers of
the $K3$-fibration $\pi_1:Y\to W_1$ described in \S \ref{sec:STP}.
This is not an easy task in general since the details depend on the
choice of $Y$ and we do not have good control of the relevant moduli
spaces as in the single $K3$ case.  So, unfortunately, there is very
little we can say at the moment. However, one easily sees that
\begin{equation}
  \Phi_0(\tau,z,\nu)=\frac{\Psi_{10,m}(\tau,z)}{\chi_{10,1}(\tau,\nu)}\,.
\end{equation}
Actually it was this observation that motivated us to consider the
meaning of $1/\chi_{10,1}(\tau,\nu)$ leading to the results in \S
\ref{subsec:D0D2inK3}.
\begin{rem}
  Let $Y_s$ be a generic (smooth) fiber of $\pi_1:Y\to W_1$. By our
  assumption, $Y_s$ is an elliptic $K3$ surface with a section. Since
  in general $Y_s$ does not satisfy the conditions in \S
  \ref{subsec:D0D2inK3}, we will need a slight perturbation of the
  complex structure of $Y_s$  in order to apply the results in \S
  \ref{subsec:D0D2inK3}.
\end{rem}

The argument given in \S \ref{subsec:D0D2inK3}
naturally suggests that $\Phi_0(\tau,z,\nu)$ counts the pertinent
$D2$-$D0$ bound states in the $K3$ fibers.  An appropriate
mathematical setting for justification of this would probably be again
coherent systems of dimension 1 in $Y$ and their moduli spaces.

We should also remark on the following point. While we have assumed
$\abs{y}<1$ so far in this section, we previously assumed that
$\abs{y}=1$ $(y\ne 1)$ when we Fourier-expand $\Phi_0(\tau,z,\nu)$ in
order to obtain the infinite product formula of the string partition
function.  This was to realize the manifest symmetry $D(n,\gamma,j)=
D(n,\gamma,-j)$ and may be regarded as the conjugation symmetry of
$D0$-brane charge.  Thus we may suppose that the Fourier coefficients
$D(n,\gamma,j)$ count (with the conjugation symmetry of $D0$-brane
charge imposed) the bound states of $D0$-branes and a $D2$-brane
moving in the fibers of $\pi_1:Y \to W_1$ where the $D0$-brane charge
is $j$ and the $D2$-brane charge specifies $(n,\gamma)$.

The cases of several coincident $D2$-branes bound to collections of
$D0$-branes are presumably taken care of by the actions of Hecke
operators $V_\ell$ on $\Phi_0$.

\section{Vertex operators and $D2$--$D0$ bound states}

In the previous section we encountered the expression
\begin{equation}
  \inv{\chi_{10,1}(\tau,\nu)}=\inv{\eta(\tau)^{24}E(\tau,\nu)^2}\,,
\end{equation}
as the enumeration function of the $D2$-$D0$ bound states in a $K3$
surface $X$.  However, as every string theorist would readily realize,
the right hand side coincides with the (unnormalized) one-loop tachyon
two-point function of bosonic open string. This fact immediately leads
to an anticipation that the $D2$-$D0$ bound states are related to the
theory of vertex operators.  In the present section we will explore this
possibility although our understanding of the relation remains
admittedly superficial.

Motivated by the observation in \cite{VW}, Nakajima \cite{Nak1,Nak2}
and independently Grojnowski \cite{Gro} showed that there exists a
geometrical realization of the Heisenberg algebra on $\oplus_n
H_*(X^{[n]})$.  See also related works \cite{Bar,Leh}.  It would be
most desirable to have similar realizations and interpretations for
what we will see below.

Almost all the technical aspects given below have been known since the
era of dual resonance model \cite{Dual} which was a precursor of
string theory.

\subsection{Heisenberg algebra and the Fock space representation}

Let $(\Lambda,\langle \ , \ \rangle)$ be an integral lattice of rank
$\ell$ and set $\mathcal{V}=\Lambda_\CC$. We extend $\langle \ , \ 
\rangle$ by $\CC$-linearity. For each $n\in \ZZ$ let $\mathcal{V}(n)$
be a copy of $\mathcal{V}$ and set
\begin{equation}
  \mathsf{h}=\mathop{\bigoplus}_{n \ne 0} \mathcal{V}(n) \oplus \CC
  \kappa\,, \qquad \tilde\mathsf{h}=\mathop{\bigoplus}_{n \in \ZZ}
  \mathcal{V}(n) \oplus \CC \kappa\,,
\end{equation}
where $\CC \kappa$ is a 1-dimensional vector space spanned by $\kappa$.
For $a\in \mathcal{V}$, let $a(n)$ denote the corresponding element in
$\mathcal{V}(n)$.
The commutation relations
\begin{equation}
  [a(m),b(n)]=m\langle a,b\rangle\delta_{m,-n}\, \kappa\,, \qquad
  [a(m),\kappa]=0\,,
\end{equation}
make $\mathsf{h}$ and $\tilde \mathsf{h}$ infinite dimensional Lie
algebras with $\mathsf{h}$ being a Heisenberg algebra.

Setting  $\mathsf{h}_\pm=\bigoplus_{n>0} \mathcal{V}(\pm n)$,
we obtain  the triangular decompositions:
\begin{equation}
  \mathsf{h}=\mathsf{h}_+\oplus \CC\kappa\oplus
  \mathsf{h}_-\,,\qquad
  \tilde \mathsf{h} =\mathsf{h}_+\oplus \CC\kappa\oplus
  \mathcal{V}(0) \oplus \mathsf{h}_-\,.
\end{equation}

Let $S(\mathsf{h}_-)$ be the symmetric algebra of $\mathsf{h}_-$.
This is isomorphic to the $\ell$-fold tensor product of the polynomial
ring $\CC[x_1,x_2,\dots]$ in infinitely many commuting variables
$x_1,x_2,\ldots$.  The {\it Fock space\/} $S(\mathsf{h}_-)$ is graded
by assigning the elements of $\mathcal{V}(-n)$ the degree $n$ and it
becomes an $\mathsf{h}$-module in the following way.  First, $a(n)$
$(n\in \ZZ_-)$ acts on $S(\mathsf{h}_-)$ by the left multiplication.
For each $n\in \ZZ_+$ let $\partial_{a(n)}:\mathsf{h}_- \to \CC$ be a
linear function determined by $b(-k) \mapsto n \langle
a,b\rangle\delta_{n,k}$ for all $b\in \mathcal{V}$ and $k\in
\ZZ_+$. We can uniquely extend $\partial_{a(n)}$ to a derivation on
$S(\mathsf{h}_-)$ for which we keep the same notation.  The action of
$a(n)$ $(n\in \ZZ_+)$ on $S(\mathsf{h}_-)$ is given by identifying
$a(n)$ with $\partial_{a(n)}$. Finally $\kappa$ acts as the identity.

Let $\CC[\Lambda]$ be the group algebra with linear basis $\{e^\alpha
\mid \alpha \in \Lambda\}$ and multiplication $e^\alpha
e^\beta=e^{\alpha+\beta}$.  The total Fock space $\mathfrak{F}$ is
defined as
\begin{equation}
  \begin{split}
    \mathfrak{F}&=S(\mathsf{h}_-)\otimes \CC[\Lambda]\\
    &=\mathop{\amalg}\limits_{\alpha\in \Lambda} \mathfrak{F}_\alpha
  \end{split}
\end{equation}
with $\mathfrak{F}_\alpha=S(\mathsf{h}_-)\otimes e^\alpha$.
Then $\mathfrak{F}$ becomes an $\tilde \mathsf{h}$-module 
by letting
\begin{equation}
  \begin{split}
   a(n)(u\otimes e^\alpha) &= (a(n)u)\otimes e^\alpha\,, \quad (n \ne 0)\,,\\
 \quad \kappa(u\otimes e^\alpha)&=u\otimes e^\alpha\,,   
  \end{split}
\end{equation}
and
\begin{equation}
  a(0)(u\otimes e^\alpha) =\langle a,\alpha\rangle u\otimes e^\alpha\,.
\end{equation}

\begin{rem}
  It is customary to introduce the twisted group algebra
  $\CC\{\Lambda\}$ instead of the group algebra $\CC[\Lambda]$ in the
  standard theory of vertex operators associated with
  lattices. However, for the purpose of the present section the
  ordinary group algebra $\CC[\Lambda]$ suffices.
\end{rem}

The conjugate linear involution ${\ }^*$ on $\mathsf{h}$ and $\tilde 
\mathsf{h}$  is defined through
 \begin{equation}
   a(n)^*=\bar a(-n)\,, \qquad \kappa^*=\kappa\,,
 \end{equation}
where $\bar {\ }$ stands for the complex conjugation.

Then one can introduce a contravariant hermitian bilinear form
$\langle \ \mid \ \rangle$ on $\mathfrak{F}$ by demanding
 \begin{equation}
   \begin{split}
     & \langle A u \mid v\rangle=
\langle u \mid A^*v\rangle\,, \quad \text{for all $A\in
       \tilde\mathsf{h}$ and for all $u,v \in \mathfrak{F}$} \\ 
& \langle  1\otimes e^\alpha \mid 1\otimes e^\beta\rangle
     =\delta_{\alpha,\beta}\,, \quad \text{for all $\alpha,\beta\in
       \Lambda$.}
   \end{split}
 \end{equation}
 In particular we set $\mathbf{1}=1\otimes e^0$.
Some useful identities can be easily obtained:
 \begin{align}
   \langle a(-n)\mid b(-n)\rangle&=n\langle \bar a,b\rangle\\
   e^{a(n)}e^{b(-n)}&=e^{n\langle \bar a,b\rangle}e^{b(-n)}e^{a(n)}\\
   \langle e^{a(-n)} \mid e^{b(-n)}\rangle&=e^{n\langle \bar a,b\rangle}\\
   e^{a(n)}e^{b(-n)}\mathbf{1}&=
e^{n\langle \bar a,b\rangle}e^{b(-n)} \mathbf{1}
 \end{align}
 
\subsection{The Virasoro algebra}
Let $\{e_i\}$ be a basis of $\mathcal{V}$ and let $\{e^i\}$ be the
dual basis with respect to $\langle\ ,\ \rangle$ so that $\langle
e^i,e_j\rangle=\delta_{i,j}$. We assume that $\bar e_i=e_i$ and $\bar
e^i=e^i$.  Then we have
 \begin{equation}\label{completeness}
   \sum_{i=1}^\ell  \langle a,e^i\rangle\langle e_i,b\rangle=
\langle a,b\rangle\,,\quad \text{for all $a,b\in
     \mathcal{V}$.}
 \end{equation}

The Virasoro operators are defined for each $n\in \ZZ$ by
\begin{equation}
  L(n)=\inv{2}\sum_{i=1}^\ell \sum_{m\in \ZZ} :e^i(n-m)e_i(m):\,,
\end{equation}
where
\begin{equation}
  :a(n)b(m):=
  \begin{cases}
    a(n)b(m) &\text{if $n\leq m$}\,,\\
    b(m)a(n) &\text{if $n>m$}\,.
  \end{cases}
\end{equation}
They satisfy the commutation relations of the Virasoro algebra with
the central charge $\ell$:
\begin{equation}
  [L(m),L(n)]=(m-n)L(m+n)+\frac{\ell}{12}(m^3-m)\, \delta_{m,-n}\, \kappa\,.
\end{equation}
Using \eqref{completeness} it is easy to see that
\begin{equation}
  L(0)(1\otimes e^\alpha)=\frac{\Abs{\alpha}^2}{2}(1\otimes e^\alpha)\,,
\end{equation}
where $\Abs{\alpha}^2=\langle \alpha,\alpha\rangle$.
It is also not difficult show that 
\begin{equation}
  [L(m),a(n)]=-m\, a(m+n)\,,
\end{equation}
from which we obtain a useful identity
\begin{equation}\label{expaLcomm}
  q^{L(0)}e^{a(-n)}q^{-L(0)}=e^{q^na(-n)}\,.
\end{equation}

\subsection{Vertex operators}
We set
\begin{align}
 X_\pm(a,y)&:=\sum_{n=1}^\infty\frac{y^{\mp n}}{\mp n} a(\pm n)\,,\\
  P_\pm(a,y)&:=y\frac{d}{dy} X_\pm(a,y)=\sum_{n=1}^\infty y^{\mp
      n} a(\pm n)\,.
\end{align}
For $\abs{w_1}>\abs{w_2}$ we  obtain commutation relations:
\begin{align}
  [X_+(a,w_1),X_-(b,w_2)]&=\langle a,b\rangle\log(1-y)\,, \label{Xpm}\\ 
  [P_+(a,w_1),P_-(b,w_2)]&=
\frac{\langle a,b\rangle}{(y^{-1/2}-y^{1/2})^{2}}\,,
\end{align}
where $y=w_2/w_1$.

The vertex operator is defined for each $\alpha\in \Lambda$ by
\begin{equation}
  V(\alpha,y)=y^{\frac{\Abs{\alpha}^2}{2}} e^{X_-(\alpha,y)}e^\alpha
  y^{\alpha(0)} e^{X_+(\alpha,y)}\,,
\end{equation}
where $e^\alpha(u\otimes e^\beta)= u\otimes e^\alpha e^\beta$ and
$y^{\alpha(0)}(u\otimes e^\beta)=y^{\langle \alpha,\beta\rangle}
u\otimes e^\beta$.

It follows from  \eqref{Xpm} that for $\abs{w_1}>\abs{w_2}$,
 \begin{equation}\label{VVop}
   \begin{split}
& V(\alpha,w_1)V(\beta,w_2)=
w_1^{\frac{\Abs{\alpha}^2}{2}}w_2^{\frac{\Abs{\beta}^2}{2}}
(w_1-w_2)^{\langle \alpha,\beta\rangle}\\
&\qquad\times e^{X_-(\alpha,w_1)+
X_-(\beta,w_2)}e^{\alpha+\beta}w_1^{\alpha(0)}w_2^{\beta(0)}
e^{X_+(\alpha,w_1)+X_+(\beta,w_2)}\,.
   \end{split}
 \end{equation}

Using \eqref{expaLcomm} we find that
 \begin{equation}\label{VLcomm}
   q^{L(0)}V(\alpha,y)q^{-L(0)}=V(\alpha,qy)\,.
 \end{equation}

\subsection{Two-point correlation functions and $D2$-$D0$ bound states}

As in \S \ref{sec:D2D0} let $X$ be a projective $K3$ surface and 
for each $h\ge 0$ let $C_h$ be a smooth curve of genus $h$ on $X$
satisfying ($\star$1) in \S \ref{subsubsec:coh.sys}.  Then we make an
identification
 \begin{equation} \label{K3lattice}
 \Lambda=H^{2*}(X,\ZZ)(-1)\cong \mathsf{E}_8^{\oplus 2}\oplus
H(-1)^{\oplus 4}\,, \qquad \langle\ ,\ \rangle=-\langle \ ,\ \rangle_X\,.
 \end{equation}
 (We will try to be general in the following so that most of the
 results are applicable to surfaces with vanishing odd cohomologies.)

 The connection between the symmetric products of a {\it smooth\/}
 curve on a surface and  vertex operators  has been pointed out by
 Grojnowski \cite{Gro} and further discussed by Nakajima
 \cite{Nak2}. Indeed it immediately follows from \eqref{VVop} that
 \begin{equation}\label{2-pt}
         \langle \mathbf{1}  \mid V(-\alpha,1)V(\alpha,y)  \mathbf{1}\rangle
=\inv{(y^{-1/2}-y^{1/2})^{\Abs{\alpha}^2}}\,, \qquad (\abs{y}<1)\,.
 \end{equation}
Consider the expansion
 \begin{equation}
   e^{X_-(\alpha,y)}=\sum_{d=0}^\infty \alpha^{(d)} y^d\,,
 \end{equation}
where 
 \begin{equation}
   \alpha^{(d)}= s_d(\alpha(-1),\alpha(-2),\dots,\alpha(-d))\,,
 \end{equation}
with $s_d(x_1,\dots,x_d)$ being the Schur polynomial of degree $d$.
Then 
 \begin{equation}
   \begin{split}
     \langle \mathbf{1} \mid V(-\alpha,1)V(\alpha,y)\,
     \mathbf{1}\rangle&=y^{\frac{\Abs{\alpha}^2}{2}}\langle \mathbf{1} 
\mid e^{X_+(-\alpha,1)}e^{X_-(\alpha,y)}\, \mathbf{1}\rangle\\
&=\sum_{m,d=0}^\infty \langle \mathbf{1} \mid (\alpha^{(m)})^* \alpha^{(d)}  
\mathbf{1}\rangle  y^{d+\frac{\Abs{\alpha}^2}{2}}\\
 &= \sum_{d=0}^\infty \langle \alpha^{(d)}  \mathbf{1} \mid  \alpha^{(d)}  
\mathbf{1}\rangle  y^{d+\frac{\Abs{\alpha}^2}{2}}\,.
   \end{split}
 \end{equation}
Take $\alpha=c_1(\mathcal{O}_X(C_h))$.
Since $\Abs{\alpha}^2=-C_h\cdot C_h=2-2h$ it follows from
\eqref{twsteulergen} and \eqref{2-pt} that
 \begin{equation}
   \chi(C_h^{(d)})=\langle \alpha^{(d)} \mathbf{1} \mid
   \alpha^{(d)} \mathbf{1}\rangle\,.
 \end{equation}
({\it cf.} \cite{Gro} and Exercise 9.18 in \cite{Nak2}.)
What we will discuss below is a more complicated relation between the
relative Hilbert schemes $\mathcal{C}_h^{[d]}$ and vertex operators.

Set $\xi'=\sum_{i=1}^\ell \xi^{(i)}e^i$ for any $\xi=\sum_{i=1}^\ell
\xi^{(i)} e_i\in \mathcal{V}$.  Let $\LL:\mathcal{V} \to \mathcal{V}$
be a linear map and let $ \mu(\LL)$ be the $\ell$ by $\ell$ matrix
whose $(i,j)$-th entry is $\langle e^i,\LL e_j\rangle$. Suppose that $
\mu(\LL)$ is diagonalizable and the real parts of its eigenvalues are
positive.  Then the Gaussian integral leads to
\begin{lem} \label{Gaussian}
\begin{equation}
  \int d \bar{\xi} d\xi\, e^{-\langle \bar{\xi'},\LL
    \xi\rangle+\langle a,\xi\rangle+\langle b,\bar{\xi'}\rangle}=
 \inv{\det \mu(\LL)}
  e^{\langle a,\LL^{-1}b\rangle}\,,
\end{equation}
where $d \bar{\xi} d\xi=\prod_{i=1}^\ell d\bar{\xi}^{(i)}
d\xi^{(i)}/(2\pi)^\ell$.
\end{lem}

The trace of an operator $\mathcal{O}$ on $\mathfrak{F}_0$ can be
conveniently expressed in terms of the coherent states \cite{Dual}:
 \begin{equation}
   \Tr_{\mathfrak{F}_0} \mathcal{O} =\prod_{n=1}^\infty \int d
   \bar{\xi'_n} d\xi_n\, e^{-\langle \bar{\xi'_n},\xi_n\rangle}
   \langle e^{\inv{\sqrt{n}}\xi'_n(-n)}\mathbf{1} \mid \mathcal{O}\,
   e^{\inv{\sqrt{n}}\xi_n (-n)}\mathbf{1}\rangle\,.
 \end{equation}
 It follows from this representation that
 \begin{prop} \label{prop:1-loop}
For $0<\abs{q}<\abs{y}<1$, we have
 \begin{equation}\label{1-loop}
   \Tr_{\mathfrak{F}_0} V(-\alpha,1)V(\alpha,y)q^{L(0)-\frac{\ell}{24}}=
  \inv{ \eta(\tau)^\ell  E(\tau,\nu)^{\Abs{\alpha}^2}}\,.
 \end{equation}
\end{prop}
\begin{proof}
  Using \eqref{expaLcomm} the left hand side can be rewritten as
  \begin{equation}
q^{-\frac{\ell}{24}}\prod_{n=1}^\infty\int\,  
d \bar{\xi'_n} d\xi_n\,
 e^{-\langle \bar{\xi'_n},\xi_n\rangle}
\langle e^{\inv{\sqrt{n}}\xi'_n(-n)}\mathbf{1} \mid 
  V(-\alpha,1)V(\alpha,y)\, 
  e^{\frac{q^n}{\sqrt{n}}\xi_n (-n)}\mathbf{1}\rangle\,.
\end{equation}
Then the integrand of each factor becomes
\begin{equation}
  \begin{split}
    &\frac{\exp[{-\langle \bar{\xi'_n},\xi_n\rangle}]}
    {(y^{-1/2}-y^{1/2})^{\Abs{\alpha}^2}}\, \langle
    e^{\inv{\sqrt{n}}\xi'_n(-n)}\mathbf{1} \mid
    e^{\frac{1-y^n}{\sqrt{n}}\langle \alpha,\bar{\xi'_n}\rangle}
    e^{\frac{(1-y^{-n})q^n}{\sqrt{n}}\langle \alpha,\xi_n\rangle}
    e^{\frac{q^n}{\sqrt{n}}\xi_n (-n)}\mathbf{1}\rangle\\ &\ \ 
    =\inv{(y^{-1/2}-y^{1/2})^{\Abs{\alpha}^2}}
    \exp\bigl[{\textstyle-(1-q^n)\langle \bar{\xi'_n},\xi_n\rangle +
      \frac{(1-y^{-n})q^n}{\sqrt{n}}\langle \alpha,\xi_n\rangle+
      \frac{1-y^n}{\sqrt{n}}\langle
      \alpha,\bar{\xi'_n}\rangle}\bigr]\,,
  \end{split}
\end{equation}
  where we used \eqref{VVop}.  By performing the integrals using Lemma
  \ref{Gaussian} we thus obtain
\begin{equation}
  \inv{\eta(\tau)^\ell
(y^{-1/2}-y^{1/2})^{\Abs{\alpha}^2}}
\exp\left[\Abs{\alpha}^2\sum_{n=1}^\infty
\frac{(2-y^n-y^{-n})q^n}{n(1-q^n)}\right]\,,
\end{equation}
which can be cast in the desired form thanks  to the identity
  \begin{equation}
    \prod_{n=1}^\infty \inv{1-tq^{n-1}}=\exp\left[\sum_{n=1}^\infty
      \frac{t^n}{n(1-q^n)}\right]\,.
  \end{equation}
\end{proof}

Suppose that we are in the situation \eqref{K3lattice}. We set
$\alpha=c_1(\mathcal{O}_X(C_0))$ so that
$\Abs{\alpha}^2=-C_0^2=2$.  Since
$\ell=24$, we see that  the right hand side of
\eqref{1-loop} reduces to $1/\chi_{10,1}(\tau,\nu)$.

Let $\mathcal{N}$ be defined by
\begin{equation}
  L(0)=\inv{2}\sum_{i=1}^\ell e^i(0)e_i(0)+\mathcal{N}\,.
\end{equation}
Consider the spectral decomposition $\mathcal{N}=\sum_{d=0}^\infty d
\mathsf{P}_d$ where $\mathsf{P}_d$ is the projection operator onto the
eigensubspace with eigenvalue $d$ of $\mathfrak{F}_0$ with the obvious
properties: $\mathsf{P}_d^2=\mathsf{P}_d$,
$\mathsf{P}_d\mathsf{P}_s=0$ if $d\ne s$, and $\sum_{d=0}^\infty
\mathsf{P}_d=id$.  Then we find that
\begin{lem}\label{1-looplemma} For $0<\abs{q}<\abs{y}<1$,
  \begin{equation}
    \begin{split}
      & \Tr_{\mathfrak{F}_0}
      V(-\alpha,1)V(\alpha,y)q^{L(0)-\frac{\ell}{24}}\\ 
      & = \sum_{h=0}^\infty \sum_{d=0}^\infty
      q^{h-\frac{\ell}{24}}
 y^{d+\frac{\Abs{\alpha}^2}{2}-h}
\Tr_{\mathfrak{F}_0}V(-\alpha,1)\mathsf{P}_d\, V(\alpha,1)\mathsf{P}_h\,.
    \end{split}
      \end{equation}
\end{lem}
\begin{proof}
Using \eqref{VLcomm} we  see that the left hand side is equal to
\begin{equation}
  \begin{split}
    &\Tr_{\mathfrak{F}_0} V(-\alpha,1)y^{L(0)}V(\alpha,1)y^{-L(0)}
    q^{L(0)-\frac{\ell}{24}}\\
   &\ \ = \Tr_{\mathfrak{F}_0} V(-\alpha,1)y^{L(0)}
\, \sum_{d=0}^\infty \mathsf{P}_d\,  V(\alpha,1)y^{-L(0)}
    q^{L(0)-\frac{\ell}{24}}\, \sum_{h=0}^\infty \mathsf{P}_h\\
&\ \ = \sum_{h=0}^\infty \sum_{d=0}^\infty \Tr_{\mathfrak{F}_0} V(-\alpha,1)
y^{\frac{\Abs{\alpha}^2}{2}+d}\,\mathsf{P}_d
V(\alpha,1)y^{-h}
    q^{h-\frac{\ell}{24}}\,\mathsf{P}_h\,.
\end{split}
\end{equation}
\end{proof}
An immediate consequence of Lemma \ref{1-looplemma} is the following
claim equivalent to Theorem \ref{relHilb}:
\begin{prop} With the identification \eqref{K3lattice} and
  $\alpha=c_1(\mathcal{O}_X(C_0))$, there exists a relation
  \begin{equation}\label{degree-genusvertex}
    \chi(\mathcal{C}_h^{[d]})=\Tr_{\mathfrak{F}_0}V(-\alpha,1)\mathsf{P}_d\,
    V(\alpha,1)\mathsf{P}_h\,,
  \end{equation}
for  each pair $(h,d)$ of  nonnegative integers.
\end{prop}
\begin{rem}
  With this expression at  hand the degree-genus duality
  \eqref{dg-duality} follows immediately from the cyclic symmetry of
  the trace and the fact that the right hand side of
  \eqref{degree-genusvertex} is invariant under the exchange $\alpha
  \leftrightarrow -\alpha$. 
\end{rem}

\subsection{Two-point correlation functions and elliptic genus}

We wish to take this opportunity to make the observation in
\cite{Kaw1} more explicit. This subsection is not logically related to
the main theme of this paper and may be skipped.

Let us recall the definition of the Weierstra{\ss} $\sigma$ function:
\begin{equation}
  \sigma(\tau,\nu)=-2\pi\sqrt{-1}\nu \prod_{\substack{\omega\in \ZZ+
      \ZZ\tau\\ \omega\ne 0}}
  \left(1-\frac{\nu}{\omega}\right)\exp\left[\frac{\nu}{\omega}+
\inv{2}\left(\frac{\nu}{\omega}\right)^2\right]\,.
\end{equation}
As is well-known this is related to
the Weierstra{\ss}  $\wp$-function
\begin{equation}
  \wp(\tau,\nu) := \frac{1}{(2\pi \sqrt{-1})^2} \left\{ \frac{1}{\nu^2} +
    \sum_{\substack{\omega\in \ZZ+ \ZZ\tau\\ \omega\ne
        0}}\left(\frac{1}{(\nu-\omega)^2} -
      \frac{1}{\omega^2}\right)\right\}\,,
\end{equation}
by the relation
\begin{equation}
  \wp(\tau,\nu)=-(y\frac{\partial}{\partial y})^2 \log \sigma(\tau,\nu)\,.
\end{equation}

The $\sigma$ function is related to the prime form 
by\footnote{In the traditional theory of elliptic functions, $E_2$ is
  usually denoted as $\eta_1$ up to a scalar multiplication.}
\begin{equation}
  \begin{split}
  \sigma(\tau,\nu)&=\exp\left( x^2 E_2(\tau)/24\right)E(\tau,\nu)\\
                &=-\sqrt{-1}x\exp\left(\sum_{k=2}^\infty
    \frac{(-1)^{k}B_{2k}}{2k(2k)!}x^{2k}E_{2k}(\tau)\right)\,.
  \end{split}
\end{equation}

In analogy to $\wp(\tau,\nu)$ let us introduce
\begin{equation}
  \Gamma(\tau,\nu):=-(y\frac{\partial}{\partial y})^2 \log E(\tau,\nu)\,.
\end{equation}
Then apparently we have a relation:
\begin{equation}
  \wp(\tau,\nu)=\Gamma(\tau,\nu)+\inv{12}E_2(\tau)\,.
\end{equation}
Note that while $\wp(\tau,\nu)$ is a (meromorphic) Jacobi form of weight
$2$ and index $1$, $\Gamma(\tau,\nu)$ is not.  Explicitly one finds that
\begin{equation}
  \begin{split}
    \Gamma(\tau,\nu)&= \inv{(y^{-1/2}-y^{1/2})^2}-
    y\frac{\partial}{\partial y} \sum_{n=1}^\infty \left(
      \frac{-yq^n}{1-yq^n} + \frac{y^{-1} q^n}{1-y^{-1} q^n} \right)\\ 
    &=\inv{(y^{-1/2}-y^{1/2})^2}+y\frac{\partial}{\partial
      y}\sum_{n=1}^\infty\sum_{k=1}^\infty (y^k-y^{-k})q^{nk} \\ 
    &=\inv{(y^{-1/2}-y^{1/2})^2}+
    \sum_{n=1}^\infty\frac{n(y^n+y^{-n})q^n}{1-q^n}\,.
  \end{split}
\end{equation}

For any $b\in \mathcal{V}$ and $\beta\in \Lambda$ such that
$\langle b,\beta\rangle=0$, define
\begin{equation}
  U_t(\beta,b,y)=V(\beta,y)e^{t P_-(b,y)}e^{tP_+(b,y)}\,,
\end{equation}
where $t$ is a formal variable. Observe that
\begin{equation}\label{vectorvertex}
  \frac{d}{dt}U_t(\beta,b,y)\vert_{t=0}= P(b,y)V(\beta,y)=:W(\beta,b,y)\,,
\end{equation}
where $P(b,y):=P_+(b,y)+P_-(b,y)$.

\begin{prop}\label{vector2-point}
 Suppose that $a,b\in \mathcal{V}$ and $\beta\in \Lambda$ satisfy
$\langle a,\beta\rangle=\langle b,\beta\rangle=0$. Then, 
 for  $0<\abs{q}<\abs{y}<1$, we obtain that
\begin{equation}
   \begin{split}
   \Tr_{\mathfrak{F}_0}
  & U_t(-\beta,a,1)U_s(\beta,b,y)q^{L(0)-\frac{\ell}{24}}=
   \inv{\eta(\tau)^\ell E(\tau,\nu)^{\Abs{\beta}^2}}\\
&\times \exp\left(ts \langle a,b\rangle\Gamma(\tau,\nu)+
\inv{24}(t^2\Abs{a}^2+  s^2\Abs{b}^2)(1-E_2(\tau))\right)\,.
\end{split}
 \end{equation}
 \begin{proof}The calculation is similar to that in the proof of Proposition
   \ref{prop:1-loop}.  The left hand side is equal to
   \begin{equation}
     \begin{split}
   &q^{-\frac{\ell}{24}} \prod_{n=1}^\infty \int  \,
d \bar{\xi'_n} d\xi_n\,
\inv{(y^{-1/2}-y^{1/2})^{\Abs{\beta}^2}}
\exp\left[\frac{ts\langle a,b\rangle}{(y^{-1/2}-y^{1/2})^2}\right]\\
&\times 
  \exp \bigl[{\textstyle -(1-q^n)\langle \bar{\xi'_n},\xi_n\rangle +
q^n\langle  
\frac{(1-y^{-n})}{\sqrt{n}} \beta + \sqrt{n}(ta+s y^{-n}b),\xi_n\rangle}\\
&\hspace{4cm}{\textstyle+\langle  \frac{1-y^n}{\sqrt{n}} \beta+ 
\sqrt{n}(ta + sy^n b),\bar{\xi'_n}\rangle}\bigr]\,.
\end{split}
   \end{equation}
By performing the Gaussian integrals we obtain that
   \begin{equation}
     \begin{split}
& \inv{\eta(\tau)^\ell E(\tau,\nu)^{\Abs{\beta}^2}} 
 \exp\Biggl[ts\langle a,b\rangle\left(\inv{(y^{-1/2}-y^{1/2})^2}+ 
\sum_{n=1}^\infty\frac{n(y^n+y^{-n})q^n}{1-q^n}\right)\\
&\qquad  +(t^2\Abs{a}^2+  s^2\Abs{b}^2) 
\sum_{n=1}^\infty\frac{n q^n}{1-q^n}  \Biggr]\,.
\end{split}
\end{equation}
This readily leads to the desired result.
 \end{proof}
\end{prop}

Suppose again that $X$ is a $K3$ surface but set
$\Lambda=H^{2*}(X,\ZZ)(-1)\oplus H(-1)$ so that $\ell=26$.  Assume
that $H(-1)$ is generated by $\alpha$ and $\beta$ where
$\Abs{\alpha}^2=2$, $\Abs{\beta}^2=0$ and $\vev{\alpha,\beta}=-1$.
Let $\{f_i\}$ be a basis of $H^{2*}(X,\ZZ)(-1)$ and let $\{f^i\}$ be
the dual basis.  Then \eqref{vectorvertex} and Proposition
\ref{vector2-point} show that
\begin{equation}
  \frac{\eta(\tau)^2\sum_{i=1}^{24}\Tr_{\mathfrak{F}_0}
    W(-\beta,f^i,1) W(\beta,f_i,y)q^{L(0)-26/24} }
  {\eta(\tau)^2\Tr_{\mathfrak{F}_0} V(-\alpha,1)
    V(\alpha,y))q^{L(0)-26/24}}=24\Gamma(\tau,\nu)E(\tau,\nu)^2\,.
\end{equation}
The left hand side is the ratio of two-point functions of vector
particles and tachyons if we make an analogy with bosonic open string.
(The expression $\eta(\tau)^2$ stems from the ghost sector.) If we
replace $\Gamma(\tau,\nu)$ by the Jacobi form $\wp(\tau,\nu)$ one
obtains the elliptic genus of $X$ in the form presented in
\cite{Kaw1}:
\begin{equation}
  \mathcal{E}_X(\tau,\nu)=24\wp(\tau,\nu)E(\tau,\nu)^2\,.
\end{equation}

\begin{rem}
  If $X$ is an elliptic $K3$ surface with a section $\sigma$ and a
  fiber $f$, we may instead set $\alpha=c_1(\mathcal{O}_X(\sigma))$
  and $\beta=c_1(\mathcal{O}_X(f))$. Then $\{f_i\}$ must be a basis of
  $(\ZZ\alpha+\ZZ\beta)^\perp$.
\end{rem}

\section{A conifold and Chern-Simons theory}

As a simple application of the infinite product representation of the
string partition function, we now reproduce some earlier obtained
results on the relation between topological type IIA string near a
conifold point and the $SU(\infty)$ Chern-Simons theory on a
3-dimensional sphere $S^3$ \cite{Wit4,P,JP,GV1,GV2,GV3}.

Let us set $\xi=q_2=pq^{-1}$. Then it is expected that the limit
$\log\xi\to 0$ corresponds to the point where a conifold singularity
arises. We first set $z=0$, then in the neighborhood of this limit
there is a factor
\begin{equation} \label{conifold}
  \prod_{j=-\infty}^\infty (1-\xi y^j)^{\frac{\abs{j}}{2}}
=\exp\left(\sum_{g=0}^\infty x^{2g-2} m_g\Li_{3-2g}(\xi)\right)\,,
\end{equation}
raised to the power of $c_0(-1,0)=-2$ in the infinite product
\eqref{mainconj}.  Here we used \eqref{mone}.  Intuitively this factor
corresponds to the bound states of a $D2$-brane and $D0$-branes (with
the charge conjugation symmetry imposed) where the $D2$-brane wraps
once around the shrinking $\PP^1$ with $\log \xi$ being its
complexified K{\" a}hler parameter.

According to the fundamental work \cite{Wit2}, up to the framing
ambiguity the partition function of the Chern-Simons theory on $S^3$
with gauge group $G$ and a positive integer coupling $k$ is equal to
$S_{k\Lambda_0,k\Lambda_0}$ where $S_{k\Lambda_0,k\Lambda_0}$ is one
of the entries of the transformation matrix of the level $k$ Weyl-Kac
characters of the affine Lie algebra of $G$ under the modular
transformation $\tau\to -\frac{1}{\tau}$.  We recall that
$S_{k\Lambda_0,k\Lambda_0}$ is expressed by the classical Weyl
denominator of $G$ \cite{Kac}.  In the case of $G=SU(N)$, the
partition function $Z_W(S^3;N,k)$ can be explicitly written down
\cite{CLZ} as
\begin{equation}
  Z_W(S^3;N,k)=\left(\frac{N}{{N'}^{N-1}}\right)^{-\frac{1}{2}}\,
 \prod_{j=1}^{N-1}\left(
    2\sin \frac{\pi j}{N'}\right)^{N-j}\,,
\end{equation}
where $N'=k+N$.
It is well-known that there exists level-rank duality:
\begin{equation}
  \sqrt{N} Z_W(S^3;N,k)=\sqrt{k} Z_W(S^3;k,N)\,,
\end{equation}
from which it follows that
\begin{equation}\label{CS}
      Z_W(S^3;N,k) = \sqrt{\frac{N'}{N}}{N'}^{-\frac{N'}{4}}{\bf e}
  \left[\rho_{N',N}\right] 
\prod_{j=1}^{N'-N-1}
\left(1 - {\bf e} \left[{\textstyle \frac{N+j}{N'}}\right]
\right)^{\frac{j}{2}}
 \prod_{j=1}^{N-1}
  \left(1 - {\bf e}\left[{\textstyle
        \frac{N-j}{N'}}\right]\right)^{\frac{j}{2}}\,,
\end{equation}
where we have set
\begin{equation}
  \begin{split}
    \rho_{N',N} &= \left\{-\frac{1}{12}\left(N/N'\right)^3 +
      \frac{1}{8}\left(N/N'\right)^2 - \frac{1}{48}\right\}{N'}^2 \\ 
    & \qquad + \left\{-\frac{1}{8}(N/N') + \frac{1}{16}\right\}N'
    + \frac{1}{12}(N/N') - \frac{1}{24}\,.
\end{split}
\end{equation}

We make the identification: $\xi={\bf e}[N/N']$ and $y={\bf e}[1/N']$.
This is a familiar choice of variables when we relate the HOMFLY
polynomial of knot theory to the $SU(N)$ Chern-Simons theory. Then we
discern the infinite product \eqref{conifold} when $N' >> N>>1$.  Note
however that since the Chern-Simons theory is an open string theory,
the symmetry under $y\leftrightarrow y^{-1}$, which is peculiar to a
closed string theory, is violated in the whole expression \eqref{CS}.

Using the formulas in Appendix A, we have\footnote{For simplicity, we
  use $\Lic_r$ instead of $\Li_r$ when $r>0$.}
\begin{align}
  m_0\Lic_3(\xi)&= -\frac{(\log\xi)^2}{2}\log(\log\xi) + \zeta(3) -
  \frac{(\log\xi)^3}{12}+\sum_{k=2}^\infty
  \frac{\zeta(3-2k)}{(2k)!}(\log\xi)^{2k}\,,\\ 
  m_1\Lic_1(\xi)&=-\frac{1}{12}\log(\log\xi)-\frac{1}{24}\log\xi+
  \sum_{k=1}^\infty \frac{\zeta(1-2k)}{12(2k)!}(\log\xi)^{2k}\,,
\end{align}
and for $g\ge 2$
\begin{equation}
  \begin{split}
    m_g\Li_{3-2g}(\xi) &= \frac{m_g(2g-3)!}{(\log
      \xi)^{2g-2}}+\sum_{k=0}^\infty
    \frac{m_g\zeta(3-2g-2k)}{(2k)!}(\log \xi )^{2k}\\
    &= \frac{(-1)^{g-1}\chi_{g,0}}{(\log \xi)^{2g-2}} +
    \sum_{k=0}^\infty \frac{(-1)^{k}2\zeta(2g-2+2k)}{(2\pi)^{2g-2+2k}}
    \chi_{g,2k}\, (\log \xi )^{2k}\,.
  \end{split}
\end{equation}
These directly reproduce the behaviors of the Gromov-Witten potentials
in the vicinity of a conifold which were discussed in \cite{P,JP,GV1}
and especially in \cite{GV2,GV3}.

\section{Discussions}

In this paper we have argued that the string partition functions of
certain elliptically and $K3$ fibered Calabi-Yau 3-folds in a
particular limit should have the infinite product representation
\eqref{mainconj}.  We have used the lifting procedure of Jacobi forms
in an essential way. It was rather ironic and somewhat against the
initial impression that purely from the viewpoint of lifting, the
amount of difficulty in computing the Gromov-Witten potential
decreases as the genus $g$ increases; in fact there is no contribution
to the ``Weyl vector'' when $g>1$.

Although we cannot be too optimistic since the lifting procedure of
Jacobi forms should be useful only for the fibered Calabi-Yau 3-folds,
it is hard to resist the temptation to make a bolder conjecture: For a
general Calabi-Yau 3-fold the string partition function may be
expressed in a form that schematically looks like:
\begin{equation}\label{boldconj}
  \mathcal{Z} = \exp\left(x^{-2}F_0^{(0)} + F_1^{(0)}\right)
  \prod_{(\{n_i\},j)>0}\left(1-\prod_{i=1}^l q_i^{n_i}\,
    y^j\right)^{D(\{n_i\},j)}\,.
\end{equation}

The major portion of this paper has been devoted to an interpretation
by $D2$-$D0$ bound states. It is obvious that one of challenging but
interesting directions for further research is to place the study of
$D2$-$D0$ bound states on a mathematically rigorous footing for
general Calabi-Yau 3-folds and ask if the Gromov-Witten theory can be
totally reformulated in that picture. This, if achieved, may shed some
light on the (homological) mirror conjecture.  We have suggested in
this work that an appropriate language toward this goal may be that of
coherent systems of dimension 1. Given our success in the $K3$ case,
this approach should merit a close scrutiny.

Also it would be most desirable to find out, if any, an organizing
theory whose partition function is directly given by \eqref{mainconj}
or \eqref{boldconj}.  The theory will presumably have some flavor of
Chern-Simons theory.  Since the infinite product representations
\eqref{mainconj} or \eqref{boldconj} have strong resemblance to the
Weyl-Kac-Borcherds denominator, it seems natural to expect the
existence of some nice algebra of $D0$-, $D2$-branes. It should be
emphasized that, while the Borcherds denominators are generally
expected to be related to enumeration problems of curves or
$D2$-branes on surfaces, in the situation of this paper where fibered
Calabi-Yau 3-folds are relevant, the analogy to the Weyl-Kac-Borcherds
denominator was most evident only after we incorporate $D0$-branes in
addition to $D2$-branes. In this analogy, the Euler characteristics of
the moduli spaces of coherent systems must in an appropriate sense be
interpreted as ``root multiplicities''.  Some aspects of the algebra
of $D$-branes were studied in \cite{HM2}.  Identifying the algebra
should help in knowing the (necessarily infinite-dimensional) gauge
symmetry of the organizing theory.

Another remaining issue, which we were unable to address in this work,
is to investigate the automorphic properties of the infinite product
which we used for the string partition function.

\section*{Appendix A}
\setcounter{equation}{0}
\renewcommand{\theequation}{A.\arabic{equation}}

We define the Bernoulli numbers $B_n$
$(n=0,1,2,\ldots)$  by
\begin{equation}
  \frac{t}{e^t-1}=\sum_{n=0}^\infty B_n\, \frac{t^n}{n!}\,.
\end{equation}
Hence we have
\begin{equation}
  B_0=1,\quad B_1=-\frac{1}{2},\quad B_2=\frac{1}{6}, \quad
  B_4=-\frac{1}{30},\ \ldots
\end{equation}
and $B_{2k+1}=0$ $ (k\ge 1)$.  The values of the Riemann zeta function
at integers can sometimes be expressed in terms of the Bernoulli
numbers:
\begin{align}
  \zeta(2k)&=\frac{(2\pi)^{2k}}{2(2k)!} \abs{B_{2k}},\quad (k\ge 0)\,,\\
  \zeta(1-2k)&=-\frac{B_{2k}}{2k},\quad (k\ge 1)\,.
\end{align}

The series
\begin{equation}
  \Omega(\xi,s)=\sum_{n=1}^\infty\frac{\xi^n}{n^s}\,,\quad (\Rea
  s>1\,,\ \abs{\xi}<1)\,,
\end{equation}
and its analytic continuation frequently appeared in the past. See for
instance \cite{T} \cite{E}.  When $s=r \in \ZZ$ we will set
\begin{equation}
  \Li_r(\xi)=\Omega(\xi,r)\,.
\end{equation}
As the notation suggests, $\Li_r(\xi)$ is the usual polylogarithm when
$r>0$. On the other hand, if $r\le 0$, $\Li_r(\xi)$ is a rational
function.  The following differential-difference equation is
well-known:
\begin{equation}
  \xi \frac{\partial}{\partial \xi}\Li_r(\xi)=\Li_{r-1}(\xi)\,.
  \end{equation}
For instance, we have
\begin{align}
  \Li_1(\xi)&=-\log(1-\xi)\,,\\
  \Li_0(\xi)&=\frac{\xi}{1-\xi}\,,\\
  \Li_{-1}(\xi)&=\frac{\xi}{(1-\xi)^2}\,.
\end{align} 

If $r>0$, the polylogarithm $\Li_r(\xi)$ can be analytically continued
to a multi-valued holomorphic function on
$\PP^1\smallsetminus\{0,1,\infty \}$. As in \cite{BL,Kaw2} we introduce
$\Lic_r(\xi)$ as $\Li_r(\xi)$ modulo any $\QQ$-linear combinations of
$S_{r-1}(\xi),S_{r-2}(\xi),\ldots,S_0(\xi)$ where
$S_{r-j}(\xi):=\frac{(2\pi\sqrt{-1})^j}{(r-j)!}(\log \xi )^{r-j}$,
$(1\le j \le r)$. This is to kill off the monodromy of $\Li_r(\xi)$
and attain the effective single-valuedness.

When  $r$ is a positive integer, we have the expansion \cite{T} \cite{E}
\begin{equation}
  \Li_r(\xi)=\frac{(\log\xi)^{r-1}}{(r-1)!}
  [\psi(r)-\psi(1)-\log(-\log\xi)]+\sideset{}{'}\sum_{j=0}^\infty
  \frac{\zeta(r-j)}{j!}(\log\xi)^j\,,
\end{equation}
where $'$ stands for the omission of the case $j=r-1$ and
$\psi(t)=\frac{d}{dt}\log \Gamma(t)$.  Note that
$\psi(r)-\psi(1)=\sum_{k=1}^{r-1}\frac{1}{k}$ when $r$ is an integer
greater than $1$.  
This expansion can be simplified for $\Lic_r(\xi)$ as
\begin{equation}
  \Lic_r(\xi)=-\frac{(\log\xi)^{r-1}}{(r-1)!}\log(\log\xi)+
  \sideset{}{''}\sum_{j=0}^\infty \frac{\zeta(r-j)}{j!}(\log\xi)^j\,,
\end{equation}
where $''$ stands for the omissions of the case $j=r-1$ as well as the cases
where the summand can be expressed as $\QQ$-linear combinations of
$S_{r-1}(\xi),S_{r-2}(\xi),\ldots,S_0(\xi)$.

If instead $r$ is $0$ or a negative integer, we have \cite{T} \cite{E}
\begin{equation}
  \Li_r(\xi)=\frac{\abs{r}!}{(-\log\xi)^{\abs{r}+1}}-\sum_{j=0}^\infty
    \frac{B_{\abs{r}+j+1}}{(\abs{r}+j+1)j!}(\log\xi )^j\,.
\end{equation}
We note that the expansion \eqref{g-to-zero} can be obtained from this
by setting $r=-1$.

\section*{Appendix B}
\setcounter{equation}{0}
\renewcommand{\theequation}{B.\arabic{equation}}

Let $A$ be an abelian surface over $\CC$. Set
$A^{[\ell]}:=\Hilb^\ell_A$.  Let $\kappa_\ell:A^{[\ell]}\to A$ be the
morphism obtained by composing the Hilbert-Chow morphism
$\pi_\ell:A^{[\ell]}\to A^{(\ell)}$ and the sum map
$\sigma_\ell:A^{(\ell)}\to A$.  Beauville \cite{Bea0} showed that
$A^{\vev{\ell-1}}:=\kappa_\ell^{-1}(0)$ is an irreducible symplectic
manifold of dimension $2\ell-2$.  In particular $A^\vev{1}$ is the
Kummer surface.

Let $\tilde \phi_{-2,1}(\tau,\nu)$ be the weak Jacobi form of weight
$-2$ and index $1$ introduced in \cite{EZ}. We have a relation $\tilde
\phi_{-2,1}(\tau,\nu)=E(\tau,\nu)^2$. 
\begin{conj}
  The elliptic genus of $A^{\vev{\ell-1}}$
  is given by
  \begin{equation}\label{Kummerconj}
    \mathcal{E}_{A^{\vev{\ell-1}}}(\tau,\nu)= \ell^4\frac{\tilde
      \phi_{-2,1}\vert_{V_\ell}(\tau,\nu)}{\tilde
      \phi_{-2,1}(\tau,\nu)}.
  \end{equation}
\end{conj}

Some evidence for this conjecture is as follows. First, the elliptic
genus $\mathcal{E}_{A^{\vev{\ell-1}}}(\tau,\nu)$ must be a weak Jacobi
form of weight $0$ and index $\ell-1$ since $c_1(A^{\vev{\ell-1}})=0$
\cite{KYY}. It is easy to see that the right hand side of
\eqref{Kummerconj} has this property.  Next, one can check the
conjecture at the level of $\chi_y$ genus: Suppose that the conjecture
holds.  Then by noting $\tilde \phi_{-2,1}(\tau,\nu)=(1-y)^2/y+\dots$,
we must have
\begin{equation}
  \begin{split}
    \chi_y(A^{\vev{\ell-1}})&=y^{\ell-1}\ell^4\frac{\tilde
      \phi_{-2,1}\vert_{V_\ell}(\tau,\nu)}{\tilde
      \phi_{-2,1}(\tau,\nu)}{\Bigg\vert}_{q^0}\\ 
    &=\frac{y^{\ell-1}}{\tilde \phi_{-2,1}(\tau,\nu)} \ell
    \sum_{\substack{ad=\ell\\ a>0}} d^2 \sum_{b=0}^{d-1}\tilde
    \phi_{-2,1}\left(\frac{a\tau+b}{d},a\nu\right){\Big\vert}_{q^0}\\ 
    &=y^{\ell-1} \ell \sum_{d\mid \ell} d^3
    \frac{(1-y^{\ell/d})^2/y^{\ell/d}}{(1-y)^2/y}\\
    &=\ell \sum_{d\mid \ell} d^3(1+y+\dots+y^{\ell/d-1})^2 y^{\ell-\ell/d}\,.
  \end{split}
  \end{equation}
  However, the last expression has already appeared in \cite{GS,G}.

\begin{rem}
  The Hilbert schemes $X^{[d]}$ of a projective $K3$ surface $X$ and
  the higher order Kummer varieties $A^{\vev{\ell-1}}$ are two
  fundamental series of irreducible symplectic manifolds
  \cite{Bea0}. If the conjectures are true, the elliptic genera of
  $X^{[d]}$ and $A^{\vev{\ell-1}}$ can be expressed respectively in
  terms of $\tilde \phi_{0,1}(\tau,\nu)$ and $\tilde
  \phi_{-2,1}(\tau,\nu)$ by using the Hecke operators. Here $\tilde
  \phi_{0,1}(\tau,\nu)$ and $\tilde \phi_{-2,1}(\tau,\nu)$ are known
  \cite{EZ} to be the generators of the ring of weak Jacobi forms of
  even weight, thus they are equally fundamental in the theory of
  Jacobi forms.
\end{rem}


\begin{thebibliography}{999}

\bibitem{AK} A.B. Altman and S.L. Kleiman, {\it Compactifying the
    Picard scheme}, Adv. in Math. {\bf 35} (1980) 50--112.

\bibitem{AM} P.S. Aspinwall and D. Morrison, {\it Topological
    field theory and rational curves}, Commun. Math. Phys. {\bf 151}
  (1993) 245--262, hep-th/9110048.


\bibitem{Bar} V. Baranovsky, {\it Moduli of sheaves on surfaces and
    action of the oscillator algebra}, math/9811092.


\bibitem{Bea0} A. Beauville, {\it Vari{\' e}t{\' e}s k{\" a}hleriennes
    dont la premi{\` e}re classe de Chern est nulle}, J. Diff. Geom.
    {\bf 18} (1983) 755--782.

\bibitem{Bea} A. Beauville, {\it Counting rational curves on K3
    surfaces}, Duke Math. J. {\bf 97} (1999)
  99--108, alg-geom/9701019.

\bibitem{Beh} K. Behrend, {\it Gromov-Witten invariants in algebraic
    geometry}, Invent. Math. {\bf 127} (1997) 601--617,
  alg-geom/9601011.

\bibitem{BF} K. Behrend and  B. Fantechi, {\it The intrinsic normal cone}, 
Invent. Math. {\bf 128} (1997) 45--88,
alg-geom/9601010.

\bibitem{BM} K. Behrend and Yu. Manin, {\it Stacks of stable maps and
    Gromov-Witten invariants}, Duke. J. Math. {\bf 85} (1996) 1--60,
  alg-geom/9506023.

\bibitem{BL} A. Beilinson and A. Levin, {\it The Elliptic
    polylogarithm}, in : {\it Motives}, (U. Jannsen, S. Kleiman, J.-P.
  Serre, eds.), Proc. Symp. Pure Math. {\bf 55}, Amer. Math. Soc.,
  1994, Part 2 123--190.


\bibitem{BCOV} M. Bershadsky, S. Cecotti, H. Ooguri and C. Vafa,
  {\it Holomorphic anomalies in topological field theories}, (with an
  appendix by S. Katz), Nucl. Phys. {\bf B405} (1993) 279--304,
  hep-th/9302103;\ {\it Kodaira-Spencer theory of gravity and exact
    results for quantum string amplitudes}, Commun. Math. Phys. {\bf
    165} (1994) 311--427, hep-th/9309140.

\bibitem{BJSV} M. Bershadsky, A. Johansen, V. Sadov and C. Vafa, {\it
    Topological reduction of 4D SYM to 2D $\sigma$-models},
  Nucl. Phys. {\bf B448} (1995) 166, hep-th/9501096.


\bibitem{Bo1} R.E. Borcherds, {\it Automorphic forms on ${\rm
      O}\sb {s+2,2}({\bf R})$ and infinite products},
  Invent. Math. {\bf 120} (1995) 161--213.


\bibitem{Bo2} R.E. Borcherds, {\it Automorphic forms with
    singularities on Grassmannians}, Invent. Math. {\bf 132} (1998)
  491--562, alg-geom/9609022.

\bibitem{Brad} S. Bradlow, {\it Vortices in holomorphic line bundles
    over closed K{\" a}hler manifolds}, Comm. Math. Phys. {\bf 135}
  (1990) 1--17.

\bibitem{BDGW} S. Bradlow, G.D. Daskalopoulos, O. Garc{\' \i}a-Prada,
  R.  Wentworth, {\it Stable augmented bundles over Riemann surfaces},
  in {\it Vector bundles in algebraic geometry (Durham, 1993)},
  (N.J. Hitchin, P.E. Newstead and W.M. Oxbury, eds), London
  Math. Soc. Lecture Note Ser., 208, Cambridge Univ. Press, 1995.

\bibitem{Bre} G.E. Bredon, {\it Sheaf theory}, 2nd edition, Graduate
  Texts in Mathematics  {\bf 170}, Springer-Verlag, 1997.

\bibitem{CLZ} M. Camperi, F. Levstein and G. Zemba, {\it The large N
    limit of Chern-Simons gauge theory}, Phys. Lett. {\bf B247} (1990)
  549--554.

\bibitem{Chee} J. Cheeger, {\it Analytic torsion and the heat
    equation}, Ann. of Math. {\bf 109} (1979) 259--300.


\bibitem{Che} J. Cheah, {\it On the cohomology of Hilbert schemes of
    points}, J. Alg. Geom. {\bf 5} (1996) 479--511.


\bibitem{CG} N. Chriss and V. Ginzburg, {\it Representation Theory and
    Complex Geometry}, Birkh{\" a}user, 1997.



\bibitem{CNF} H.R.~Christiansen, C. N{\' u}\~ nez and F.A. Schaposnik,
  {\it Uniqueness of Bogomol'nyi equations and Born-Infeld like
    supersymmetric theories}, Phys. Lett. {\bf B441} (1998) 185--190,
  hep-th/9807197.



\bibitem{Del} P.~ Deligne, {\it Th\'{e}orie de Hodge III},
Inst. Hautes Etudes Sci. Publ. Math. {\bf 44}
(1974) 5--77.


\bibitem{Dijk} R. Dijkgraaf, {\it Mirror symmetry and elliptic
    curves}, in {\it The moduli space of curves}, (R. Dijkgraaf,
  C. Faber and G. van der Geer, eds.), Progr. Math. {\bf 129}, Birkh{\"
    a}user, 1995.

\bibitem{DMVV} R. Dijkgraaf, G. Moore, E. Verlinde and
  H. Verlinde, {\it Elliptic genera of symmetric products and second
    quantized strings}, Commun. Math. Phys. {\bf 185} (1997) 197--209.


\bibitem{DKL} L.J. Dixon, V.  Kaplunovsky and J.  Louis, {\it Moduli
    dependence of string loop corrections to gauge coupling
    constants}, Nucl. Phys. {\bf B355} (1991) 649--688.

\bibitem{ENF} J. Edelstein, C. N{\' u}\~ nez and F.A. Schaposnik, {\it
    Supersymmetry and Bogomol'nyi equations in the Abelian Higgs
    Model}, Phys. Lett. {\bf B329} (1994) 39--45, hep-th/9311055.


\bibitem{EZ} M. Eichler and D. Zagier, {\it The Theory of Jacobi
    Forms}, Progress in Mathematics {\bf 55}, Birkh\"{a}user, 1985.

\bibitem{EGL} G. Ellingsrud, L. G{\" o}tsche, M. Lehn, {\it On the
    cobordism class of the Hilbert scheme of a surface}, math/9904095.

\bibitem{E} A. Erdelyi, W. Magnus, F. Oberhettinger and  F.G. Tricomi,
{\it Higher transcendental functions}, Vol.1, McGraw-Hill, 1953.


\bibitem{F} C. Faber, {\it A conjectural description of the
    tautological ring of the moduli space of curves}, preprint 1996.


\bibitem{FP} C. Faber and R. Pandharipande, {\it Hodge integrals and
    Gromov-Witten theory}, math.AG/9810173.


\bibitem{FGS} B. Fantechi, L. G{\" o}ttsche, D.  van Straten, {\it
    Euler number of the compactified Jacobian and multiplicity of
    rational curves}, J. Alg. Geom. {\bf 8} (1999) 115--133,
  alg-geom/9708012.


\bibitem{FS} R. Fintushel and R.J. Stern, {\it Knots, links, and
    4-manifolds}, Invent. Math. {\bf 134} (1998) 363--400,
  dg-ga/9612014.

\bibitem{Ful} W. Fulton, {\it Intersection Theory}, 2nd edition,
  Springer-Verlag, 1998.



\bibitem{Fri} R. Friedman, {\it Algebraic surfaces and holomorphic
    vector bundles}, Springer-Verlag,  1998.


\bibitem{Gar} O. Garc{\' \i}a-Prada, {\it Seiberg-Witten invariants and
vortex equations}, in {\it Sym{\' e}tries quantiques} (A. Connes,
K. Gawedzki and J. Zinn-Justin, eds), North-Holland, 1998.


\bibitem{GP} E. Getzler and R. Pandharipande, {\it Virasoro
    constraints and the Chern classes of the Hodge bundle},
  Nucl. Phys. {\bf B530} (1998) 701--714, math.AG/9805114.



\bibitem{G} L. G{\" o}ttsche, {\it Hilbert schemes of
    zero-dimensional subschemes of smooth varieties}, Lecture Notes in
  Mathematics {\bf 1572}, Springer-Verlag, 1994.



\bibitem{GS} L. G{\" o}ttsche and W. Soergel, {\it Perverse
    sheaves and the cohomology of Hilbert schemes of smooth algebraic
    surfaces}, Math. Ann. {\bf 296} (1993) 235--245.


\bibitem{GZ} L. G{\" o}ttsche and D. Zagier, {\it Jacobi forms and the
    structure of Donaldson invariants for 4-manifolds with $b_+=1$},
  Selecta Math. (N.S.) {\bf 4} (1998) 69--115, alg-geom/9612020.


\bibitem{GV1} R. Gopakumar and C. Vafa, {\it Topological gravity as
    large N topological gauge theory}, Adv. Theor. Math. Phys. {\bf 2}
  (1998) 413--442, hep-th/9802016.

\bibitem{GV3} R. Gopakumar and C. Vafa, {\it On the gauge
    theory/geometry correspondence}, hep-th/9811131.
 

\bibitem{GV2} R. Gopakumar and C. Vafa, {\it M-theory and topological
    strings--I, II}, hep-th/9809187, 9812127.

\bibitem{GP2} T. Graber and R. Pandharipande, {\it Localization of
    virtual classes}, Invent. Math. {\bf 135} (1999) 487--518,
  alg-geom/9708001.

\bibitem{GN} V.A.~Gritsenko and V.V.~Nikulin, {\it Siegel automorphic
    form corrections of some Lorentzian Kac--Moody Lie algebras},
  Amer. J. Math. {\bf 119} (1997) 181--224, alg-geom/9504006.

\bibitem{Gro} I. Grojnowski, {\it Instantons and affine algebras I:
    the Hilbert scheme and vertex operators}, Math. Res. Letters {\bf
    3} (1996) 275--291.

\bibitem{Grot} A. Grothendieck, {\it Techniques de construction et
    th{\' e}or{\` e}mes d'existence en g{\' e}om{\' e}trie alg{\'
      e}brique IV: Les sch{\'e}mas de Hilbert}, S{\' e}minaire
  Bourbaki, 1960/61, no.221.

\bibitem{HZ} J. Harer and D. Zagier, {\it The Euler characteristic of
    the moduli space of curves}, Invent. Math. {\bf 85} (1986) 457--485.


\bibitem{HM1} J.A. Harvey and G. Moore, {\it Algebras, BPS states, and
    strings}, Nucl. Phys. {\bf B463} (1996) 315--368, hep-th/9510182.


\bibitem{HM2} J.A. Harvey and G. Moore, {\it On the algebras of BPS
    states}, Commun. Math. Phys. {\bf 197} (1998) 489--519,
  hep-th/9609017.

\bibitem{He} M. He, {\it Espaces de modules de syst{\` e}mes coh{\' e}rents},
      Internat. J. Math. {\bf 9} (1998) 545--598.


\bibitem{Hic} D. Hickerson, {\it A proof of the mock theta conjectures},
Invent. Math. {\bf 94} (1988) 639--660.


\bibitem{HST} S. Hosono, M.-H. Saito and A. Takahashi, {\it
    Holomorphic anomaly equation and BPS state counting of rational
    elliptic surface}, Adv. Theor. Math. Phys. {\bf 3} (1999)
  177--208, hep-th/9901151.



\bibitem{HS} B. Hunt and R. Schimmrigk, {\it K3-fibered Calabi-Yau
    threefolds I, the twist map}, math/9904059.

\bibitem{Huy1} D. Huybrechts, {\it Birational symplectic manifolds and
    their deformations}, J. Diff. Geom. {\bf 45} (1997) 488--513,
  alg-geom/9601015.


\bibitem{HL} D. Huybrechts and M. Lehn, {\it The geometry of moduli
    spaces of sheaves}, Vieweg, 1997.

\bibitem{Dual} M. Jacob (ed.), {\it Dual Theory}, North-Holland, 1974.


\bibitem{JP} D.P. Jatkar and B. Peeters, {\it String theory near a
    conifold singularity}, Phys. Lett. {\bf B362} (1995) 73--77,
  hep-th/9508044.


\bibitem{Kac} V.G. Kac, {\it Infinite dimensional Lie algebras}, 3rd
  edition, Cambridge University Press, 1990.

\bibitem{Kac2} V.G. Kac, {\it Vertex algebras for beginners}, 2nd
  edition,  American Mathematical
  Society, 1998.


\bibitem{KZ} M. Kaneko and D. Zagier, {\it A generalized theta
    function and quasimodular forms}, in {\it The moduli space of curves},
  (R. Dijkgraaf, C.Faber and G. van der Geer, eds.), Progr. Math. {\bf 129},
  Birkh{\" a}user 1995.

\bibitem{KKV} S. Katz, A. Klemm and C. Vafa, {\it M-theory,
    topological strings and spinning black holes}, hep-th/9910181.

\bibitem{Kaw0}  T. Kawai, {\it $N=2$ heterotic string threshold
    correction, $K3$ surface and generalized Kac-Moody superalgebra},
  Phys. Lett. {\bf B372} (1996) 59--64.

\bibitem{Kaw1} T. Kawai, {\it K3 surfaces, Igusa cusp form and string
    theory}, in {\it Topological field theory, primitive forms and
    related topics}, (M. Kashiwara, A. Matsuo, K. Saito and I. Satake,
  eds.), Progr. Math. {\bf 160}, Birkh{\" a}user 1998, hep-th/9710016.

\bibitem{Kaw2} T. Kawai, {\it String duality and enumeration of curves
    by Jacobi forms}, in {\it Integrable systems and algebraic
    geometry}, (M.-H. Saito, Y. Shimizu and K. Ueno, eds.), World
  Scientific 1998, hep-th/9804014.


\bibitem{KYY} T. Kawai, Y. Yamada and S.-K. Yang, {\it Elliptic
    genera and N=2 superconformal field theory}, Nucl. Phys. {\bf
    B414} (1994) 191--212.


\bibitem{KlZa} A. Klemm and E. Zaslow, {\it Local mirror symmetry at
    higher genus}, hep-th/9906046.




\bibitem{Kon1} M. Kontsevich, {\it Homological algebra of mirror symmetry},
 Proceedings of the International Congress of Mathematicians, Vol. 1,
  2 (Z{\" u}rich, 1994), 120--139, Birkh{\" a}user, Basel, 1995.



\bibitem{KM} M. Kontsevich and Yu. Manin, {\it Gromov-Witten
    classes, quantum cohomology, and enumerative geometry},
  Commun. Math. Phys. {\bf 164} (1994) 525--562, hep-th/9402147.


\bibitem{KM2} M. Kontsevich and Yu. Manin, {\it Relations between the
    correlators of the topological sigma-model coupled to
    gravity}, Commun. Math. Phys. {\bf 196} (1988) 385--398,
  alg-geom/9708024.


\bibitem{Leh} M.~Lehn, {\it Chern classes of tautological sheaves on
    Hilbert schemes}, math/9803091.


\bibitem{LeP} J. Le Potier, {\it Syst{\` e}mes coh{\' e}rents et
    structures de niveau}, Ast{\' e}risque {\bf 214} (1993).

\bibitem{LT} J. Li and G. Tian, {\it Virtual moduli cycles and
    Gromov-Witten invariants of algebraic varieties}, Jour, AMS {\bf
    11} (1998) 119--174, alg-geom/9602007.


\bibitem{LuTe} M. L{\" u}bke and A. Teleman, {\it The Kobayashi-Hitchin
    correspondence}, World Scientific, 1995.


\bibitem{Mac} I.G. Macdonald, {\it Symmetric products of an algebraic
    curve}, {\it Topology}, {\bf 1} (1962) 319--343.


\bibitem{M} Yu. Manin, {\it Generating functions in algebraic geometry
    and sums over trees}, in {\it The moduli space of curves}, (R.
  Dijkgraaf, C. Faber and G. van der Geer, eds.), Progr. Math. {\bf
    129}, Birkh{\" a}user 1995, alg-geom/9407005.

\bibitem{Mar} E. Markman, {\it Brill-Noether duality for moduli spaces
    of sheaves on K3 surfaces}, math/9901072.

 
\bibitem{MM} M. Mari{\~ n}o and G. Moore, {\it Counting higher genus
    curves in a Calabi-Yau manifold}, Nucl. Phys. {\bf B543} (1999)
  592--614, hep-th/9808131.


\bibitem{MM2} M. Mari{\~ n}o and G. Moore, {\it 3-manifold topology
    and the Donaldson-Witten partition function}, Nucl. Phys. {\bf
    B547} (1999) 569--598, hep-th/9811214.


\bibitem{MT} G. Meng and C. Taubes, {\it \underline{SW}=Milnor torsion},
  Math. Res. Lett. {\bf 3} (1996) 661--674.

\bibitem{MW} G. Moore and E. Witten, {\it Integration over the u-plane
    in Donaldson theory}, Adv. Theor. Math. Phys. {\bf 1} (1997)
  298--387, hep-th/9709193.


\bibitem{MS} J.W. Morgan and Z. Szab{\' o}, {\it Embedded tori in
    four-manifolds}, Topology {\bf 38} (1999) 479--496.


\bibitem{MV} D.R. Morrison and C. Vafa, {\it Compactifications of
    F-theory on Calabi--Yau threefolds -- II} Nucl. Phys. {\bf B476}
  (1996) 437--469, hep-th/9603161.

\bibitem{Mull} W. M{\" u}ller, {\it Analytic torsion and R-torsion on
    Riemannian manifolds}, Adv. in Math. {\it 28} (1978) 233--305.


\bibitem{Muk1} S. Mukai, {\it Symplectic structure of the moduli space
    of sheaves on an abelian or $K3$ surface}, Invent. Math. {\bf 77}
  (1984) 101--106.

\bibitem{Muk2} S. Mukai,
 {\it On the moduli space of bundles on $K3$
    surfaces I}, in {\it Vector bundles on algebraic varieties}, Tata
  Inst. Fund. Res., 1987.

\bibitem{Muk3} S. Mukai,
 {\it Moduli of vector bundles on $K3$
    surfaces and symplectic manifolds}, Sugaku Expositions {\bf 1}
  (1988) 139--174.



\bibitem{Mu} D. Mumford, {\it Towards an enumerative geometry of the
    moduli space of curves}, in {\it Arithmetic and geometry},
  Vol. II, 271--328, Progr. Math. {\bf 36}, Birkh{\" a}user, 1983.


\bibitem{Nak1} H. Nakajima, {\it Heisenberg algebra and 
    Hilbert schemes of points on projective surfaces}, Ann. Math. {\bf
    145} (1997) 379--388.

\bibitem{Nak2} H. Nakajima, {\it Lectures on Hilbert schemes of points
    on surfaces}, American Mathematical Society, 1999.

\bibitem{O'Grady1} K. O'Grady,
{\it The weight-two Hodge structure of moduli spaces of sheaves
on a K3 surface},
J. Alg.  Geom. {\bf 6} (1997)  599--644.

\bibitem{O'Grady2}
K. O'Grady,
{\it Desingularized moduli spaces of sheaves on a $K$3}, 
J. reine angew. Math. {\bf 512} (1999) 49--117. 

\bibitem{Pan} R.~Pandharipande, {\it Hodge integrals and degenerate
    contributions}, Commun. Math. Phys. {\bf 208} (1999) 489--506,
  math.AG/9811140.


\bibitem{P} V. Periwal, {\it Topological closed-string interpretation
    of Chern-Simons theory}, Phys. Rev. Lett. {\bf 71} (1993) 1295--1298,
  hep-th/9305115.

\bibitem{RS} D. Ray and I. Singer, {\it R-torsion and the Laplacian in
    Riemannian manifolds}, Adv. in Math. {\bf 7} (1971) 145--210.


\bibitem{T} C. Truesdell, {\it On a function which occurs in the
    theory of the structure of polymers}, Ann. of Math. {\bf 46}
  (1945) 144--157.


\bibitem{Tur} V. Turaev, {\it Reidemeister torsion in knot theory},
  Russ. Math. Surv. {\bf 41} (1986) 119--182.


\bibitem{VW} C. Vafa and E. Witten, {\it A strong coupling test for
    S-duality}, Nucl. Phys. {B431} (1994) 3--77, hep-th/9408074.

\bibitem{Wei} A. Weil, {\it Elliptic functions according to Eisenstein and
Kronecker}, Ergebnisse der Mathematik und ihrer Grenzgebiete, Band 88,
Springer-Verlag, 1976.

\bibitem{Wir} K. Wirthm{\" u}ller, {\it Root systems and Jacobi
    forms}, Compositio Math. {\bf 82} (1992)  293--354.
  

\bibitem{Wit2} E. Witten, {\it Quantum field theory and the Jones
    polynomial}, Commun. Math. Phys. {\bf 121} (1989) 351.


\bibitem{Wit3} E. Witten, {\it On quantum gauge theories in two
    dimensions}, Commun. Math. Phys. {\bf 141} (1991) 153--209.


\bibitem{Witmirror} E. Witten, {\it Mirror manifolds and topological
    field theory}, in {\it Essays on mirror manifolds}, (S.-T. Yau ed.),
  Internat. Press, 1992, hep-th/9112056


\bibitem{Wit4} E. Witten, {\it Chern-Simons gauge theory as a string
    theory}, in {\it The Floer memorial volume}, (H. Hofer, C.H.
Taubes, A. Weinstein and E. Zehnder eds.), Progr. Math. {\bf
    133}, Birkh{\" a}user 1995, hep-th/9207094.


\bibitem{Witphase} E. Witten, {\it Phases of $N=2$ theories in two
    dimensions}, Nucl. Phys. {\bf B403} (1993) 159-222,
  hep-th/9301042.

\bibitem{Witmon} E. Witten, {\it Monopoles and four-manifolds},
Math. Res. Lett. {\bf 1} (1994) 769--796, hep-th/9411102.


\bibitem{YZ} S-T. Yau and E. Zaslow, {\it BPS states, string
    duality, and nodal curves on K3}, Nucl. Phys. {\bf B471} (1996)
  503--512, hep-th/9512121.


\bibitem{Yos1} K.~Yoshioka, {\it Some examples of Mukai's reflections
    on K3 surfaces}, J. reine angew. Math. {\bf 515} (1999) 97--123.

\bibitem{Yos2} K.~Yoshioka, {\it Irreducibility of moduli spaces of
    vector bundles on K3 surfaces}, math.AG/9907001.

\bibitem{Z} D. Zagier, {\it The Bloch-Wigner-Ramakrishnan
    polylogarithm function}, Math. Ann. {\bf 286} (1990) 613--624.

\bibitem{Z2} D. Zagier, {\it Periods of modular forms and Jacobi theta
    functions}, Invent. Math. {\bf 104} (1991) 449--465.

\end{thebibliography}
\end{document}